\documentclass[a4paper,11pt]{article}
\pdfoutput=1
\usepackage{jheppub}

\usepackage{graphicx}
\usepackage{placeins}
\usepackage{url}
\usepackage{multirow}
\usepackage{paralist}

\usepackage{amssymb}
\usepackage{amsmath}
\usepackage{xspace}
\usepackage{booktabs} 

\usepackage[symbol]{footmisc}

\usepackage{lineno}
\newcommand{\GeVsq}{\ensuremath{\mathrm{GeV}^2}\xspace}
\newcommand{\GeV}{\ensuremath{\mathrm{GeV}}\xspace}

\newcommand{\Qsq}{\ensuremath{Q^{2}}\xspace}
\newcommand{\xbj}{\ensuremath{x}\xspace}

\newcommand{\Qxy}{\Qsq, \ensuremath{y} and \xbj}

\newcommand{\ptbal}{\ensuremath{p_T^{\rm bal}}}
\newcommand{\pzbal}{\ensuremath{p_z^{\rm bal}}}

\begin{document}

\title{Reconstructing the Kinematics of Deep Inelastic Scattering with Deep Learning}
\begin{flushright}
Nuclear Inst. and Methods in Physics Research, A 1025 (2022) 166164 \\
10.1016/j.nima.2021.166164 \\
\texttt{MPP-2021-174}
\end{flushright}

\affiliation[a]{Department of Physics and Astronomy, University of California, Riverside, CA 92521, USA}
\affiliation[b]{Thomas Jefferson National Accelerator Facility, Newport News, VA 23606, USA}
\affiliation[c]{Max-Planck-Institut f{\"u}r Physik, F{\"o}hringer Ring 6, 80805 M{\"u}nchen, Germany}
\affiliation[d]{Physics Division, Lawrence Berkeley National Laboratory, Berkeley, CA 94720, USA}
\affiliation[e]{Berkeley Institute for Data Science, University of California, Berkeley, CA 94720, USA}

\author[a,b]{Miguel Arratia,}
\author[c]{Daniel Britzger,}
\author[a]{Owen Long,\footnote{Corresponding author.}}
\author[d,e]{and Benjamin Nachman}

\abstract{
  We introduce a method to reconstruct the kinematics of
  neutral-current deep inelastic scattering (DIS) using a deep neural
  network (DNN). Unlike traditional methods, it exploits the full kinematic information of both
  the scattered electron and the hadronic-final state, and it accounts
  for QED radiation by identifying events with radiated photons and
  event-level momentum imbalance.
  The method is studied with simulated events at HERA and the future
  Electron-Ion Collider (EIC). We show that the DNN method outperforms all the traditional methods over the full phase space, improving resolution and reducing bias. 
  Our method has the potential to extend the kinematic reach of future
  experiments at the EIC, and thus their
  discovery potential in polarized and nuclear DIS. 
}

\maketitle

\section{Introduction}
\label{sec:intro}
The process of deep-inelastic
scattering (DIS) is governed by the four-momentum transfer squared of the
exchanged boson \Qsq, the
inelasticity $y$, and the Bjorken scaling variable
$\xbj$~\cite{ellis_stirling_webber_1996,Devenish:2004pb,ParticleDataGroup:2020ssz}.
These kinematic variables are related via the relation $\Qsq=s\xbj y$, where $s$ is the square of the center-of-mass energy.

Conservation of momentum and energy over constrain the DIS kinematics and leads to a freedom to calculate \Qsq, $y$ and \xbj from measured quantities~\cite{JB:1979,Blumlein:1990dj,Hoeger:1991wj,Bentvelsen:1992fu,Bassler:H1intnote93,Bassler:1994uq,ZEUS:1996uid,Bassler:1997tv}. 
Each of these methods has advantages and disadvantages and no single approach is optimal over the entire phase space. In addition, each method exhibits different sensitivity to QED radiative
effects, which further complicates the choice of
an optimal approach.  It is a critical time to re-examine reconstruction techniques given ongoing analyses of data from HERA and the future electron-ion colliders in the USA (EIC)~\cite{Accardi:2012qut,AbdulKhalek:2021gbh} and China (EicC)~\cite{Anderle:2021wcy}, as well as the proposed Large Hadron electron Collider (LHeC) at CERN~\cite{LHeCStudyGroup:2012zhm,LHeC:2020van}.

Machine learning is a promising tool for kinematic reconstruction in DIS because of its potential to automatically synthesize many dimensions at once.  Deep learning has been proposed for a variety of tasks in hadronic final-state (HFS, $h$) reconstruction, including particle identification~\cite{deOliveira:2018lqd,Belayneh:2019vyx,ATL-PHYS-PUB-2020-018,MicroBooNE:2020hho,Aurisano:2016jvx}, jet-energy reconstruction~\cite{CMS:2019uxx, ATL-PHYS-PUB-2018-013, ATL-PHYS-PUB-2020-001, Baldi:2020hjm, Kasieczka:2020vlh, 1910.03773}, jet tagging~\cite{deOliveira:2015xxd, CMS:2019dqq, ATLAS:2018wis, Kasieczka:2019dbj}, unfolding~\cite{Andreassen:2019cjw,Datta:2018mwd,Bellagente:2019uyp,Glazov:2017vni,Vandegar:2020yvw,Howard:2021pos,Baron:2021vvl,Andreassen:2021zzk,H1:2021wkz}, and more~\cite{2102.02770,Larkoski:2017jix,Guest:2018yhq,Albertsson:2018maf,Carleo:2019ptp,Bourilkov:2019yoi}.

We develop a method to reconstruct DIS kinematic
variables and account for QED radiation that relies on deep-neural networks (DNNs).  
This paper is organized as follows.
Section~\ref{sec:methods} briefly reviews the current methods for
kinematic reconstruction in DIS.  A DNN trained using fast simulations
of the proposed ATHENA detector at the future EIC is studied in
Section~\ref{sec:DNN} and the same approach is
demonstrated in full simulations of the H1 detector at HERA in
Section~\ref{sec:H1}.
Section~\ref{sec:calo-noise} explores the impact of additional
acceptance and resolution effects on a fast simulation in comparison
with a full detector simulation.
The paper ends with conclusions and outlook in Section~\ref{sec:conclusion}. 
Concurrently to our proposal, M. Diefenthaler et al.~\cite{Diefenthaler:2021rdj} studied the application of DNNs for the combination of the input and output variables of three reconstruction methods for $Q^2$ and $x$ in NC DIS.

\section{Basic kinematic reconstruction in DIS}
\label{sec:methods}
\begin{figure}[tb!]
    \centering
    \includegraphics[width=0.22\textwidth]{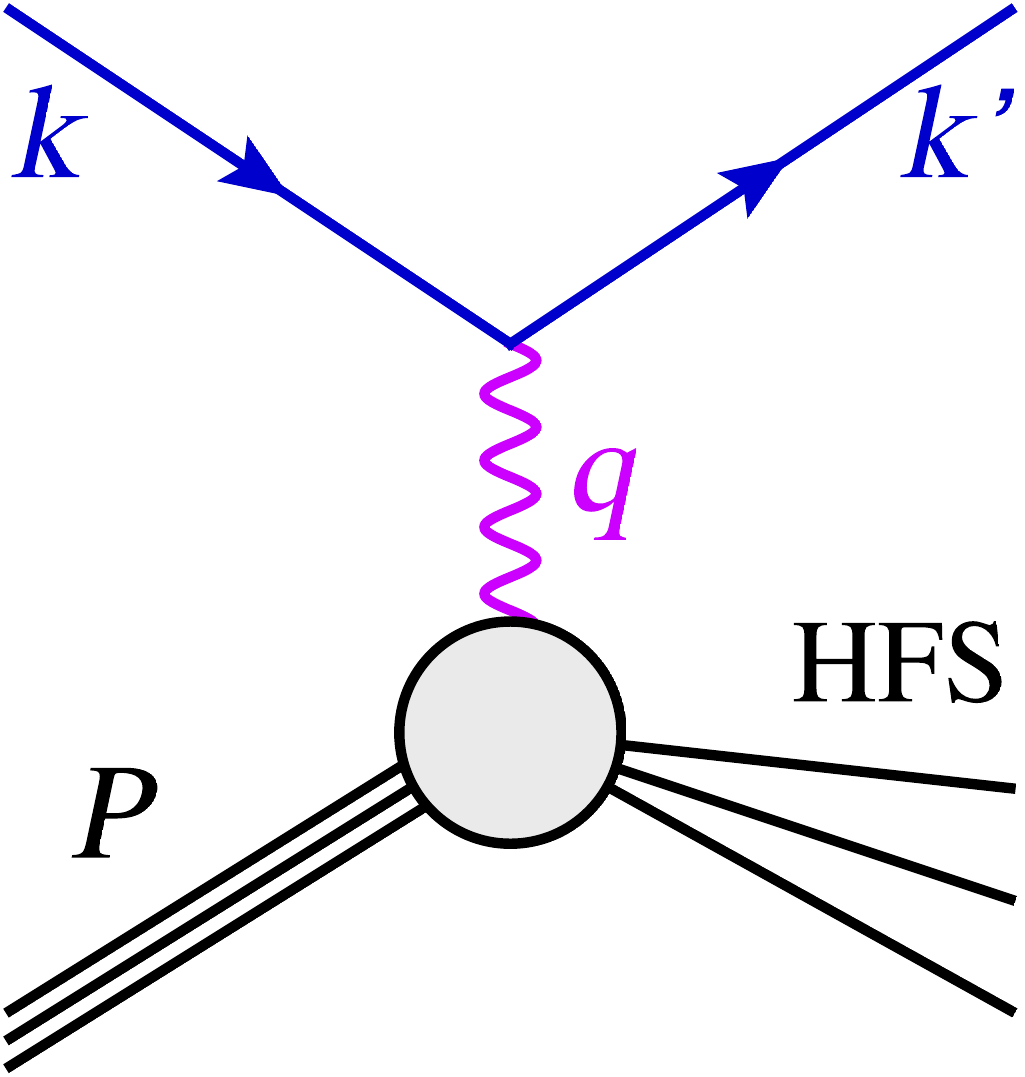}
    \hspace{0.10\textwidth}
    \includegraphics[width=0.22\textwidth]{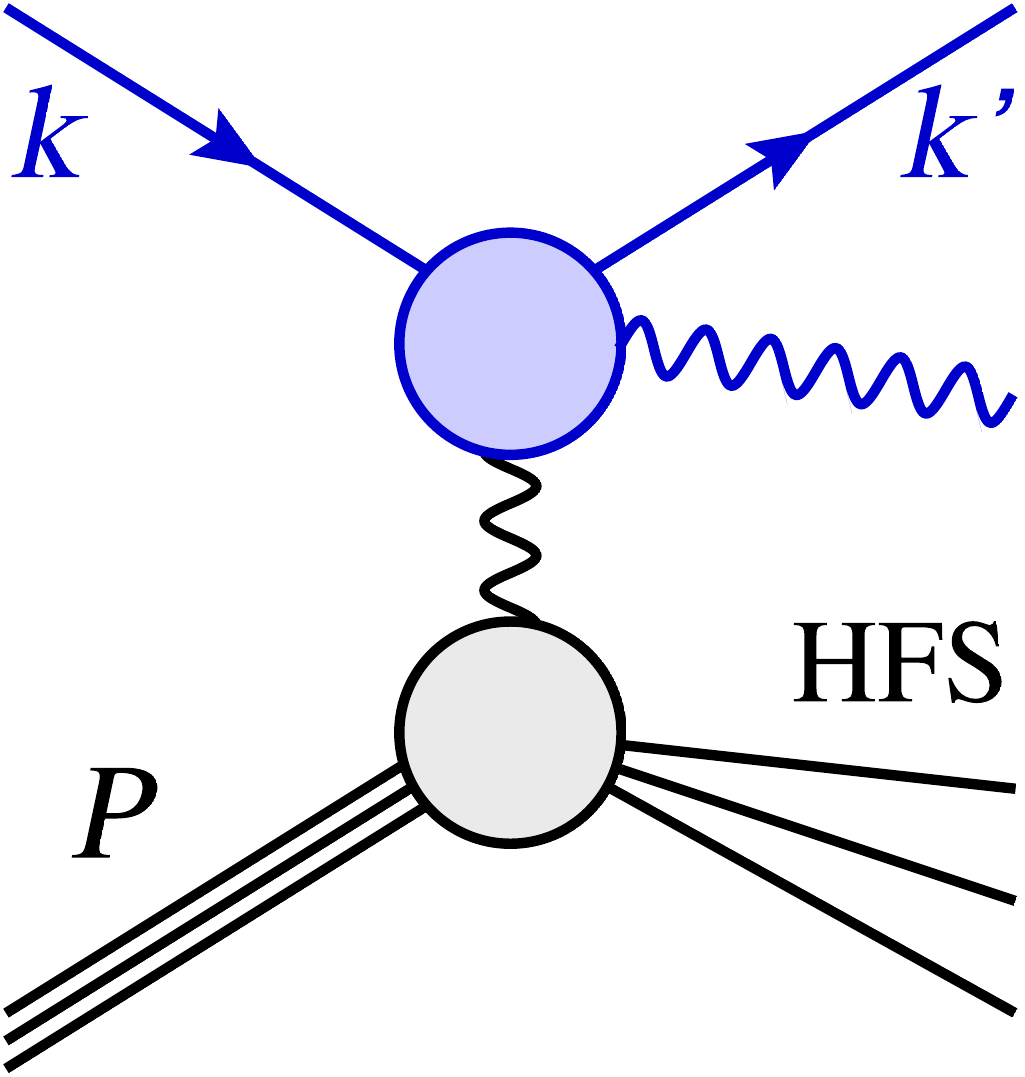}
    \hspace{0.10\textwidth}
    \includegraphics[width=0.22\textwidth]{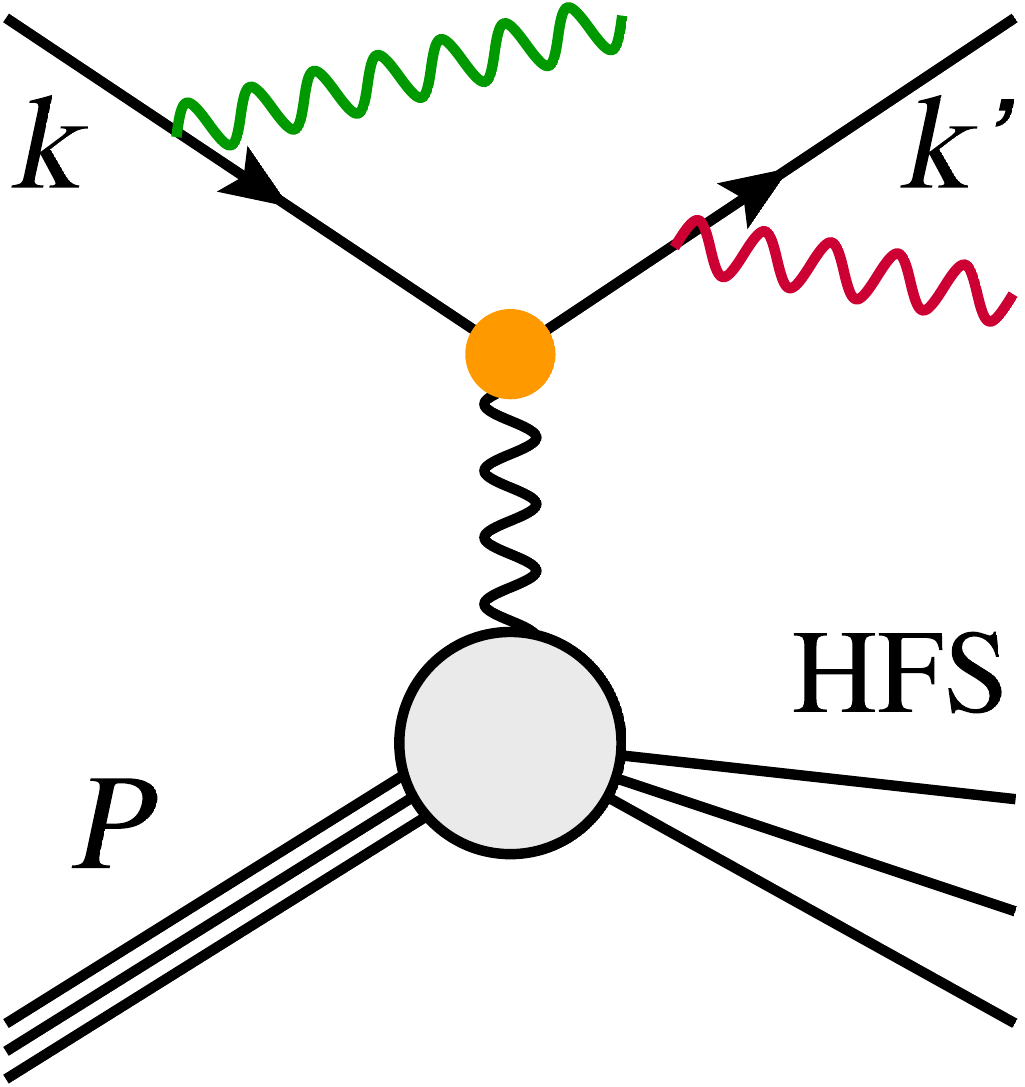}
    \caption{Illustration of $ep$ scattering.
      Left: (QED) Born-level diagram. Middle: generic illustration of a radiative leptonic tensor including higher-order QED  corrections at the lepton vertex (the box-diagrams are additive and not shown).
      Right: practical implementation of QED higher-order corrections
      in MC event generators in terms of initial state radiation (ISR, green), final state radiation (FSR, red), and an effective coupling (orange).
    }
    \label{fig:feynman}
\end{figure}

Figure~\ref{fig:feynman} illustrates the process of lepton-proton scattering\footnote{In the following, the lepton can be an
$e^-$ or $e^+$ but will generically just be referred to as an electron.}, which is defined by the incoming electron four-vector $k$,
the outgoing electron four-vector $k^\prime$, the incoming proton
four-vector $P$, and the four-vector of the HFS defined as the sum of all four-vectors that originate from the hadron vertex.
Using the photon four-vector $q=k-k^\prime$, the QED Born-level kinematics are described by:
\begin{equation}
  s = (k+P)^2\,,~~~~\Qsq = -q^2\,,~~~~y=\frac{q\cdot P}{k\cdot P}\,,~~~~\text{and}~~~~\xbj=\Qsq/(sy)\,.
\end{equation}

Due to azimuthal symmetry and ignoring mass effects, only two variables of the outgoing four-vectors are of relevance.
The usual choice is
\begin{itemize}
\item the scattered electron energy $E$ and its polar angle
  $\theta$,
  and
\item the energy-momentum balance of the HFS, $\Sigma=E_h-p_{z,h}$, and
  the inclusive angle of the HFS $\gamma$.
\end{itemize}
The HFS quantity $\Sigma$ can be calculated as the sum of
all HFS particles $\Sigma=\textstyle{\sum_i^\text{HFS}}(E_i-p_{z,i})$,
and $\gamma$ is defined using the transverse momentum of the
HFS, $T$,
through $\tan\frac{\gamma}{2}=\frac{\Sigma}{T}$. Together with the electron-beam energy $E_0$, five observables are
known, while three of them suffice to define a \emph{basic
reconstruction method} for \Qxy.
Only for \xbj one further needs the proton beam energy $E_p$.

Table~\ref{tab:methods} summarizes some of the most common reconstruction methods, which use derived quantities from the scattered electron and HFS including $p_{T,e}^2=E^2\sin^2\theta$, $p_{z,e}=E\cos\theta$, $\Sigma_e=E-p_{z,e}=E(1-\cos\theta)$, $\tan\frac{\theta}{2}=\frac{\Sigma_e}{p_{T,e}}$ and $T^2 = p_{x,h}^2 + p_{y,h}^2
    = (\textstyle{\sum_i^\text{HFS}}p_{x,i})^2 + (\textstyle{\sum_i^\text{HFS}}p_{y,i})^2
    = E^2_h\sin^2\gamma
    = \Sigma^2\cot^2\frac{\gamma}{2}
    = \Sigma(2 E_h-\Sigma)$.

\begin{table}[tbhp!]
  \footnotesize
  \centering
    \begin{tabular}{lcccc}  
      \toprule
      Method name & Observables  &  $y$   &  \Qsq  &  $x\cdot E_p$ \\
      \midrule
      Electron ($e$) &
      $[E_0$,$E$,$\theta]$ &
      {$1-\frac{\Sigma_e}{2E_0}$} &
      {$\frac{E^2\sin^2\theta}{1-y} $} &
      {$\frac{E(1+\cos\theta)}{2y}$}      \\
      \addlinespace
      Double angle (DA)~\cite{Hoeger:1991wj,Bentvelsen:1992fu} & 
      $[E_0$,$\theta$,$\gamma]$ &
      {$\frac{\tan\tfrac{\gamma}{2}}{\tan\tfrac{\gamma}{2}+\tan\tfrac{\theta}{2}}$} &
      {$4E_0^2\cot^2\tfrac{\theta}{2}(1-y)$} &
      {$\frac{\Qsq}{4E_0 y}$}
      \\
      \addlinespace
      Hadron ($h$, JB)~\cite{JB:1979} &
      $[E_0$,$\Sigma$,$\gamma]$ &  
      {$\frac{\Sigma}{2E_0}$} &
      {$\frac{T^2}{1-y}$} &
      {$\frac{\Qsq}{2\Sigma}$}
      \\
      \addlinespace
      ISigma (I$\Sigma$)~\cite{Bassler:1994uq} &
      $[E$,$\theta$,$\Sigma]$&
      {$\frac{\Sigma}{\Sigma+\Sigma_e}$} &
      {$\frac{E^2\sin^2\theta}{1-y} $} &
      {$\frac{E(1+\cos\theta)}{2y}$}
      \\
      \addlinespace
      IDA~\cite{Bentvelsen:1992fu} & 
      $[E$,$\theta$,$\gamma]$ &
      {$y_\text{DA}$} &
      {$\frac{E^2\sin^2\theta}{1-y}$} &
      {$\frac{E(1+\cos\theta)}{2y}$}
      \\

      \addlinespace
      $E_0E\Sigma$ & 
      $[E_0$,$E$,$\Sigma]$ &
      {$y_h$} &
      {$4E_0E-4E_0^2(1-y)$} &
      {$ \frac{\Qsq}{2\Sigma}$}
      \\
      \addlinespace
      $E_0\theta\Sigma$&  
      $[E_0$,$\theta$,$\Sigma]$ &
      {$y_h$} &
      {$4E_0^2\cot^2\tfrac{\theta}{2}(1-y)$} &
      {$ \frac{\Qsq}{2\Sigma}$}
      \\
      \addlinespace
      $\theta\Sigma\gamma$~\cite{Bassler:H1intnote93} &
      $[\theta$,$\Sigma$,$\gamma]$ &
      {$y_\text{DA}$} &
      {$\frac{T^2}{1-y}$} &
      {$\frac{\Qsq}{2\Sigma}$}
      \\
      \addlinespace
      Double energy (A4)~\cite{Bentvelsen:1992fu} & 
      $[E_0$,$E$,$E_h]$ & 
      $\frac{E-E_0}{(xE_p)-E_0}$ &
      $4E_0y(xE_p)$  &
      $E+E_h-E_0$
      \\
      \midrule
      $E\Sigma T$ & 
      $[E$,$\Sigma$,$T]$ & 
      $\frac{\Sigma}{\Sigma + E\pm\sqrt{E^2+T^2}}$ & 
      $\frac{T^2}{1-y}$ &
      $\frac{\Qsq}{2\Sigma}$
      \\
      \addlinespace
      $E_0ET$ & 
      $[E_0$,$E$,$T]$ & 
      $\frac{2 E_0-E \mp \sqrt{E^2-T^2}}{2E_0}$ &
      $\frac{T^2}{1-y}$  &
      $\frac{\Qsq}{4E_0 y}$
      \\
      \midrule
      Sigma ($\Sigma$)~\cite{Bassler:1994uq} &
      $[E_0$,$E$,$\Sigma$,$\theta]$ &
      {$y_{\text{I}\Sigma}$} &
      {$Q^2_{\text{I}\Sigma}$} &
      {$\frac{Q^2}{4E_0y}$} 
      \\
      \addlinespace
      $e$Sigma ($e\Sigma$)~\cite{Bassler:1994uq} &
      $[E_0$,$E$,$\Sigma$,$\theta]$ &
      $\frac{2E_0\Sigma}{(\Sigma+\Sigma_e)^2}$ &
      $2E_0E(1+\cos\theta)$ &
        $\frac{E(1+\cos\theta)(\Sigma+\Sigma_e)}{2\Sigma}$
      \\
      \bottomrule
    \end{tabular}
    \caption{
      Summary of basic reconstruction methods that employ only
      three out of five quantities:
      $E_0$ (electron-beam energy),  $E$ and $\theta$ (scattered electron
      energy and polar angle), $\Sigma$ and $\gamma$ (longitudinal energy-momentum
      balance, $\Sigma= \textstyle\sum_\text{HFS}(E_i-p_{z,i})$, and
      the inclusive angle of the HFS). 
      Alternatively, the A4 method makes use of the HFS total energy $E_h$.
      Shorthand notations are used for the longitudinal energy-momentum balance
      of the electron, $\Sigma_e$,
      and for the transverse momentum of the HFS, $T$. 
      The $E\Sigma T$ and $E_0ET$ methods are under-constrained and
      have two solutions, referring to two possible electron polar
      angles, and several more are existent when using $E_h$ (see
      Ref.~\cite{Bentvelsen:1992fu} for two examples).  
      The two bottom rows provide the equations of the $\Sigma$ and $e\Sigma$-methods, which combine quantities of
      different basic reconstruction methods, while further methods
      (like the PT (rD$\Sigma$), D$\Sigma$, r$e\Sigma$ or mixed method)  are found,
      e.g., in Ref.~\cite{Bassler:1994uq,Bassler:1997tv}. 
    }
    \label{tab:methods}
\end{table}

Each of these methods
has pros and cons, and yield good performance in limited kinematic
ranges~\cite{Bentvelsen:1992fu,Bassler:1994uq,Bassler:1997tv}.
For example, the methods that mostly rely 
on the scattered electron yield the best resolution in events with large
$y$, but their resolution on \xbj~quickly diverges at low
$y$.
In contrast, the methods that rely mostly on the HFS
variables yield better performance at low $y$, but are rather limited at
high $y$.
Consequently, the H1
and ZEUS collaborations have used different methods in different
kinematic ranges (see Refs.~\cite{Klein:2008di,H1:2015ubc} and references therein). For
example, in Refs.~\cite{H1:2009bcq,H1:2012qti,H1:2013ktq}, the electron method is
used for $y\gtrsim0.19$, while the  $\Sigma$ or $e\Sigma$ method are
used at lower $y$, and DA method is employed for calibration.

In a massless, Born-level calculation, all methods yield
equivalent results because of momentum and energy conservation ($2E_0=\Sigma+\Sigma_e$, $p_{T,e}=T$
and $\Qsq=sxy$).
However, once (real)
higher-order QED effects are considered, the various methods yield different results and the calculated quantities for \Qxy\ are not
representative for the $\gamma/Z +p$ scattering process at the hadronic vertex.

Higher-order QED effects at the lepton vertex are generically
represented as a correction to the leptonic tensor, as displayed in the middle diagram of
Figure~\ref{fig:feynman}.
Such radiative corrections include QED bremsstrahlung off the lepton,
photonic lepton-vertex corrections, self-energy contributions at the
external lepton lines, and fermionic contributions to the running of
the fine-structure constant, and additional box-diagrams representing
multi-boson exchange.
The complete first-order corrections are calculable
semi-analytically~\cite{Spiesberger:237380,Kwiatkowski:1990cx,Kwiatkowski:1990es,Blumlein:1994ii,Arbuzov:1995id}.
For an implementation in Monte-Carlo (MC) event generators, these
calculations are split by partial-fraction decomposition into 
initial-state and final-state photon 
radiation (ISR and FSR) and using effective couplings, as displayed
in the right diagram of Figure~\ref{fig:feynman}.

Two techniques are commonly applied to reduce sensitivity to QED radiation.  The first technique is to merge the FSR photons  with the electron, thus providing the four-vector linked to the exchanged boson, and the second technique is to take the ISR radiation to be collinear, which implies
that $(\Sigma+\Sigma_e)/2$ provides the incoming electron beam energy
that contributes in the interaction.
For soft and collinear FSR, the first is done implicitly also at
detector level, e.g.\ when the photon is measured in the same
calorimeter cell as the electron. ISR photons often escape undetected
through the beam hole of the detector.

In cross-section measurements, the radiative particle-level is commonly just an
intermediate step, and additionally requires the application of 
well-defined QED correction factors.
For precision measurements, these factors should be small.
Common  definitions for cross sections in DIS are:
\begin{compactitem}
  \item Structure-function measurements are made as a function of
    \Qsq\ and \xbj\ and are quoted at the `Born-level' and thus have
    to correct for all higher-order QED effects, such that the
    fine-structure constant factorizes from the calculation of the
    structure functions.  
  \item At HERA, measurements of the HFS were quoted
    as non-radiative $\gamma^* p$ cross sections, which are corrected for
    first-order QED and electroweak effects.
\end{compactitem}

A \emph{radiative cross section} can be defined by merging any radiated photon with the scattered electron that is closer to the scattered electron than to the electron beam. By specifying a single reconstruction method, a
    well-defined, meaningful and almost model-independent definition of \Qsq,
    $y$ and \xbj  is obtained. 

In the following, we will deal carefully with radiated photons at
the particle level and the detector level to determine kinematic quantities \Qsq, $y$ and \xbj, in an optimal way.
These observables can then be used for subsequent cross-section
measurement with small and well-defined QED corrections.

\section{Method}
\label{sec:DNN}

We use \textsc{TensorFlow}~\cite{tensorflow} to construct and train a DNN to estimate the DIS kinematics using both the scattered electron and the HFS. To demonstrate our methodology, we use a fast simulation of the proposed EIC experiment ATHENA using the \textsc{Delphes} package~\cite{deFavereau:2013fsa,miguel_arratia_2021_4592887}.  After presenting results for ATHENA, we will show results applying the same methodology to a full simulation of the H1 experiment~\cite{Brun:1987ma,H1:1996prr,H1:1996jzy}.
Both studies use the \textsc{Rapgap} MC generator~\cite{Jung:1993gf}, which employs routines from Refs.~\cite{Bengtsson:1987kr,Bentvelsen:1992fu,Ingelman:1996mq,Dobbs:2001ck}.
For H1, \textsc{Rapgap} version 3.1 is used, while verion 3.3 is used for the ATHENA studies.
In addition, the MC generator \textsc{Djangoh} 1.4~\cite{Charchula:1994kf} is used to test prior dependence.
Both MC generators employ the \textsc{Heracles} routines~\cite{Spiesberger:237380,Kwiatkowski:1990cx,Kwiatkowski:1990es} for the simulation of higher-order QED effects.

We restrict our study to events with $Q^2>200$~\GeVsq.
This kinematic region is well measured, since the electron is scattered into the central regions of the detector.
However, no single reconstruction method gives optimal performance over the full phase space~\cite{Bassler:1994uq}.

\subsection{Fast simulation of the ATHENA experiment}
Neutral-current DIS events
are generated with the \textsc{Rapgap}~3.3 event generator
for electron-proton scattering with beam energies of $E_0=18\,\GeV$
and $E_p=275\,\GeV$
and processed with the \textsc{Delphes} fast simulation of the ATHENA detector at the EIC.
The scattered electron is selected as the highest-$p_T$ track that satisfies the following criteria:
correct charge, $p_T > 5$~GeV, electromagnetic fraction in the calorimeter $>0.80$, and isolation  $<0.20$, where isolation is defined as the scalar $p_T$ sum of all other tracks and neutral hadrons within a cone of $\Delta R < 0.5$ around the electron direction divided by the electron $p_T$. The HFS is reconstructed from the sum of all Energy-Flow candidates (tracks, photons, and neutral
hadrons), excluding the
scattered electron and any photon candidates that is within a cone of $\Delta R<0.4$
around the scattered electron.
We require $\Sigma + \Sigma_e$ to be within $\pm4\,\GeV$ of $2 E_0$ to suppress ISR events.

\subsection{QED radiation and categorization of events }
\label{sec:QEDrad}
We introduce a practical categorization of events, which is closely related to our proposed radiative cross-section definition above:
\begin{compactitem}
\item if the radiated photon is closer to the
  electron-beam direction, it is an Initial-State Radiation event (ISR);
\item if the radiated photon is closer to the scattered-electron direction, it is a Final-State Radiation event (FSR);
  \item if no photon radiation is emitted by the generator, it is a non-radiative event (NoR).
\end{compactitem}
Within \textsc{Rapgap}, which implements first-order QED corrections, the radiated photon either branches 
off from the beam electron before it interacts with the proton, or it branches off of the scattered 
electron after interacting with the proton.  This leads to the natural interpretation of the former
as ISR and the latter as FSR, which agrees with our practical definitions in 94\% of QED
radiation events.

To define the target values of \Qxy\ for
events with QED radiation, we use the generated beam electron after radiation for ISR events and the 
generated scattered
electron prior to radiation for FSR events.  
While this definition can be considered as the \emph{true} kinematic quantities\footnote{Note that learning the true value from detector-level quantities introduces a prior dependence~\cite{ATL-PHYS-PUB-2018-013}.}, alternative definitions in terms
of particle-level observables are also possible.
For example, by applying FSR merging and using the equations of ISR insensitive reconstructions methods for calculating \Qxy.  
In our studies, these different definitions yield indistinguishable
results. The generator-level quantities are generically denoted with a subscript `gen' in the following.

\subsection{DNN inputs}
For our main goal of determining \Qsq, $y$ and \xbj, the task of the
DNN is in fact two-fold.
In the absence of QED radiation, the task of the DNN would
be to learn to compute \Qsq, $y$ and \xbj from the input quantities subject to finite-resolution and acceptance effects.
In events with QED radiation, the DNN needs to learn to quantify
the extent of QED radiation and account for it while calculating \Qxy. 
Hence, the regression DNN must learn to treat ISR, FSR, and
NoR events separately in order to achieve optimal
performance. This is a key feature of our DNN method, which is absent in traditional methods. 

We define the following variables to characterize the strength of QED radiation in the event:
\begin{equation}
    \ptbal \ = \ 1 - \frac{ p_{T,e} } { T } \ = \ 1 - \frac{\Sigma_e\tan\frac{\gamma}{2}}{\Sigma\tan\frac{\theta}{2}}
    \ \ \ ~~\text{and}~~  \ \ \ 
    \pzbal \ = \ 1 - \frac{ \Sigma_e + \Sigma } {2 \ E_0}\,.
\end{equation}
When calculating them at particle level, both quantities  \ptbal~and \pzbal, are zero for events with no QED radiation, but positive for events with FSR and ISR, respectively. 
The \ptbal\ (\pzbal\ ) value indicates the strength of FSR (ISR).

The following observables that help indicate QED radiation in the
event are included as inputs to the regression DNN for \Qsq, $y$ and \xbj:
\begin{itemize}
    \item The values of \ptbal~and \pzbal.

    \item The energy, $\eta$, and $\Delta\phi$ of the reconstructed photon in the event that is closest to the electron-beam direction, where $\Delta\phi$ is with respect to the scattered electron.
    \item The sum ECAL energy within a cone of $\Delta R<0.4$ around the scattered electron divided by the 
            scattered-electron track momentum.
    \item The number of ECAL clusters within a cone of $\Delta R<0.4$ around the scattered electron.
\end{itemize}
These seven observables are combined with the following eight:
\begin{itemize}
    \item Scattered-electron quantities $p_{T,e}$, $p_{z,e}$ and $E$.
    \item HFS four-vector quantities $T$, $p_{z,h}$ and $E_h$.
    \item $\Delta \phi(e,h)$ between the scattered electron and the HFS momentum vector.
    \item The difference $\Sigma_e - \Sigma$.
\end{itemize}

The transverse momenta of the scattered electron and the HFS are highly correlated.  We replace the pair of $p_T$ values with the difference and the sum of the $p_T$ values, which removes the correlation and aids the DNN training.  
The sum $\Sigma_e + \Sigma$ appears in the definition of \pzbal\ and is sharply peaked at twice the electron-beam energy, making the $\Sigma$ values anti-correlated; hence, we include the orthogonal combination  $\Sigma_e - \Sigma$ in addition.

\subsection{DNNs for the classification or quantification of QED radiation}

To determine the ability of the DNN to identify and quantify QED radiation, we investigated two approaches: a classification network and a regression network for \ptbal~and~\pzbal.  Both DNNs differ only in the activation function for the final layer, the learning targets, and the loss function for the training.

We followed a heuristic approach in designing the DNN, guided by prior experience.
We chose to have the number of nodes per hidden layer grow in the first layers, reach a peak size, and then fall at a rate that is symmetric about the peak layer.  
Several trial configurations were tested, each with a different number of hidden layers and/or number of nodes per layer, though we did not perform a thorough scan.  
We also tried a few sets of optimizer hyperparameters before finding values that gave good results in a reasonable amount of training time.  
Various standard activation functions were tested with no appreciable differences in the results.  
This basic DNN design is effective for a variety of tasks, including classification and regression, as we explain below.  
The same design works well for both the ATHENA fast simulation and the H1 full simulation.  
Our design explorations were terminated once we achieved DIS reconstruction performance that exceeded the conventional methods. 
We have not carried out thorough hyperparameter scans, so further improvements may be possible.

The QED-classification DNN consists of a sequential network with 8 layers.  The 15 DNN inputs, defined in the previous section, are transformed prior to training to have zero mean and unit RMS using the {\tt sklearn} {\tt StandardScaler}~\cite{scikit-learn}.
The learning targets are three binary (0 or 1) state variables to tag the events as ISR, FSR, or NoR.
The activation function is a rectified linear unit (\texttt{relu}) for the first layer, scaled exponential linear unit ({\tt selu})~\cite{NIPS2017_5d44ee6f} for the middle layers, and the {\tt softmax} function for the final output layer.  
The numbers of nodes per layer are 64, 128, 512, 1024, 512, 128, and 64, with 3 outputs in the final layer.  
The outputs are three numbers, each between 0 and 1, that sum up to 1.
The loss function for the training is categorical cross entropy.
The training is performed with the {\tt Adam} optimizer~\cite{kingma2017adam} with a learning rate of $10^{-4}$.  
The total number of parameters of the DNN model is 1,199,555.
The training and validation are performed using over 28 million
simulated events with half for training and half for validation.  
The batch size for the training is 128.  
The training terminates after 38 epochs, finding no further improvement in the validation loss function.

\begin{figure}[tbh!]
    \centering
    \includegraphics[width=0.95\textwidth]{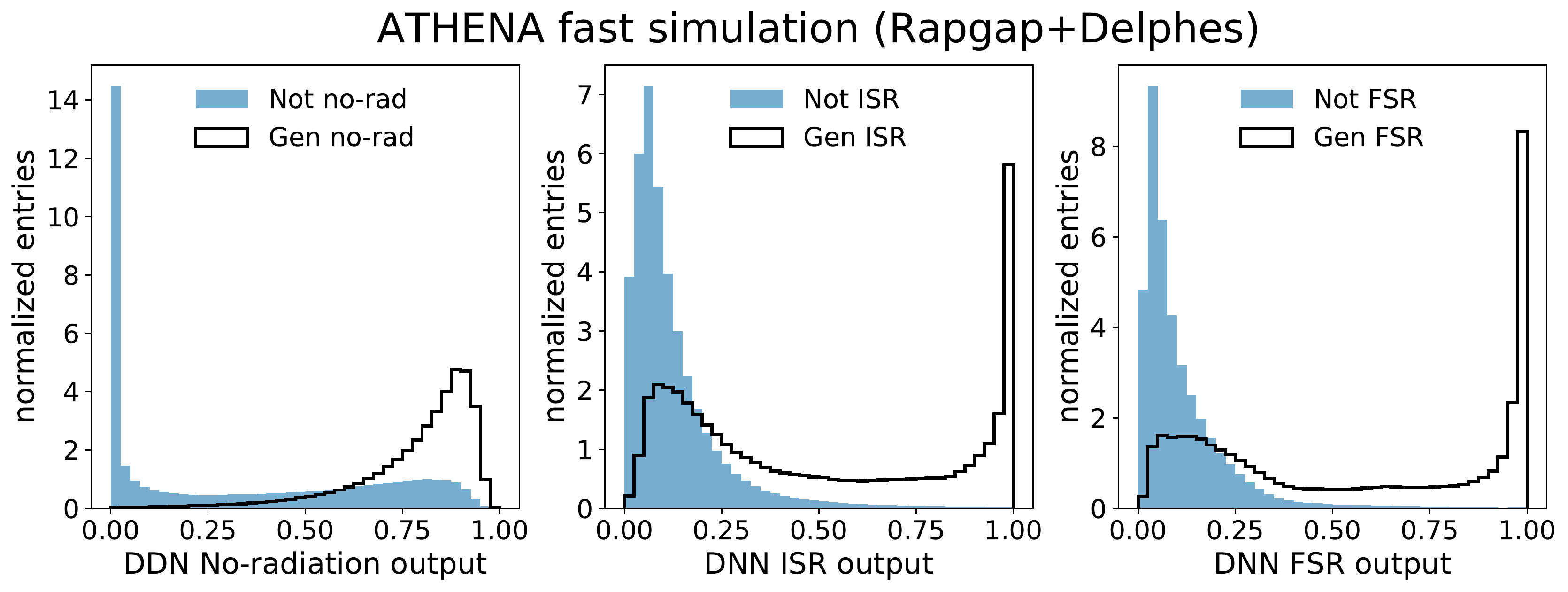}
    \caption{Distributions of the QED-classification DNN predictions for the ATHENA fast simulation. The distributions are normalized to equal area.}
    \label{fig:classifier-output}
\end{figure}

Figure~\ref{fig:classifier-output} shows distributions for the three outputs of the QED-classification DNN.  Some events have clear evidence of QED radiation and are strongly identified.
Some degree of mis-classification is to be expected, since neither soft ISR/FSR, nor collinear FSR induce a measurable signal in the detector.

In the second study, the QED regression for \ptbal~and~\pzbal, the DNN has the same structure as the QED-classification DNN except for the following differences.  
The activation function is the {\tt linear} function for the final output layer, which has 2 nodes instead of 3.  
The learning targets are the particle-level values of \ptbal~and~\pzbal, which are transformed to have zero mean and unit RMS prior to training.
The loss function for the training is \texttt{Huber} loss~\cite{10.1214/aoms/1177703732} with a transition between quadratic and linear loss at $\pm 0.01$.  
The batch size for the training is 1024.
The training and validation are done using the same sample as for the QED-classification DNN.

\begin{figure}[tbh!]
    \centering
    \includegraphics[width=0.87\textwidth]{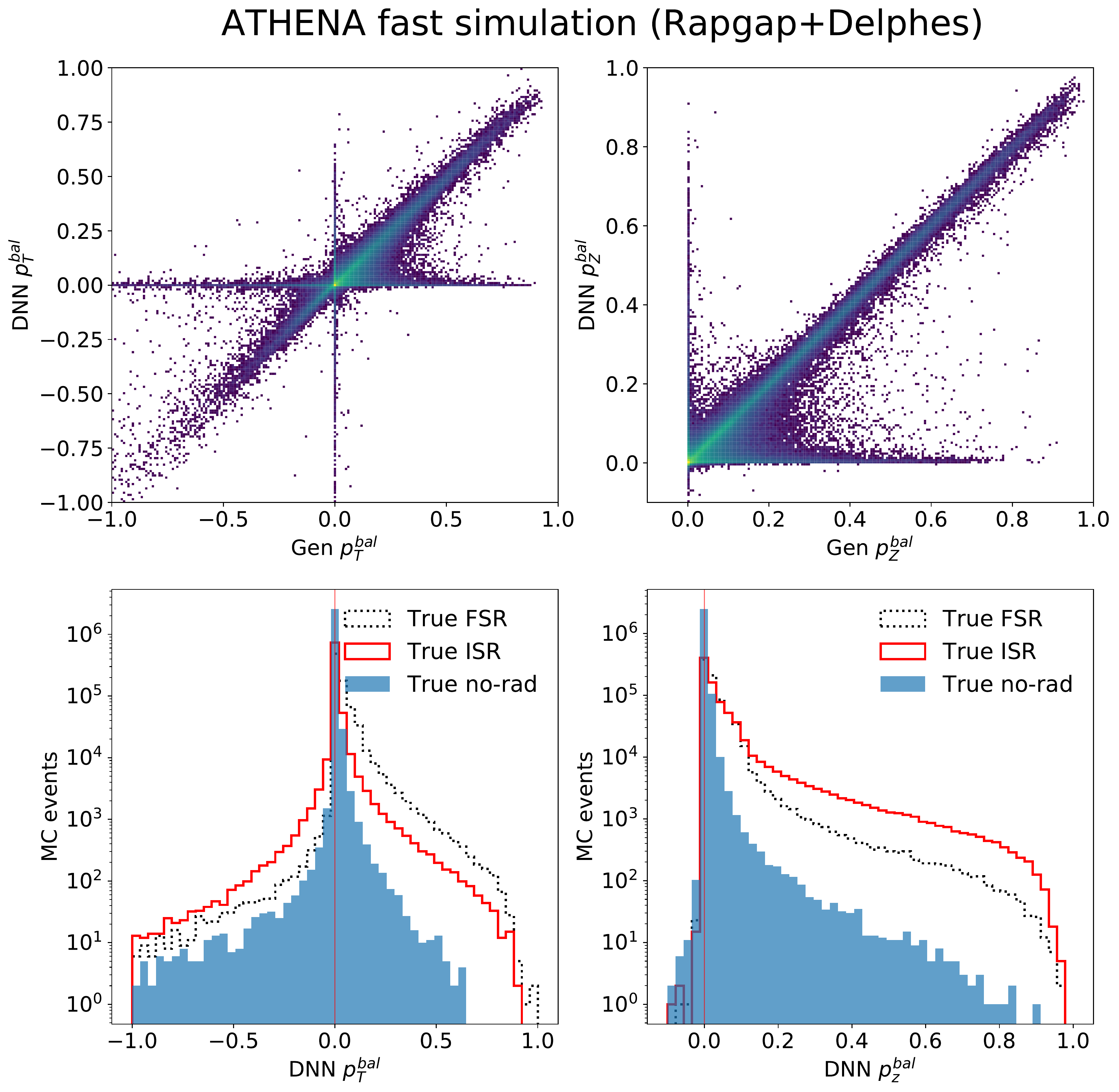}
    \caption{Distributions of the QED-regression DNN predictions and particle-level values of \ptbal\ and \pzbal\ for the ATHENA fast simulation.}
    \label{fig:ptbal-pzbal}
\end{figure}

Figure~\ref{fig:ptbal-pzbal} shows distributions of the DNN predictions for \ptbal~and~\pzbal, separately for ISR, FSR, and NoR events, as well as scatter plots of the predicted values vs the particle-level values.  The predicted \ptbal~distribution for FSR events is  shifted to positive values for FSR events, while the predicted \pzbal~distribution is  shifted to positive values for ISR events, as expected.  For many events, the DNN is able to accurately estimate both \ptbal~and~\pzbal.  There are also QED-radiation events where the prediction is zero, which correspond to cases where the radiated photon is either out of acceptance or not clearly identified.

\subsection[Regression DNN for DIS kinematic variables \Qxy]{\boldmath Regression DNN for DIS kinematic variables \Qxy} 
We estimate \Qxy using a regression DNN that has a similar structure as the QED
regression DNN described above, except for the final output layer that has
three nodes corresponding to the target variables \Qxy.
The learning rate is $10^{-5}$ and the batch size is 1024.  
Since the distributions of  \Qxy\ are approximately exponential, the DNN is trained to predict the logarithm of each
variable. The training and validation are performed using over 28 million
events, using half for validation and half for training.

As a pilot study, we consider three different choices for the
input variables to the DNN:
\begin{compactitem}
    \item add the three QED-classification DNN outputs (FSR, ISR, NoR) as inputs, in addition to the 15 variables. 
    \item add the QED-regression DNN predictions \ptbal~and~\pzbal\ as
inputs, in addition to the 15 variables. 
\item use the same 15 inputs as in the QED-classification and regression DNNs.
\end{compactitem}
We found that the results of these three approaches are essentially the same.  This suggests that the QED-classification and the QED-regression DNNs do not provide any additional information beyond what the regression DNN for \Qxy learned.  In light of this, we chose the simplest (third) option, which uses only the original 15 variables as inputs.
The total number of parameters for the DNN model is 1,199,619.
The training terminates after 103 epochs, finding no further improvement in the validation loss function.

\subsection{Benchmark of the DNN reconstruction vs. standard methods}
\label{sec:Result}
\begin{figure}[h]
    \centering
    \includegraphics[width=0.80\textwidth]{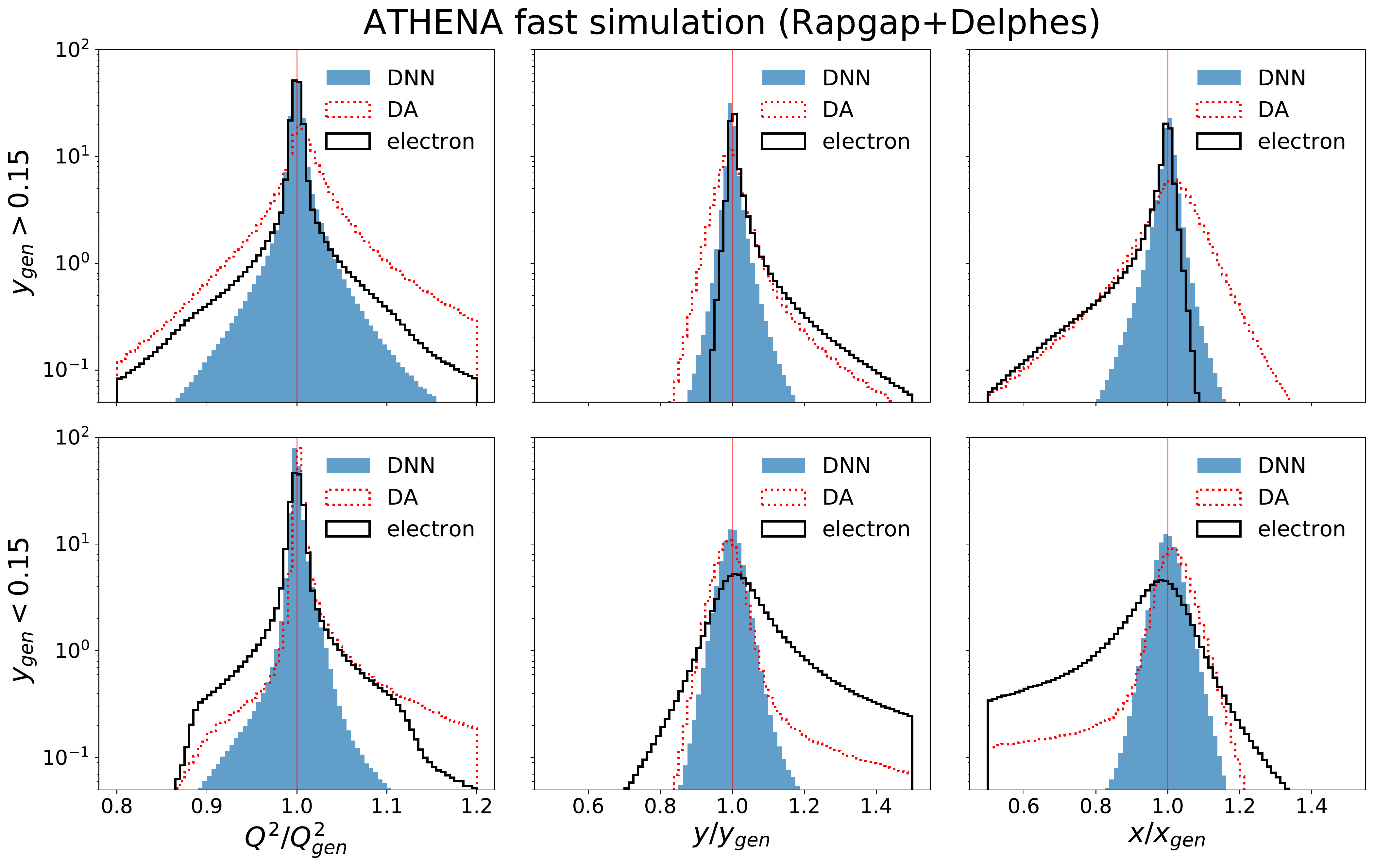}
    \caption{Resolution for $Q^2$ (left), $y$ (middle), and $x$ (right) for the DNN, electron, and double-angle (DA) methods for the fast simulation of the ATHENA experiment. The top (bottom) row is for events with $y_{\rm gen}>0.15$ ($y_{\rm gen}<0.15$).  All distributions are normalized to the same area.}
    \label{fig:athena-xyQ2-resolution_main}
\end{figure}

Figure~\ref{fig:athena-xyQ2-resolution_main} shows the results for the DNN and traditional methods using events with
{$Q^2>200~{\rm GeV}^2$} and two $y$ regions, obtained with the ATHENA fast simulation.
The resolutions from the DNN exhibit a peak at unity and
mostly Gaussian-like tails; in contrast, classical methods yield a peak but larger tails caused by their limited use of the
reconstructed quantities and the presence of ISR or FSR.

\begin{figure}[tbh!]
    \centering
        {\large \fontfamily{lmss}\selectfont ATHENA fast simulation (Rapgap+Delphes)} \\
    \vspace{1mm} 
    \includegraphics[width=0.32\textwidth]{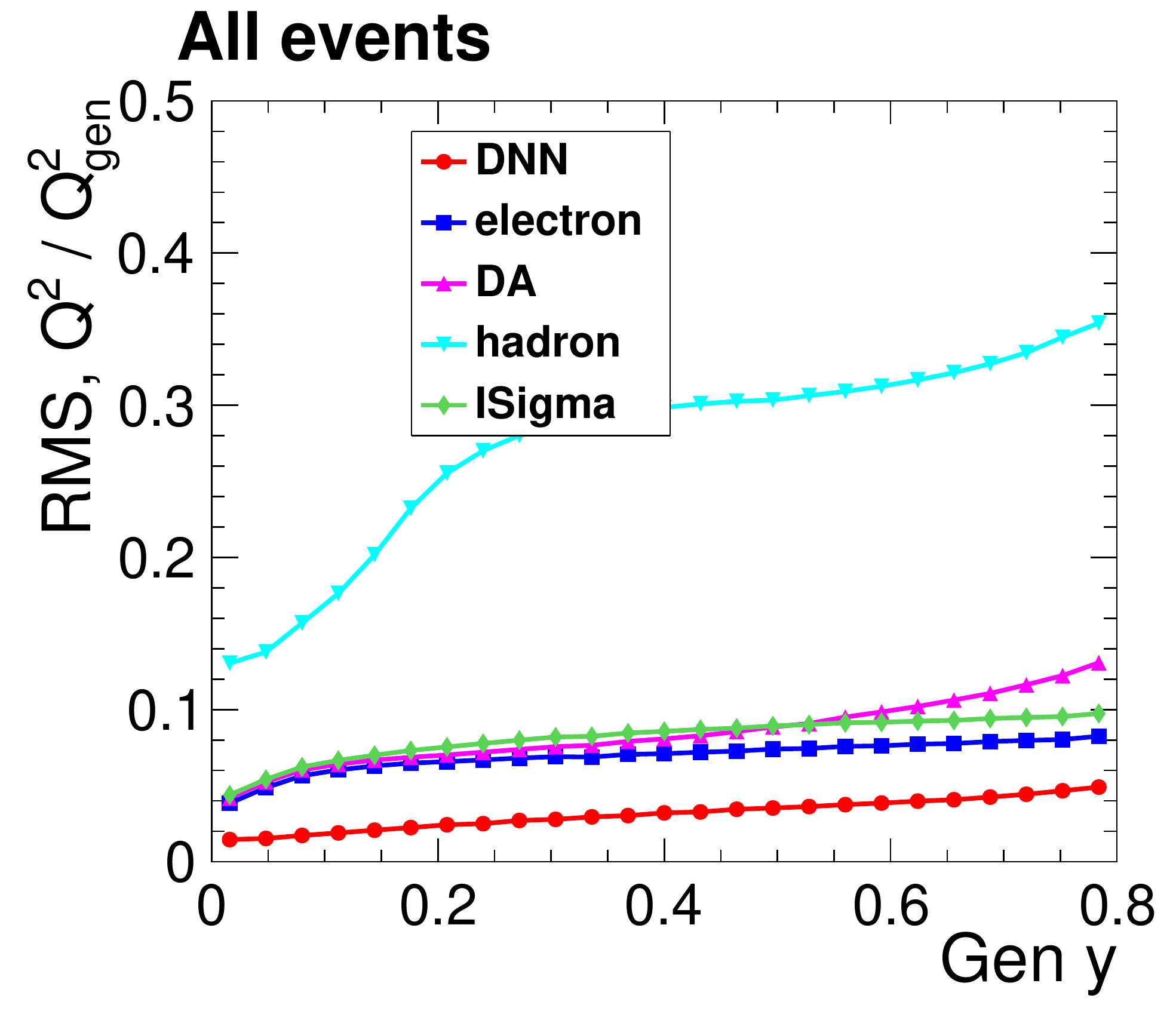}
        \includegraphics[width=0.32\textwidth]{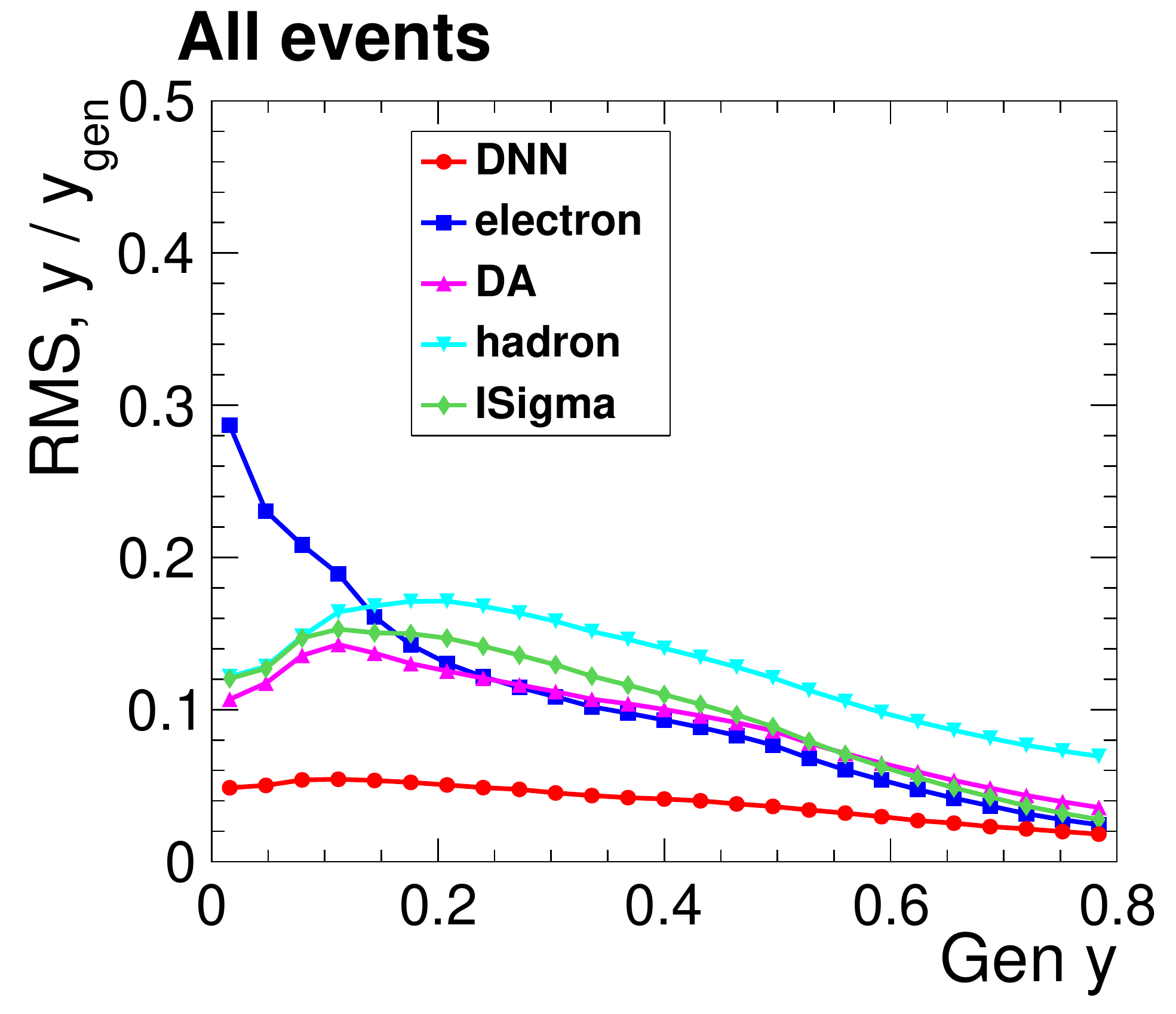}   
            \includegraphics[width=0.32\textwidth]{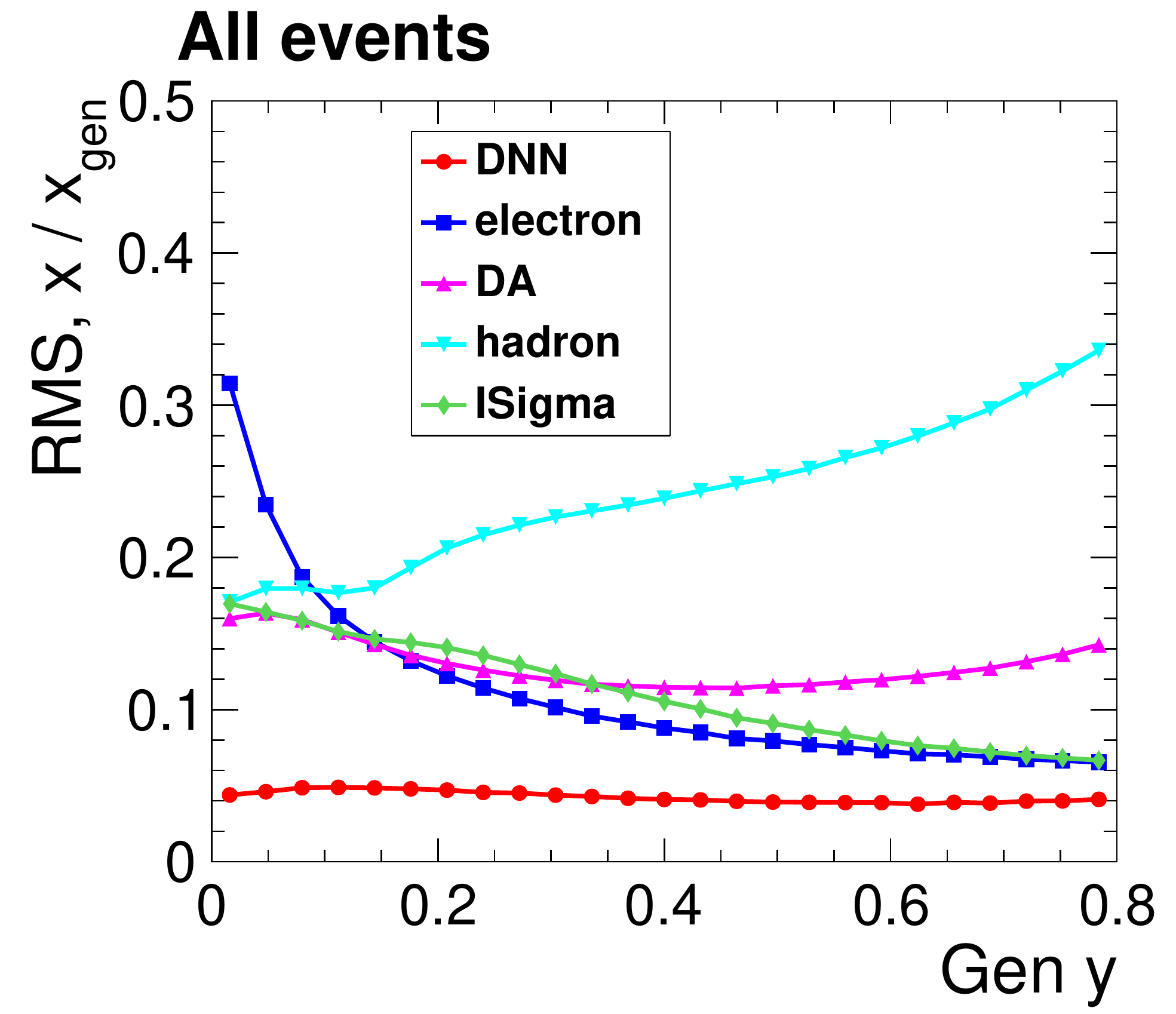} \\
    \includegraphics[width=0.32\textwidth]{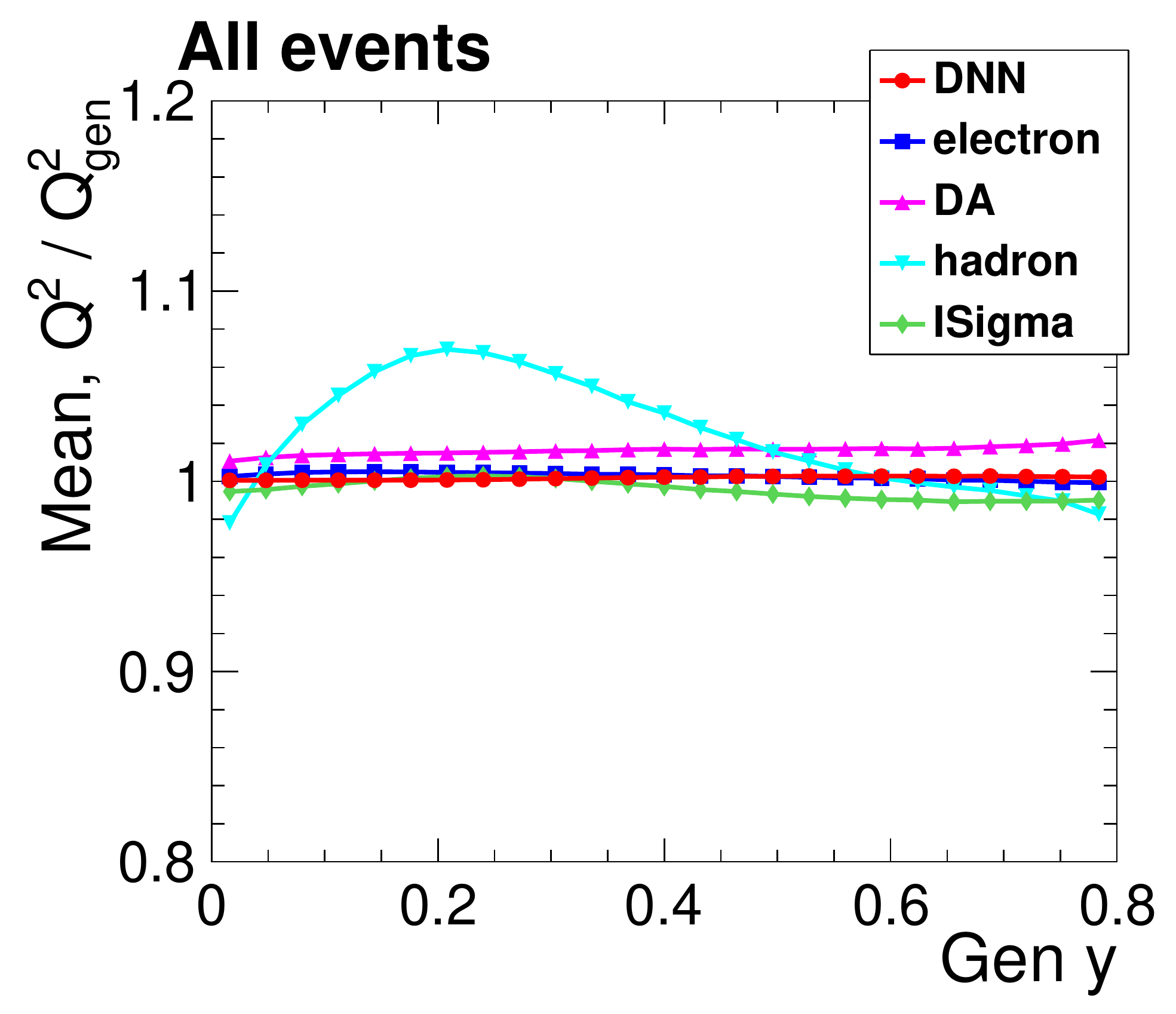}
    \includegraphics[width=0.32\textwidth]{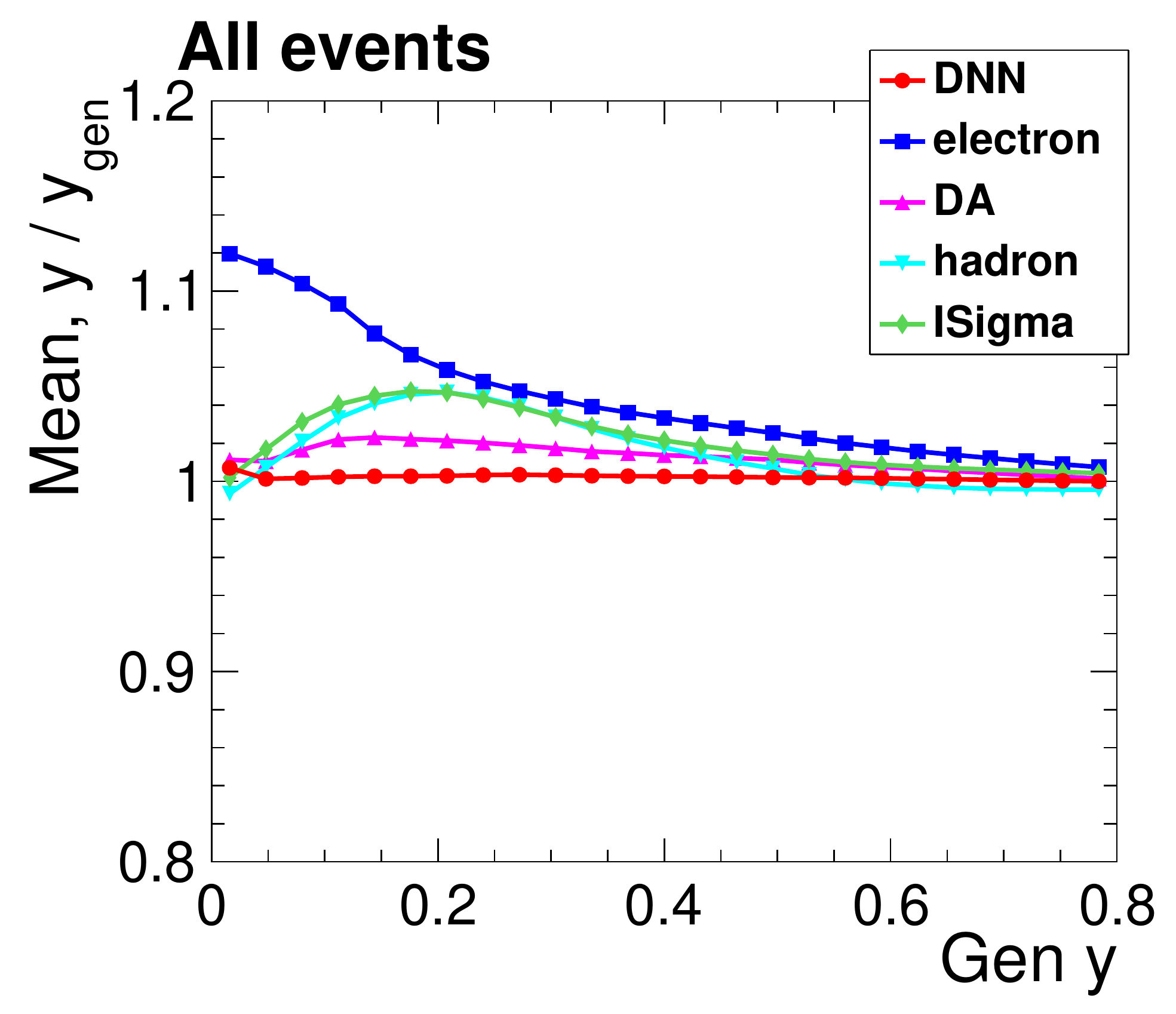}    
    \includegraphics[width=0.32\textwidth]{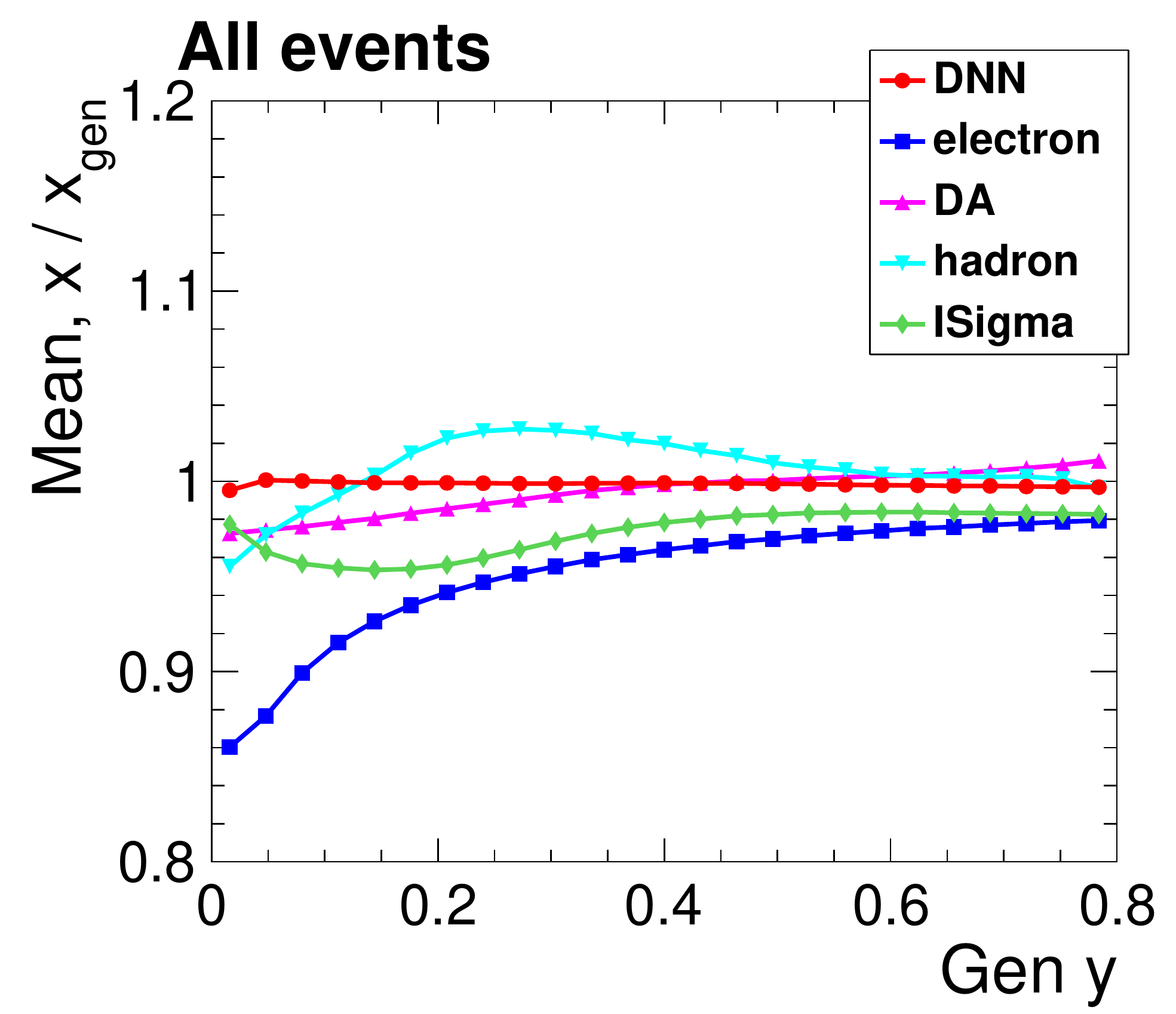} \\    
    \caption{
    Resolution on the reconstruction of $Q^2$ (left), $y$ (middle),
    and $x$ (right) as a function of the generated $y$ for the fast simulation of ATHENA.
    The top (bottom) row shows the RMS (mean) of the measured-over-generated distribution
    as a function of generated $y_\text{gen}$.  
    The RMS and mean are calculated using events with the measured-over-generated ratio within the interval 0 to 2.
    }
    \label{fig:athena-xyQ2-resolution-vs-y}
\end{figure}
The properties of the resolutions, their mean and RMS, are displayed in
Figure~\ref{fig:athena-xyQ2-resolution-vs-y} for small intervals of
$y$~\footnote{More detailed representations of the resolutions for the different methods are shown in Appendix~\ref{sec:additionalfigsAthena}.}. The DNN reconstruction has the smallest RMS among
all methods, for all three kinematic variables
and all $y$ intervals. Also, the mean distributions are unbiased for \Qxy\ for all $y$
intervals, while the classical methods exhibit large biases.

\begin{figure}[tbh!]
    \centering
        {\large \fontfamily{lmss}\selectfont ATHENA fast simulation (Rapgap+Delphes)} \\
    \vspace{2mm} 
    \includegraphics[width=0.32\textwidth]{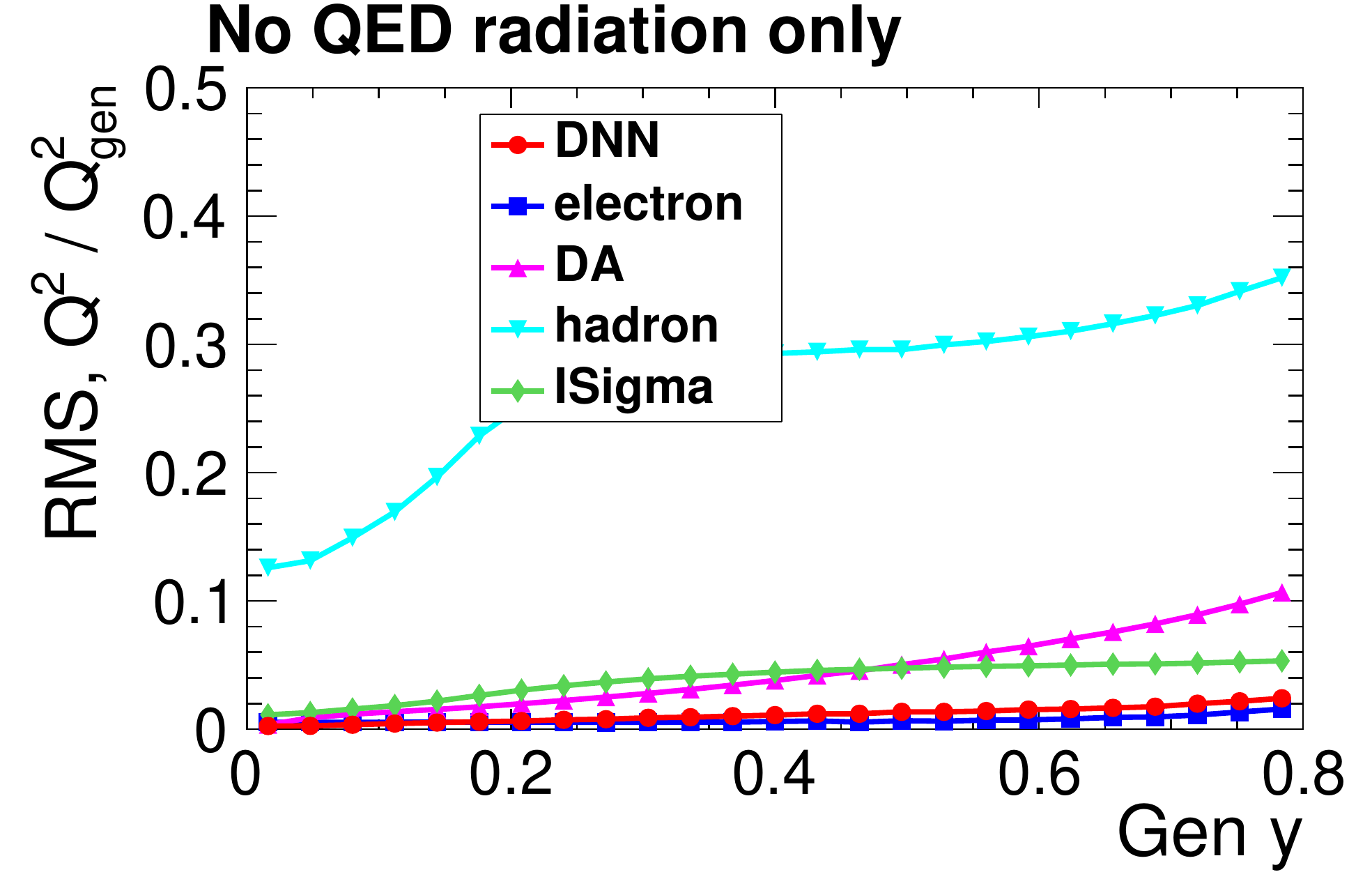}
    \includegraphics[width=0.32\textwidth]{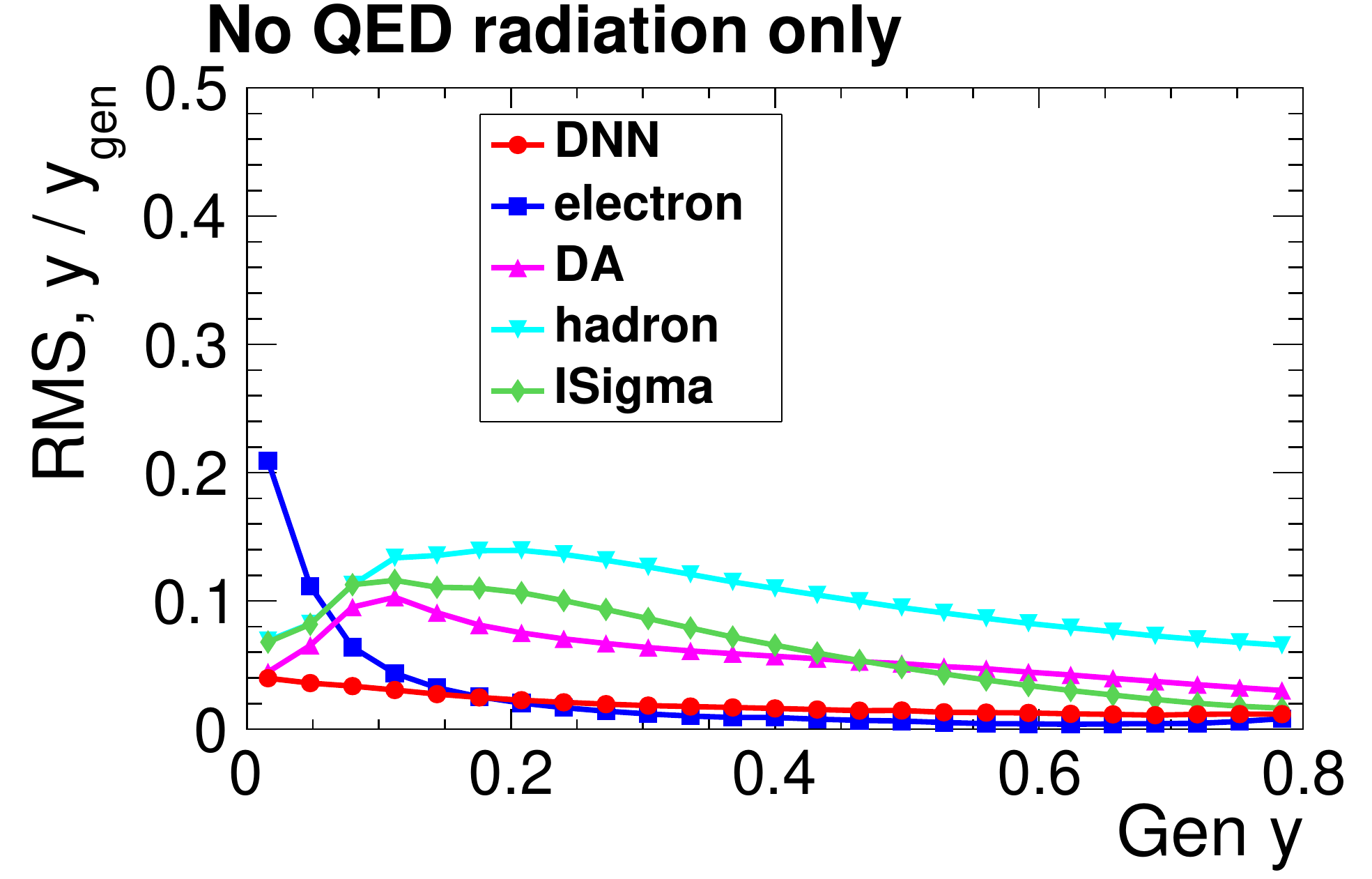}
    \includegraphics[width=0.32\textwidth]{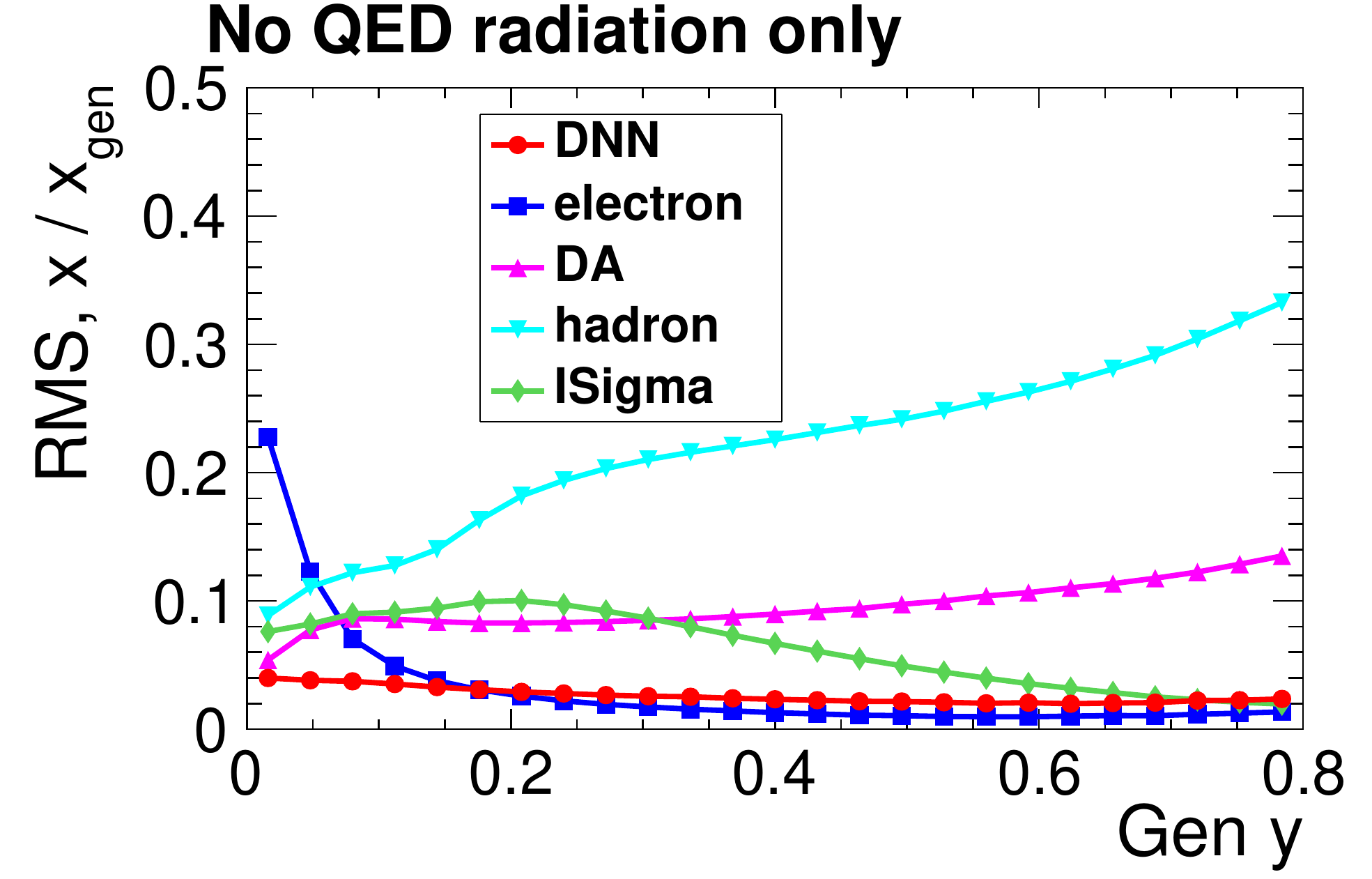}
    \includegraphics[width=0.32\textwidth]{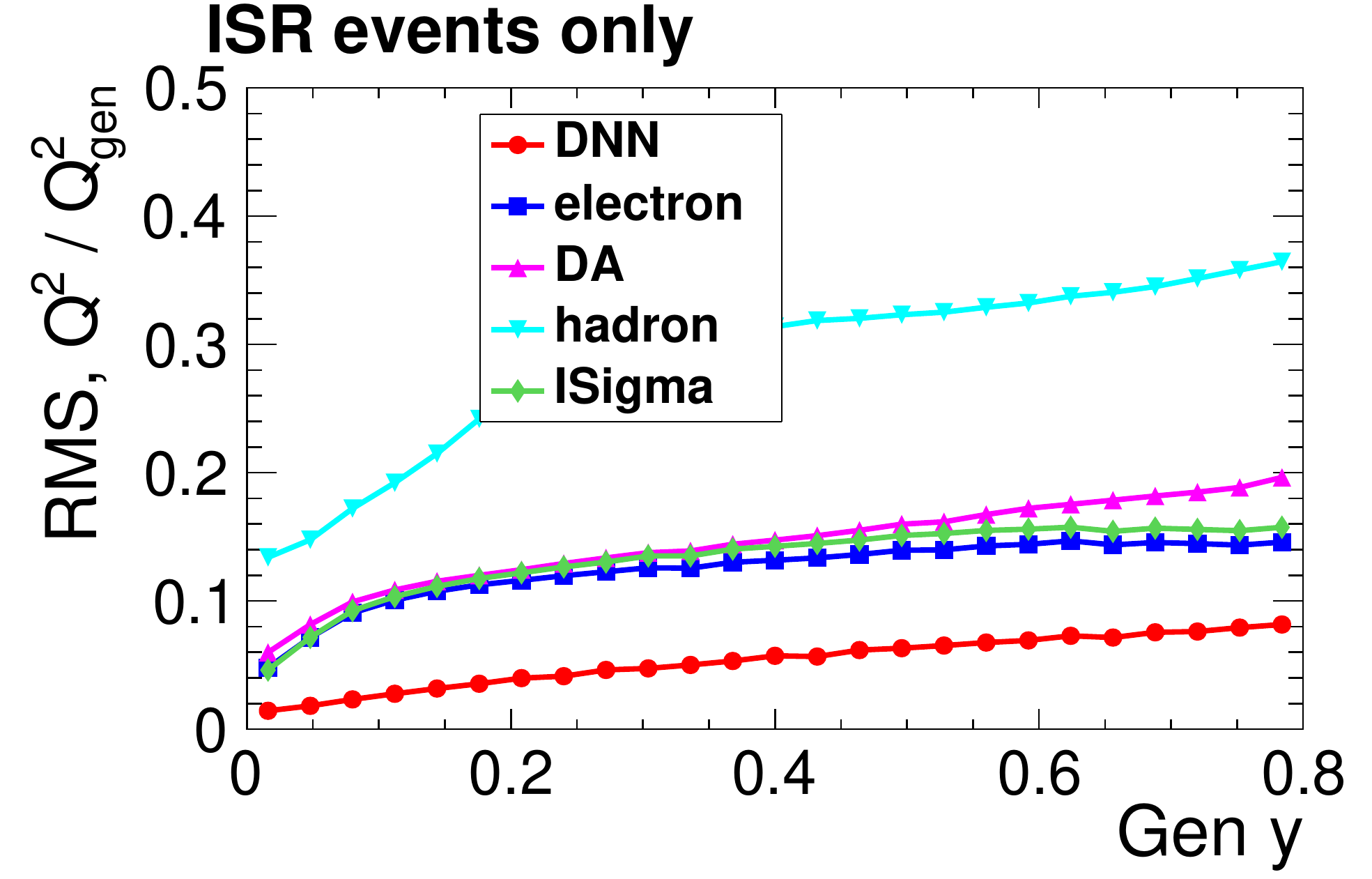}
    \includegraphics[width=0.32\textwidth]{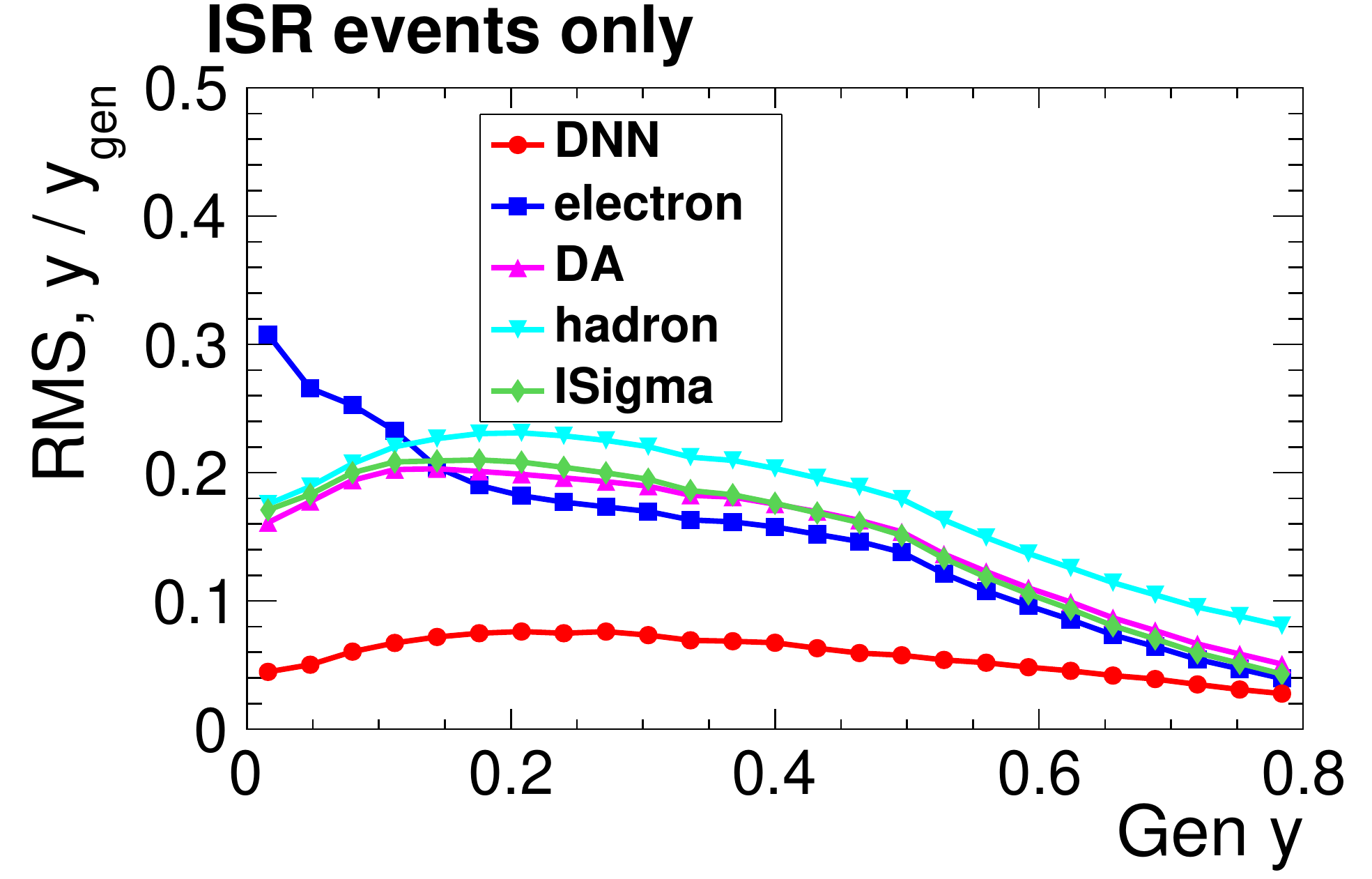}
    \includegraphics[width=0.32\textwidth]{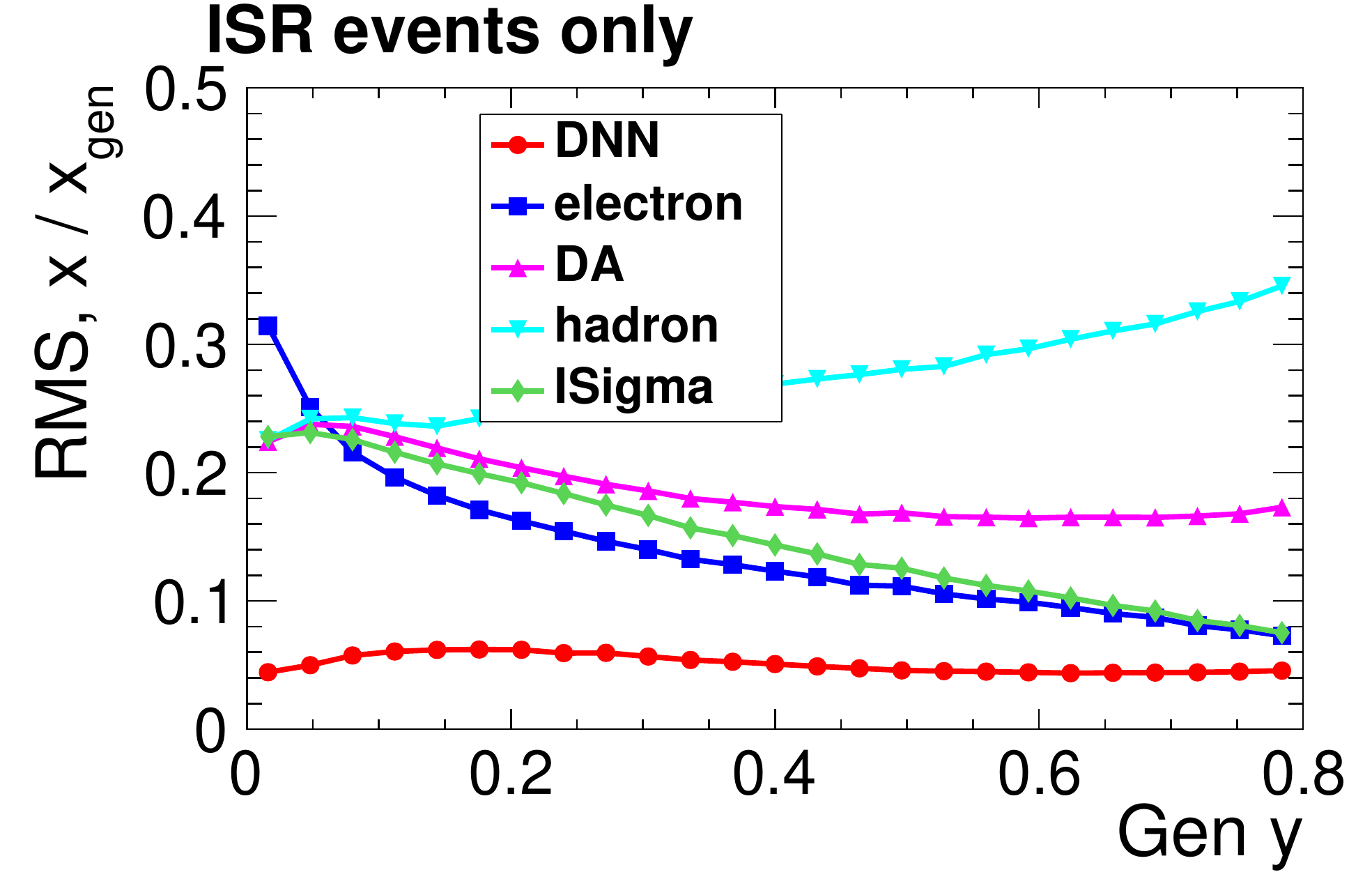}
    \includegraphics[width=0.32\textwidth]{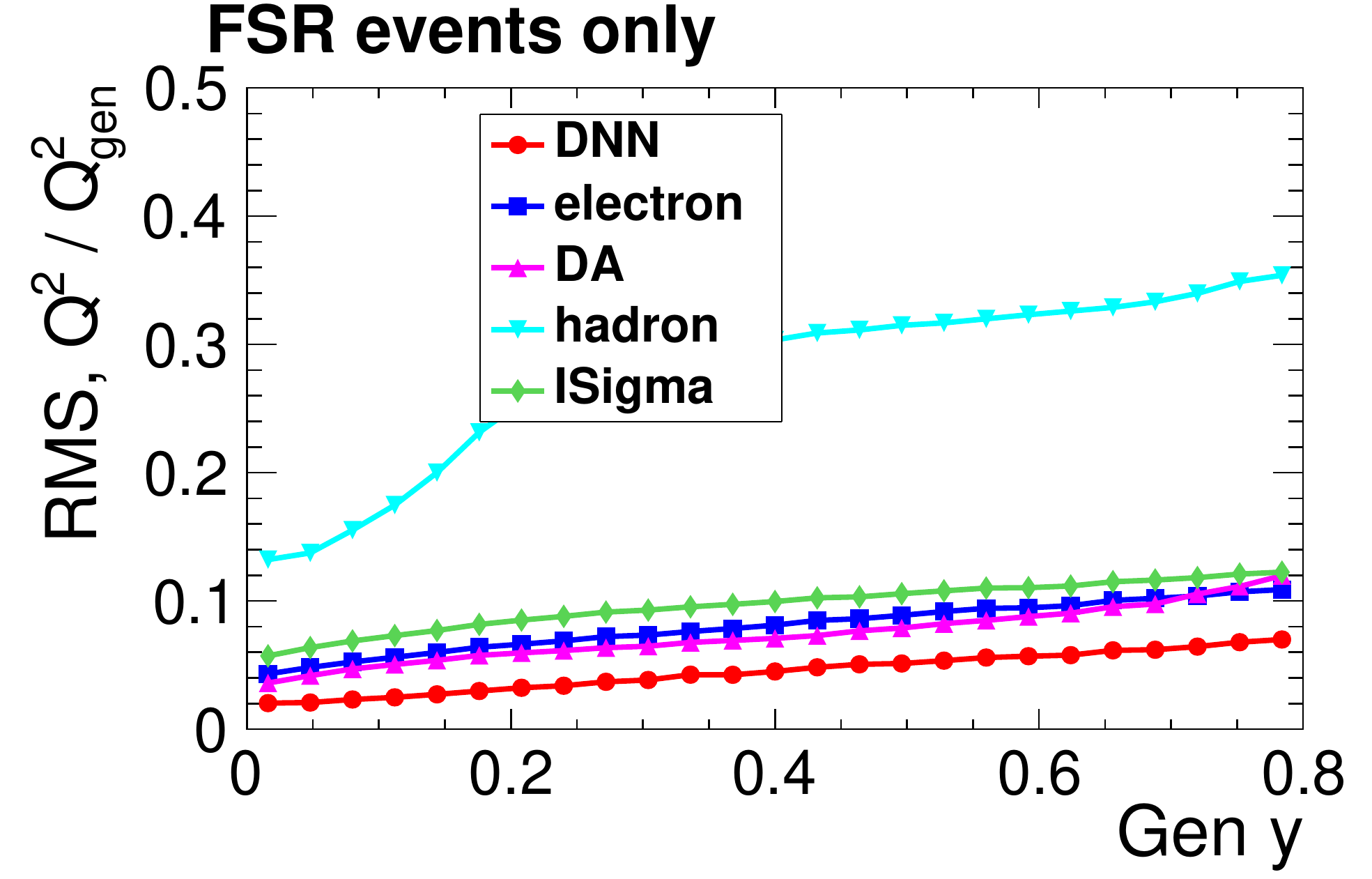}
    \includegraphics[width=0.32\textwidth]{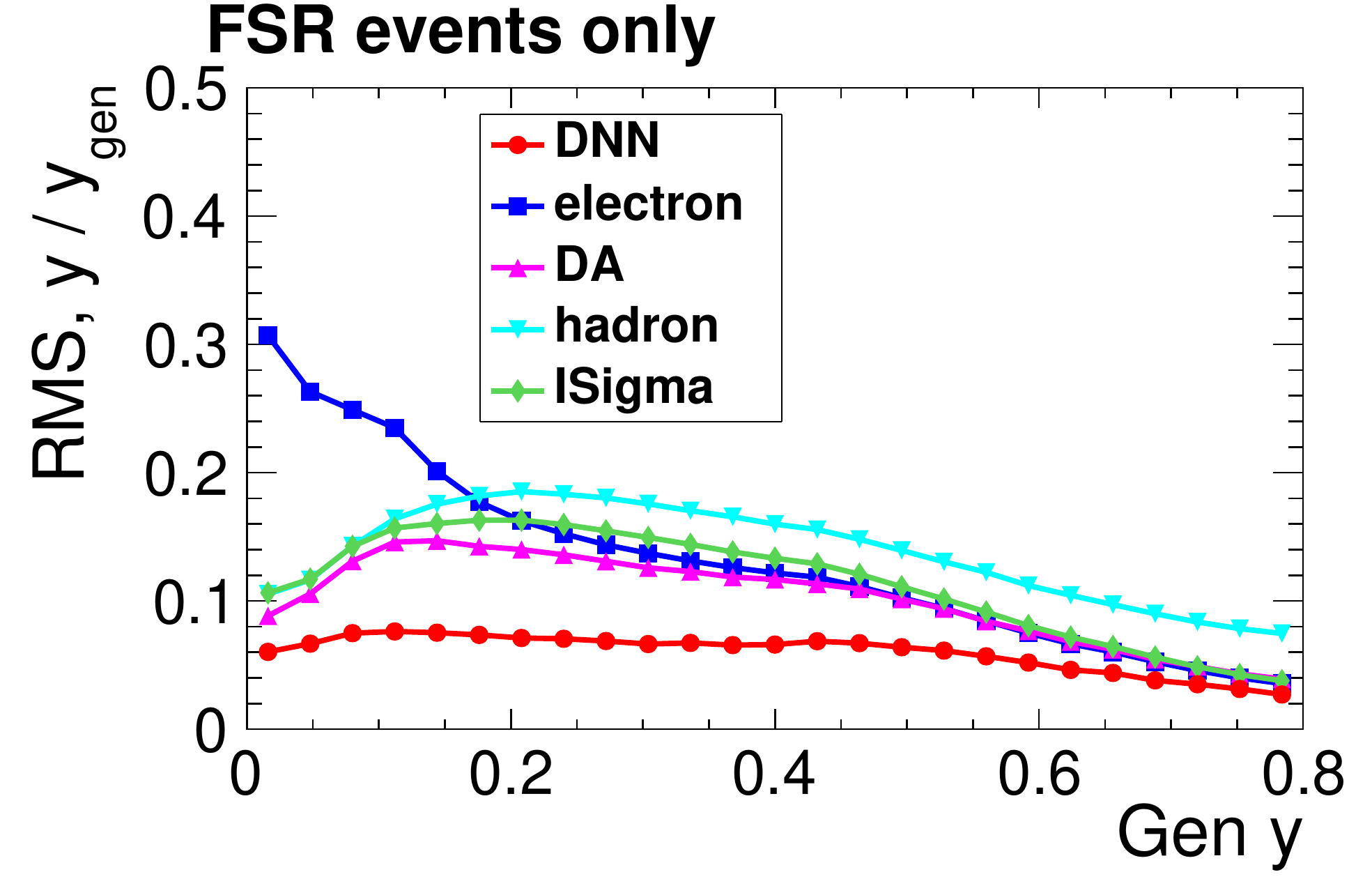}
    \includegraphics[width=0.32\textwidth]{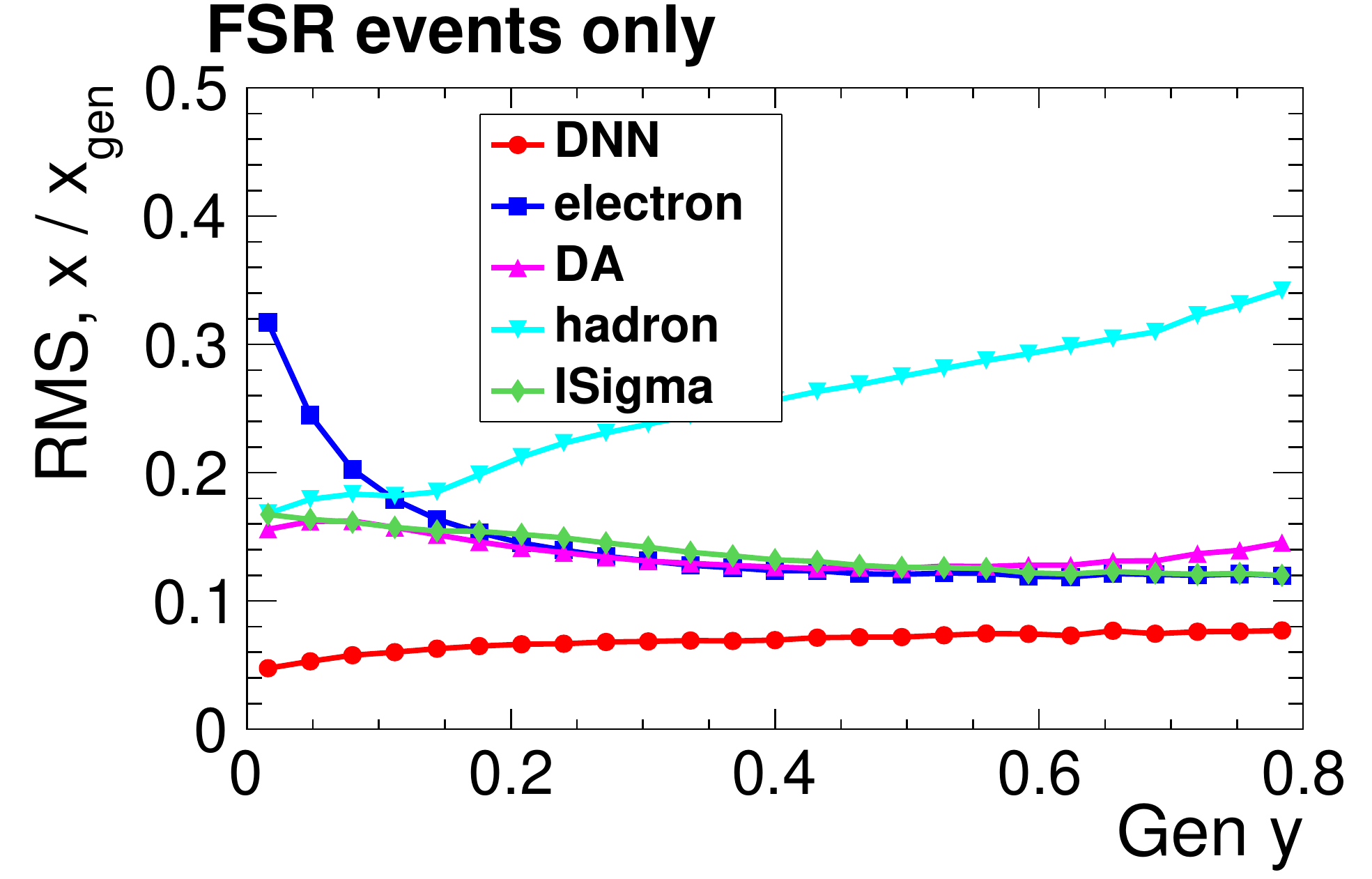}
    \caption{
      Comparison of the RMS of the resolution of \Qxy (from left to right)
      for NoR (top), ISR (middle), and FSR (bottom) events. The RMS is calculated using events with the measured-over-generated ratio within the interval 0 to 2.
    }
    \label{fig:rms-rad-events}
\end{figure}

\begin{figure}[tbhp!]
    \centering
        {\large \fontfamily{lmss}\selectfont ATHENA fast simulation (Rapgap+Delphes)} \\
\vspace{2mm} 
    \includegraphics[width=0.32\textwidth]{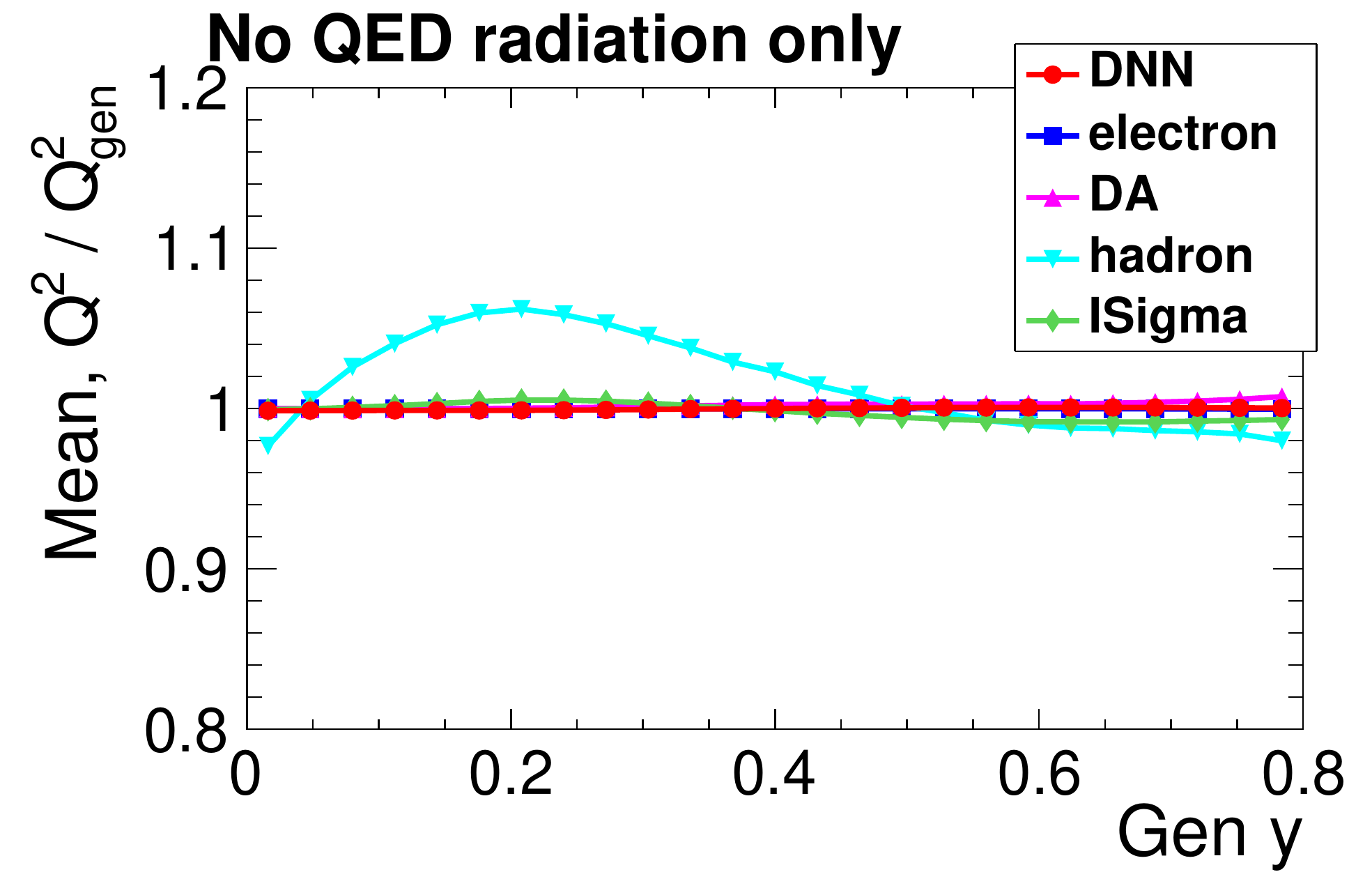}
    \includegraphics[width=0.32\textwidth]{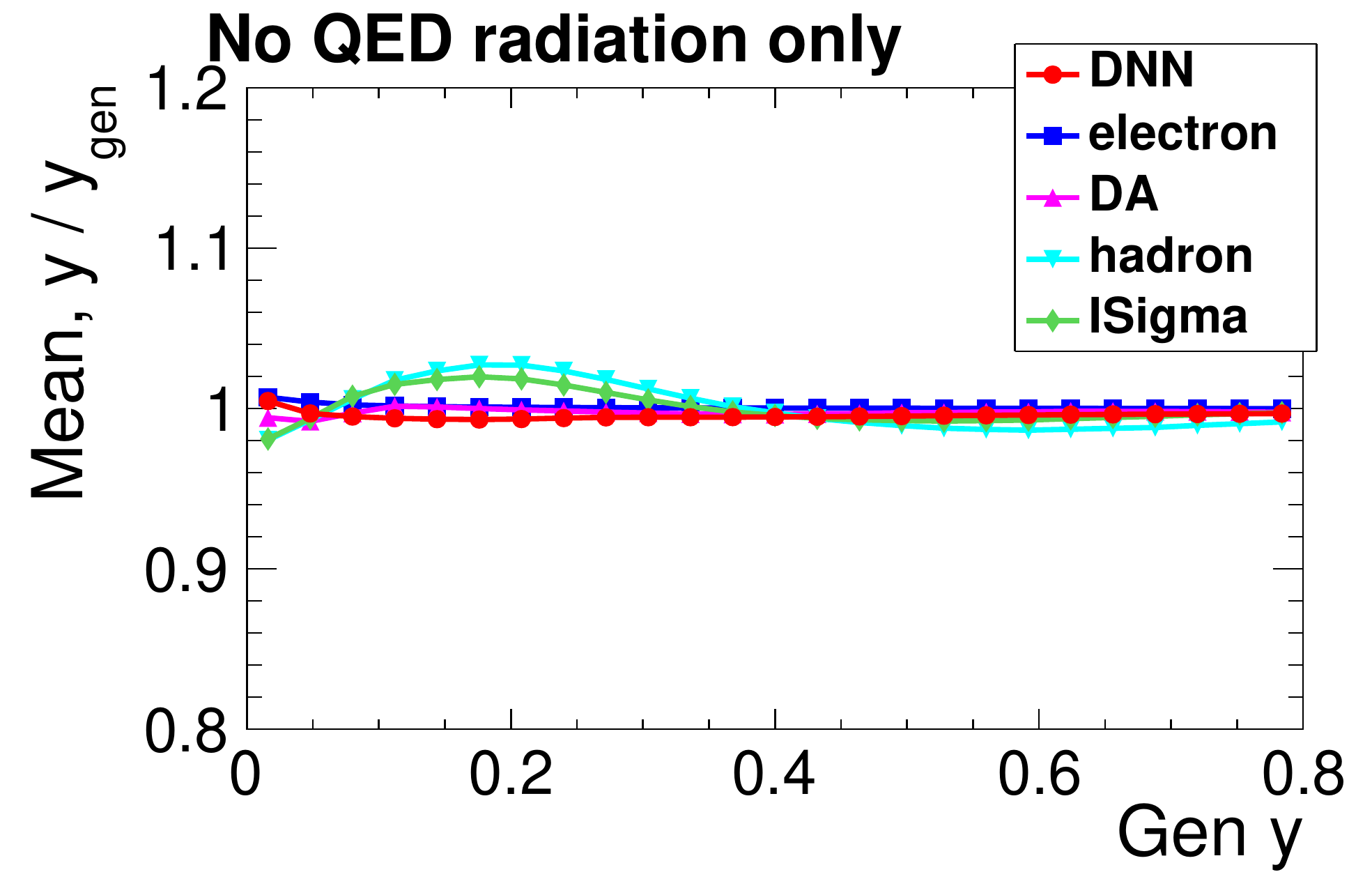} 
    \includegraphics[width=0.32\textwidth]{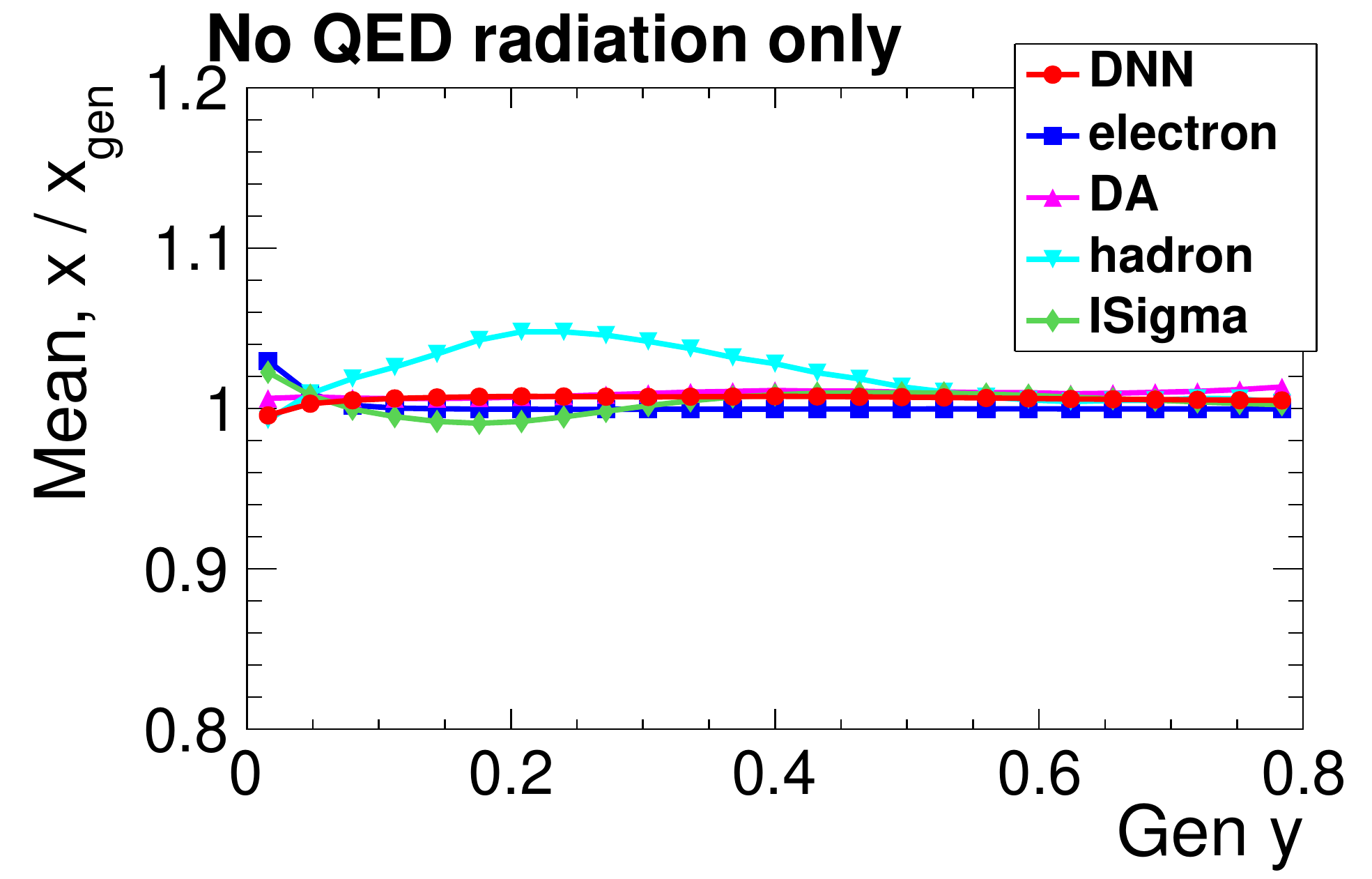}
    \includegraphics[width=0.32\textwidth]{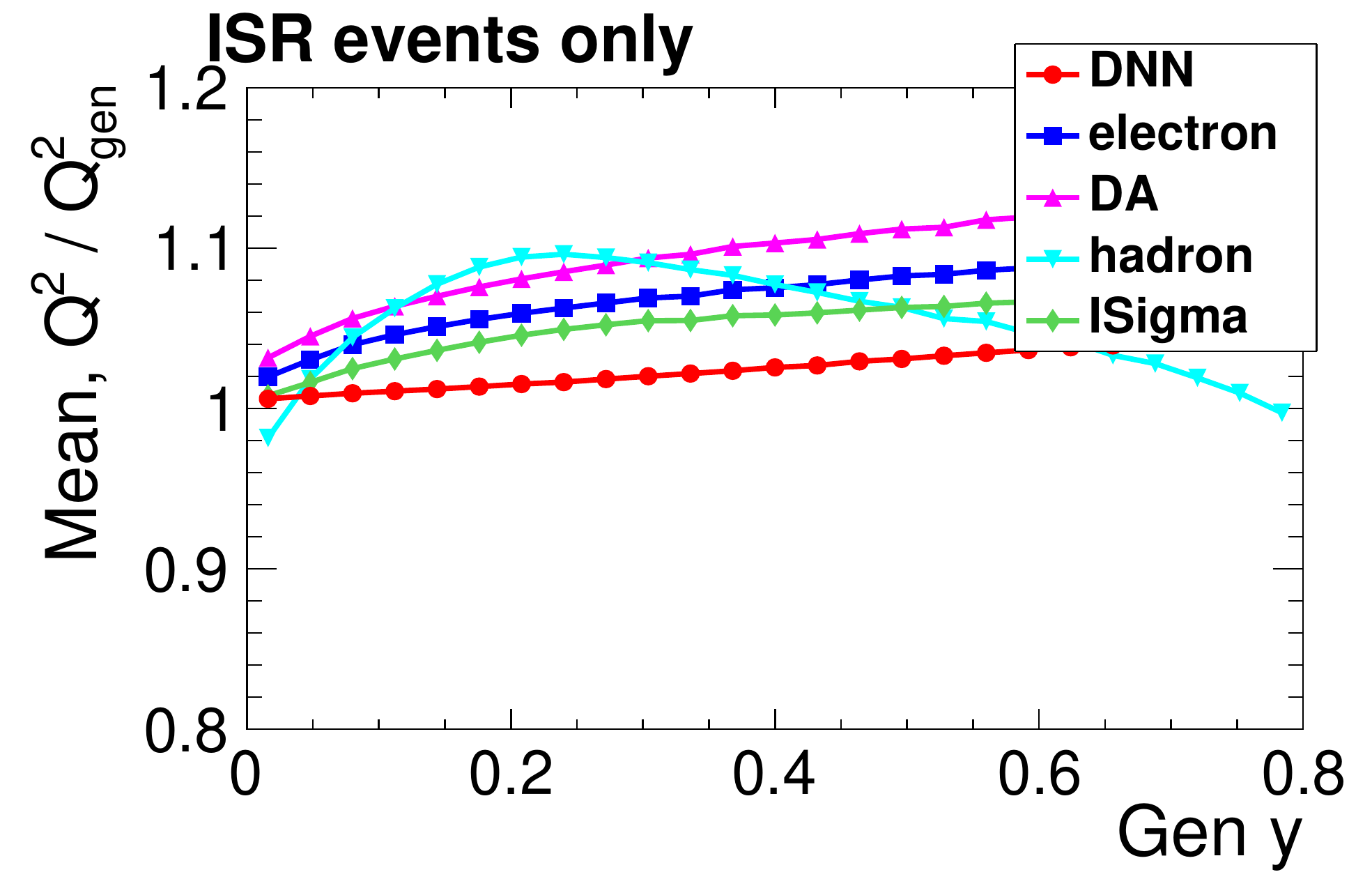}
    \includegraphics[width=0.32\textwidth]{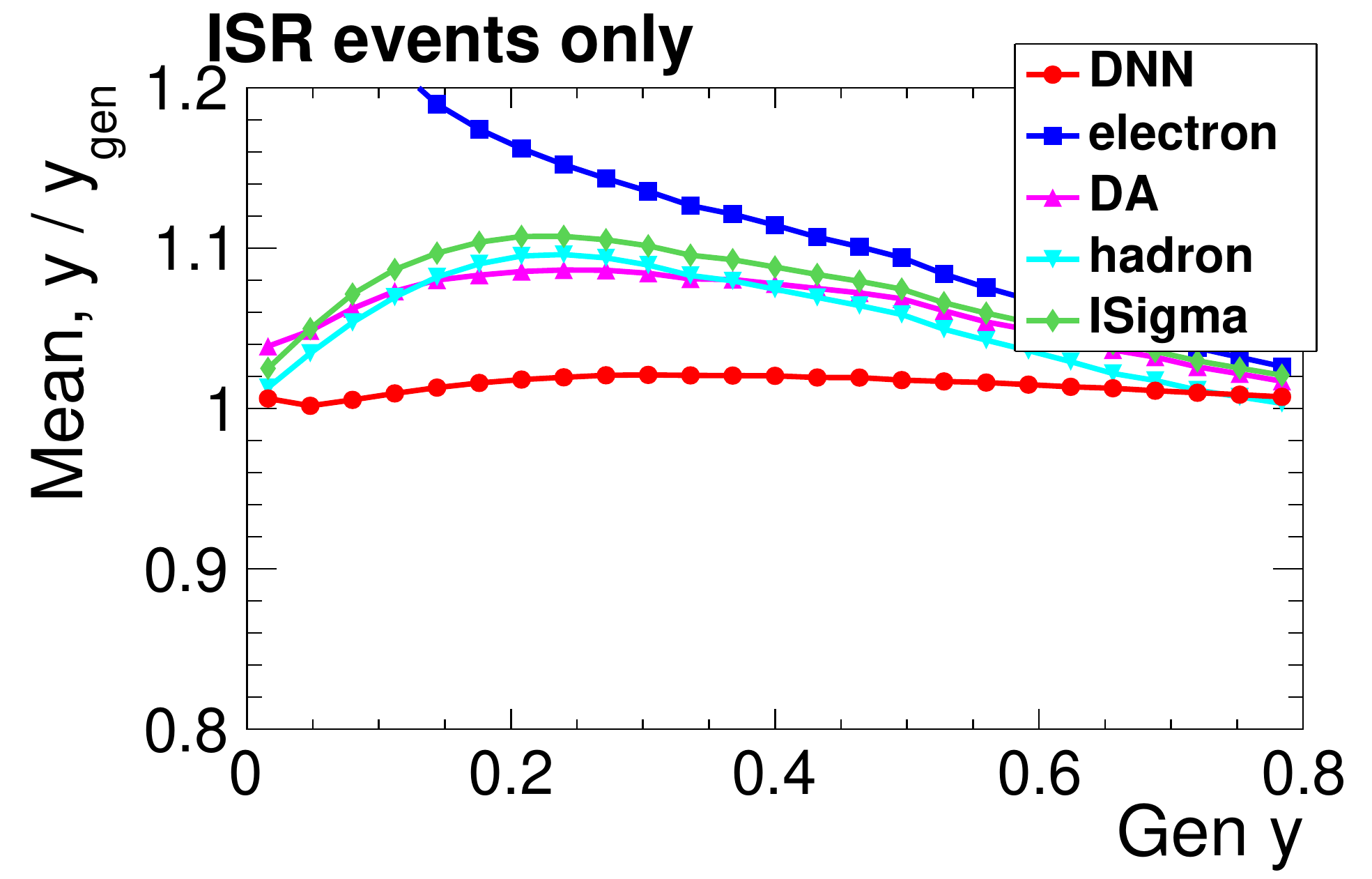} 
    \includegraphics[width=0.32\textwidth]{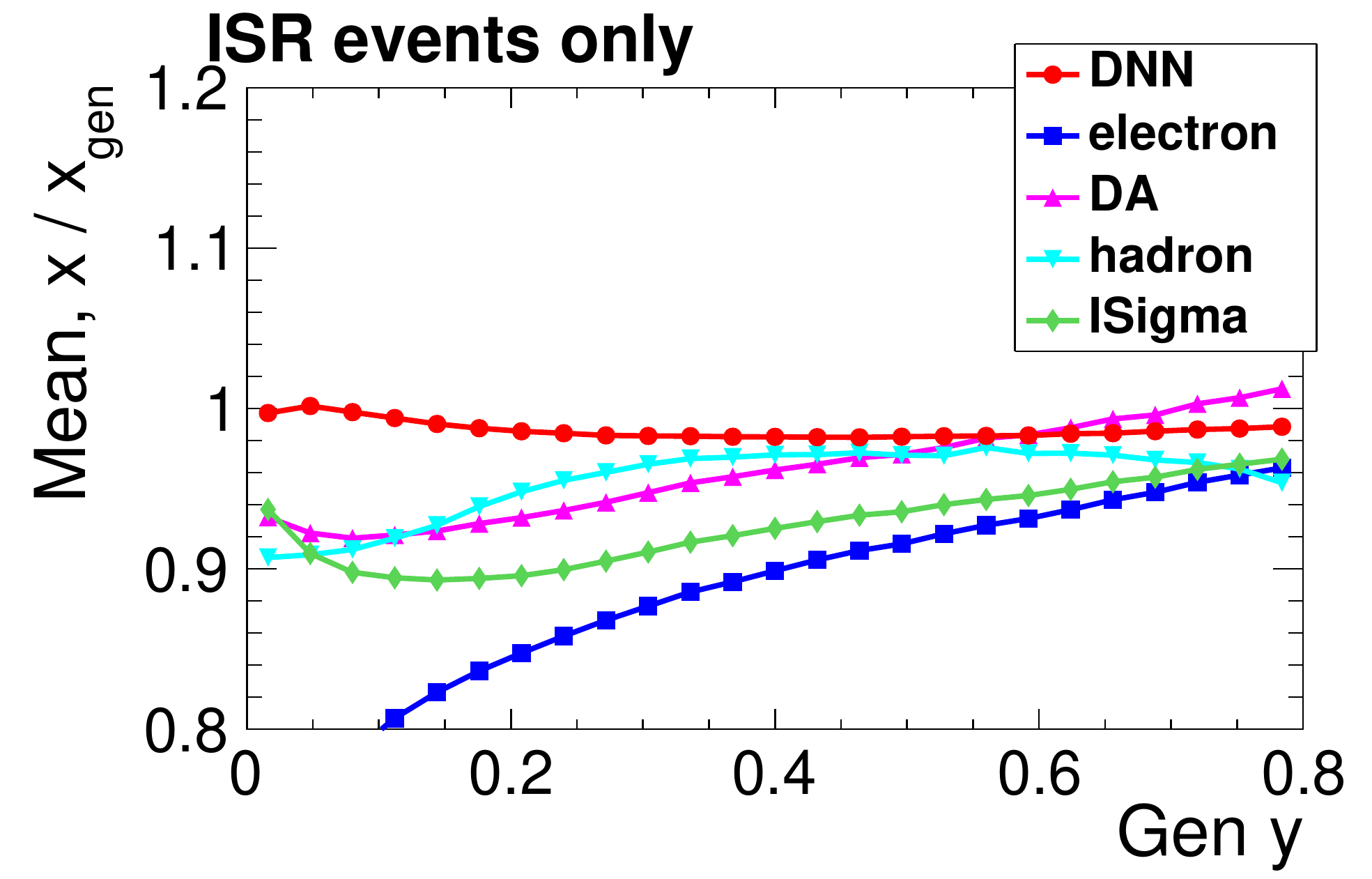} 
    \includegraphics[width=0.32\textwidth]{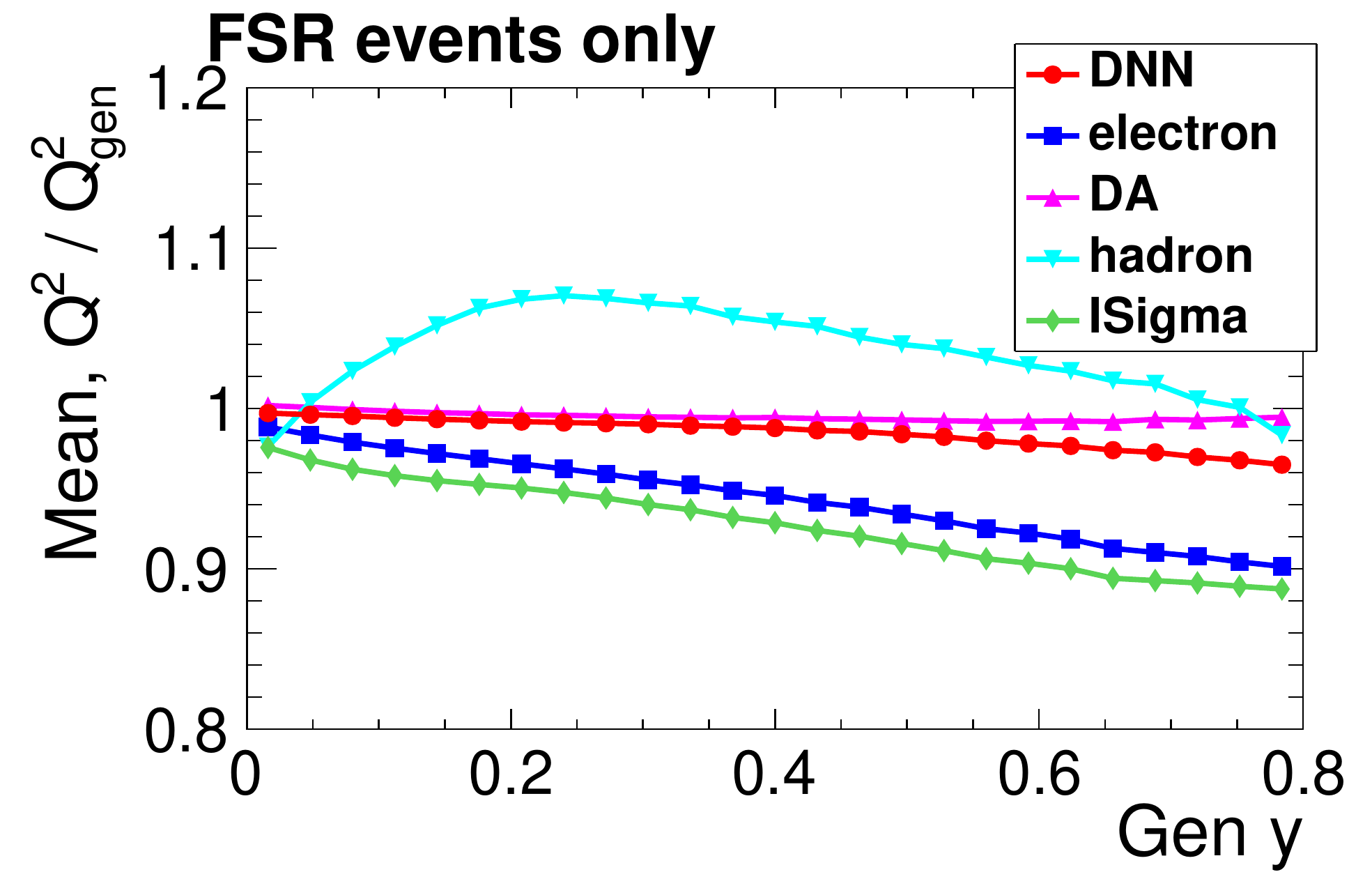}
    \includegraphics[width=0.32\textwidth]{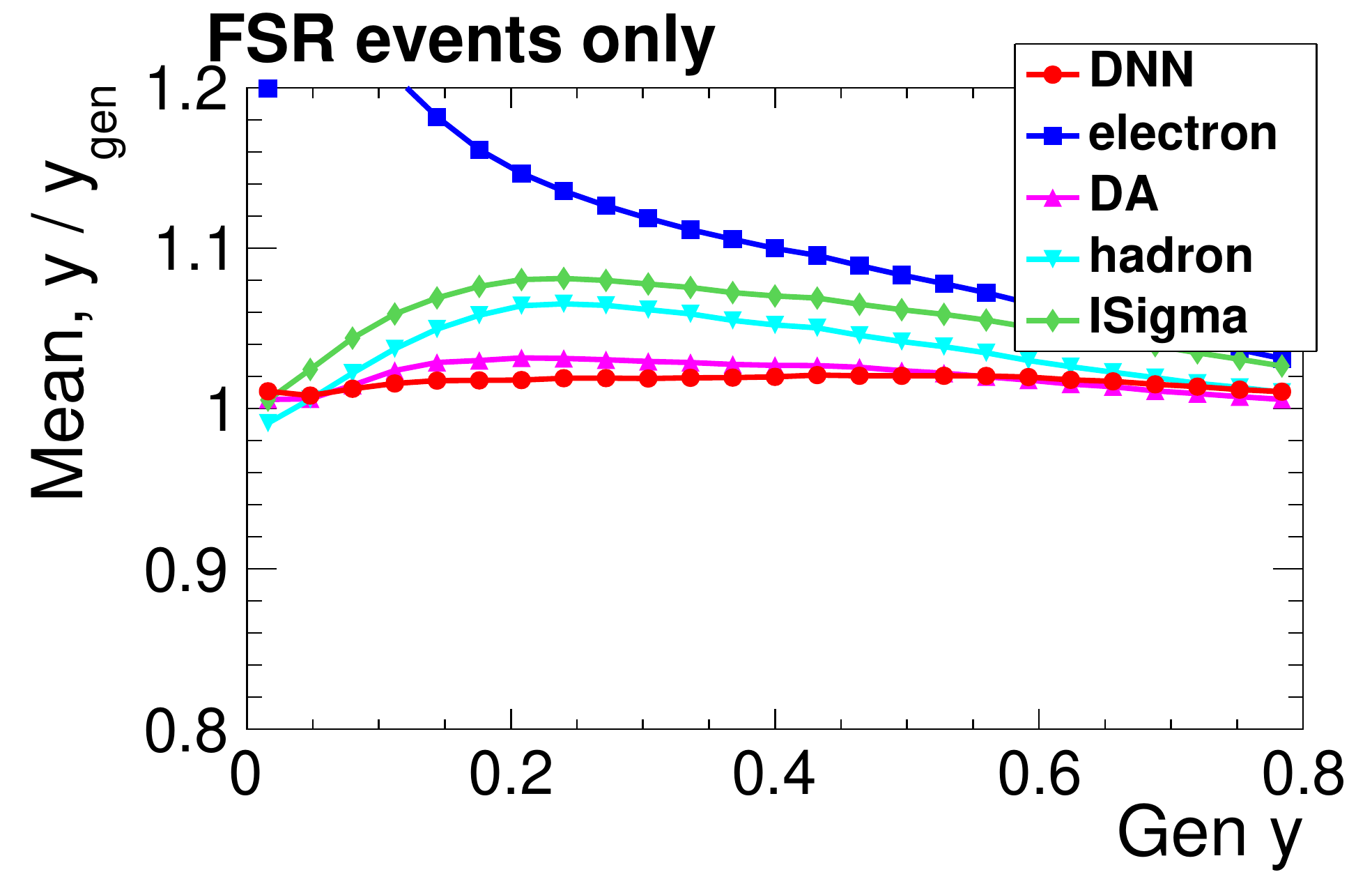} 
    \includegraphics[width=0.32\textwidth]{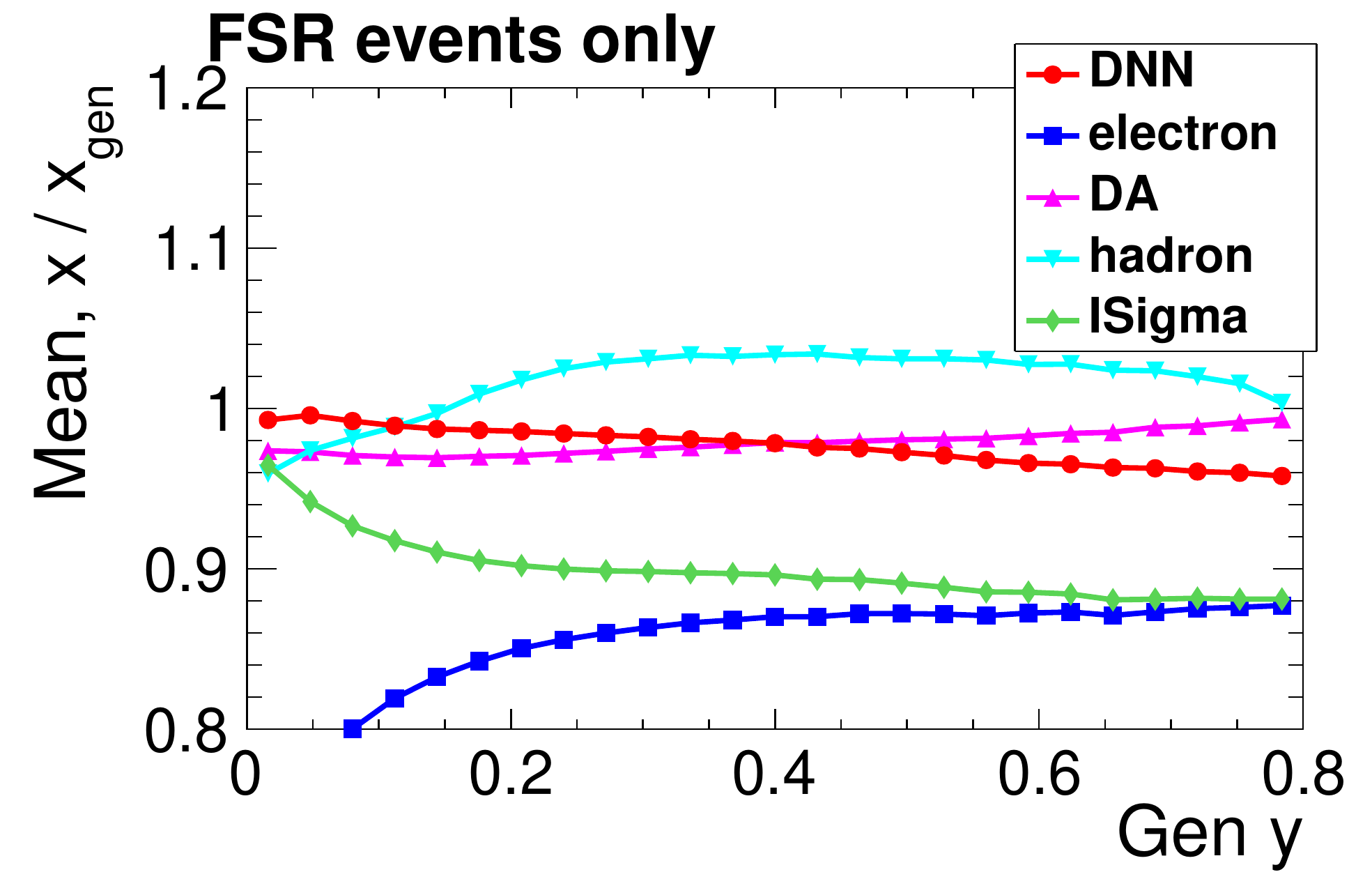} 
    \caption{
      Comparison of the mean of the resolution distributions  of \Qxy (from left to right)
      for NoR (top), ISR (middle), and FSR (bottom) events. The mean is calculated using events with the measured-over-generated ratio within the interval 0 to 2.
    }
    \label{fig:mean-rad-events}
\end{figure}

We examine more closely the resolution and bias for events with and without QED radiation in Figures~\ref{fig:rms-rad-events} and~\ref{fig:mean-rad-events}, where we use the definition of NoR, ISR and FSR events from Section~\ref{sec:QEDrad}.
The RMS for events with no QED radiation gives a measure of the core resolution,
free from the tails that are visible in the distributions of Figure~\ref{fig:athena-xyQ2-resolution_main} for the conventional methods.
All methods, except the hadron method, show no bias in NoR events.
For $x$ and $y$, the electron method has a better core resolution than the DNN for $y>0.15$; however, it suffers from poorer resolution
and a strong bias in events with QED radiation.
The DNN reconstruction has some loss in performance in QED-radiation events,
but it is about a factor of two better than the electron method in these events, and shows no bias. We conclude that the DNN has successfully learned to mitigate the effects of QED radiation that spoil the resolution and bias the calculations of the conventional methods.

%


\section{Demonstration using the full simulation of the H1 experiment}
\label{sec:H1}
We apply our DNN methodology to
simulated events of the H1 experiment at HERA. The events were simulated by the H1 Collaboration using the
\textsc{Rapgap}~3.1~\cite{Jung:1993gf}
and
\textsc{Djangoh}~1.4~\cite{Charchula:1994kf} generators for the beam energies $E_0=27.6\,\GeV$ and $E_p=920\,\GeV$.
The generators employ the
\textsc{Heracles} routines~\cite{Spiesberger:237380,Kwiatkowski:1990cx,Kwiatkowski:1990es}
for QED radiation, the CTEQ6L PDF set~\cite{Pumplin:2002vw},
and the Lund hadronization
model~\cite{Andersson:1983ia} with parameters fitted by the ALEPH
Collaboration~\cite{Schael:2004ux}.
The simulation of the H1 experiment~\cite{H1:1996prr,H1:1996jzy}
employs the \textsc{Geant}~3 package~\cite{Brun:1987ma} and 
includes real calorimeter noise and fast shower simulations~\cite{Fesefeldt:1985yw,Grindhammer:1989zg,Gayler:1991cr,Kuhlen:1992ey,Grindhammer:1993kw,Glazov:2010zza}.
The simulation includes time-dependent properties (`run-specific'), where the detector state and beam properties correspond altogether to the HERA-II data
taking periods.  

The simulated events are reconstructed just like data, in particular, an energy-flow
algorithm~\cite{energyflowthesis,energyflowthesis2,energyflowthesis3}
is used to define objects whose sum yields the HFS four-vector,
 and the scattered electron candidate are defined using the same 
approach as Refs.~\cite{H1:2012qti,Andreev:2014wwa,H1:2021wkz}.
The simulated events also undergo the same (in-situ) calibration procedure
as real data, using the latest calibration by the H1
Collaboration~\cite{H1:2012qti,Kogler:2011zz,Andreev:2014wwa}. 
Some technical selections and fiducial cuts are applied as it
would be done similarly to real data.
In particular, events are required to have $45 <\Sigma+\Sigma_e< 65$\,GeV to suppress
ISR events; a veto on QED Compton events is imposed; and since a trigger simulation is included, our study is
limited to $E\gtrsim11\,\GeV$~\cite{Andreev:2014wwa}.  
The simulated events are processed within H1's computing
environment~\cite{Britzger:2021xcx} and altogether several $10^8$
events were simulated, and after 
`run'-selection, acceptance effects and our selection of
$\Qsq>200\,\GeVsq$ yield about $7\cdot10^7$ simulated and
reconstructed events for our DNN studies.

We use the same regression DNN structure, input variables, and training methods for the
simulated H1 events as previously for the ATHENA events. We take a sample of over 12 million \textsc{Rapgap} events and use half of the
events for training and half for validation. 
The training terminates after 125 epochs, finding no further improvement in the validation loss function.

\begin{figure}[tbhp!]
    \centering
    \includegraphics[width=0.8\textwidth]{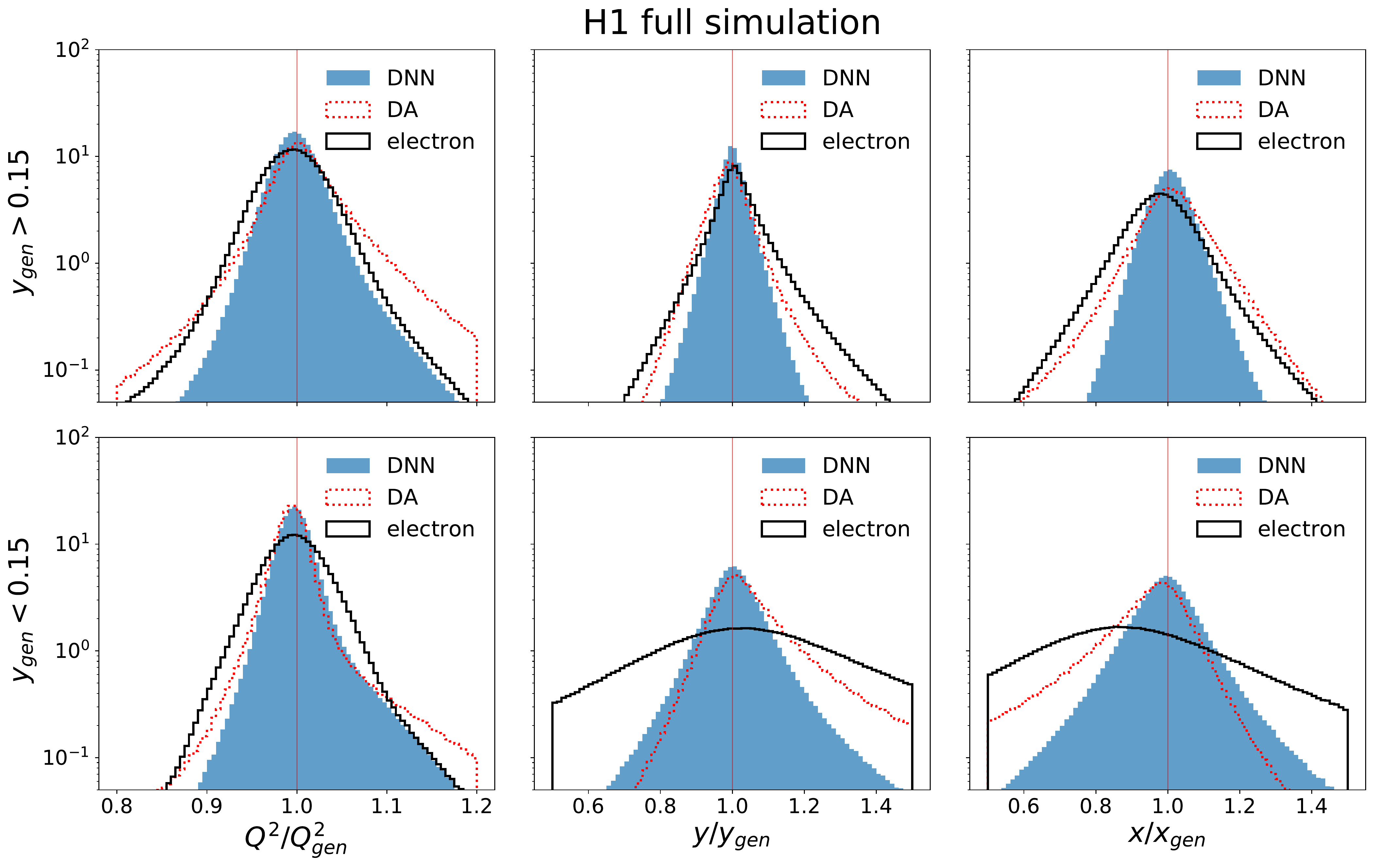}
    \caption{Resolution for $Q^2$ (left), $y$ (middle), and $x$ (right) for the DNN, electron, and double-angle (DA) reconstruction methods for the full simulation of the H1 experiment.  The top (bottom) row is for events with $y_{\rm gen}>0.15$ ($y_{\rm gen}<0.15$).  All distributions are normalized to the same area.}
    \label{fig:h1-xyQ2-resolution}
\end{figure}
Figure~\ref{fig:h1-xyQ2-resolution} shows the resolutions for the DNN and two classical methods in two intervals 
of $y$. Similarly to the results obtained with the ATHENA simulation, the DNN yields a peak at unity and no asymmetric
tails for all three quantities, in contrast with classical methods.
\begin{figure}[tbhp!]
    \centering
        {\large \fontfamily{lmss}\selectfont H1 full simulation} \\
    \vspace{1mm} 
    \includegraphics[width=0.32\textwidth]{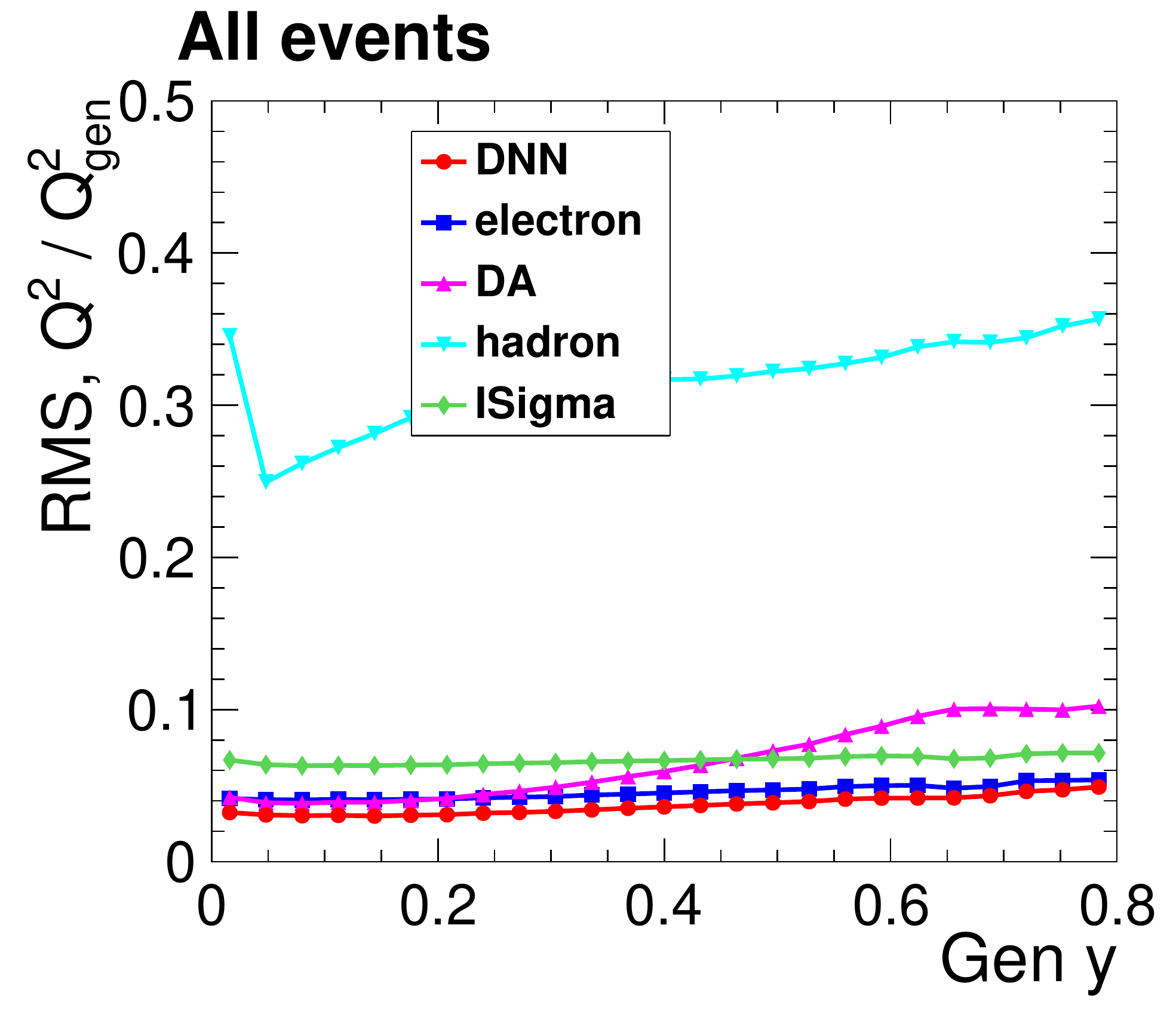}
    \includegraphics[width=0.32\textwidth]{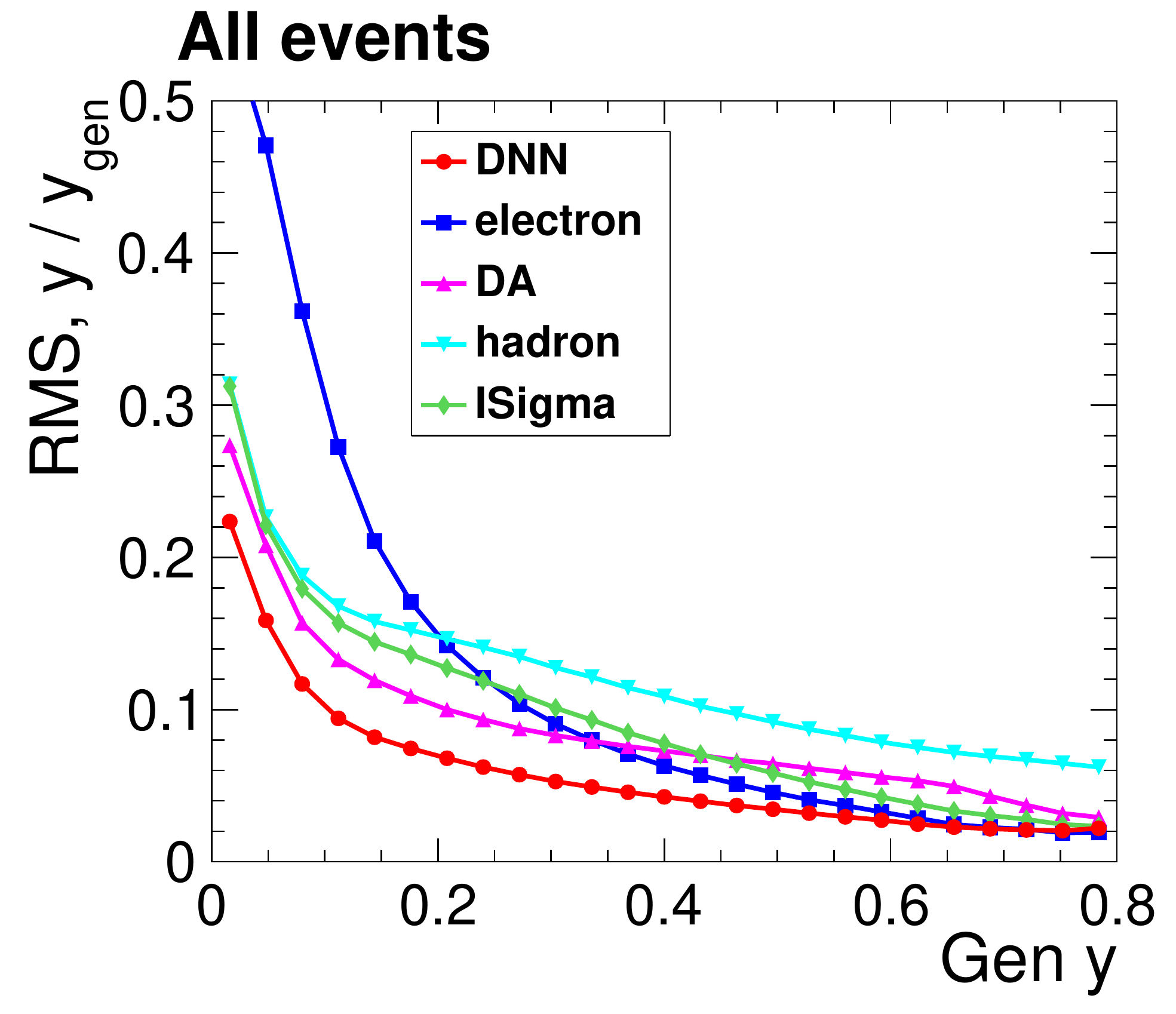}    
    \includegraphics[width=0.32\textwidth]{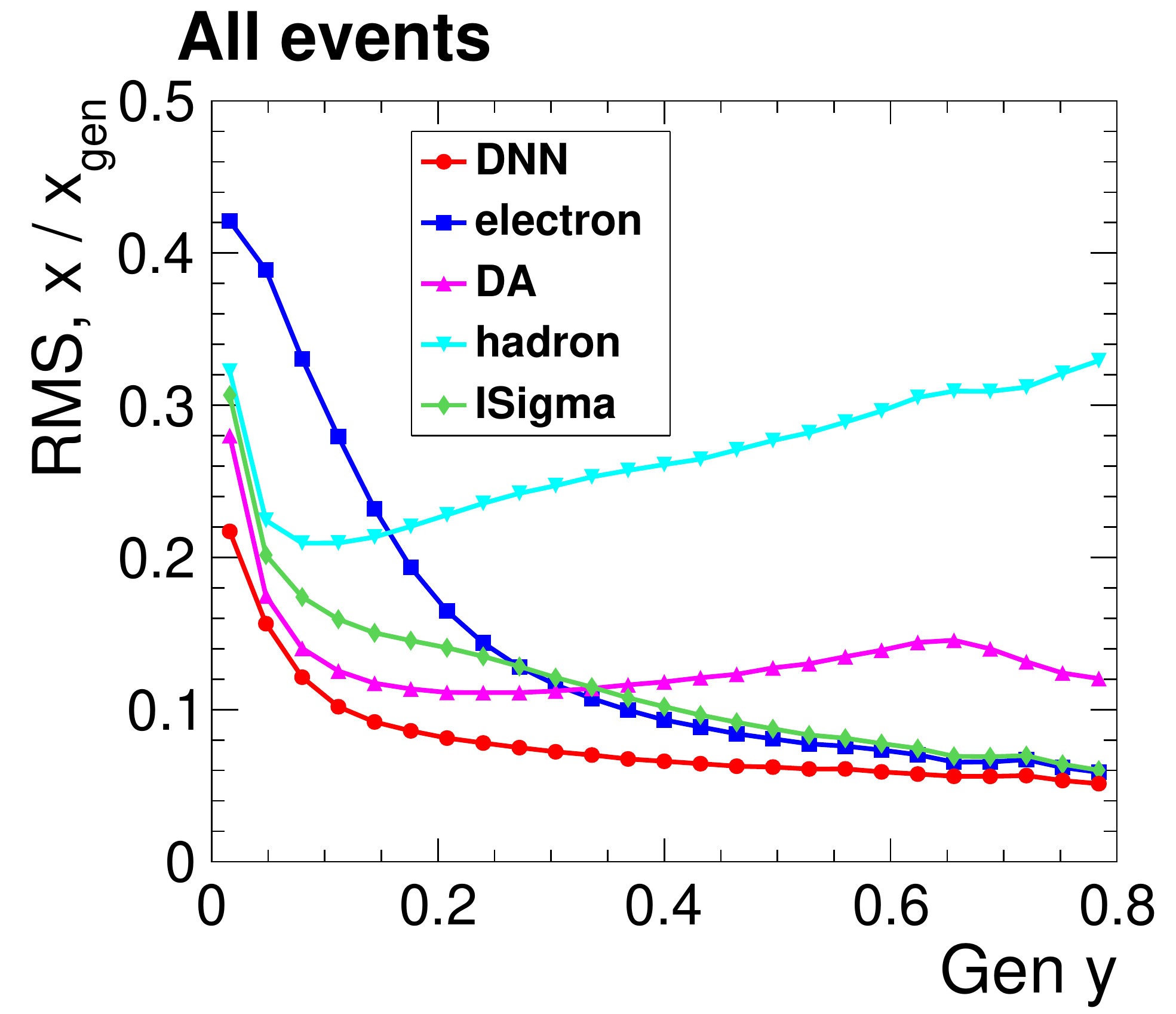} \\    
    \includegraphics[width=0.32\textwidth]{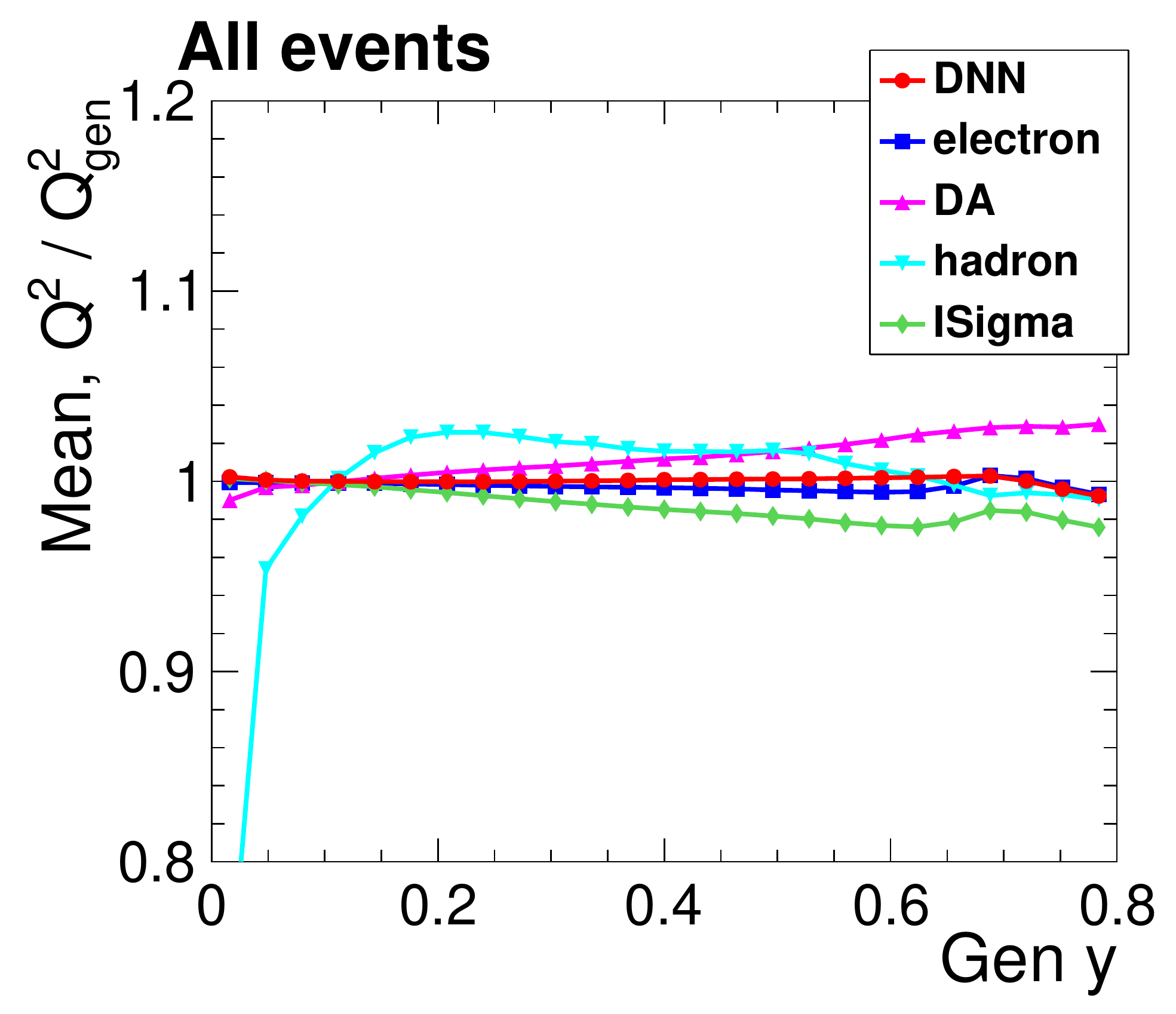}
    \includegraphics[width=0.32\textwidth]{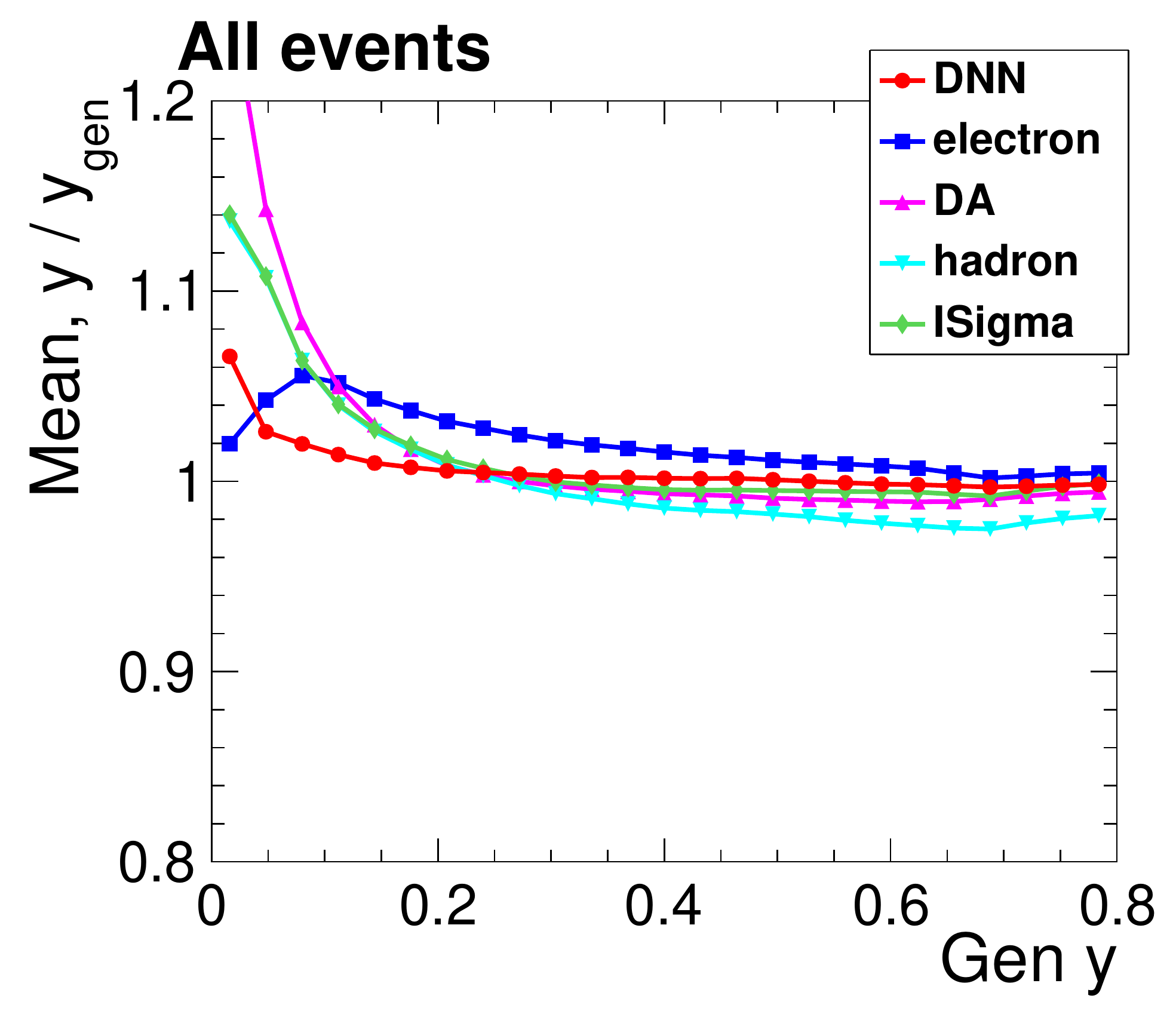}    
    \includegraphics[width=0.32\textwidth]{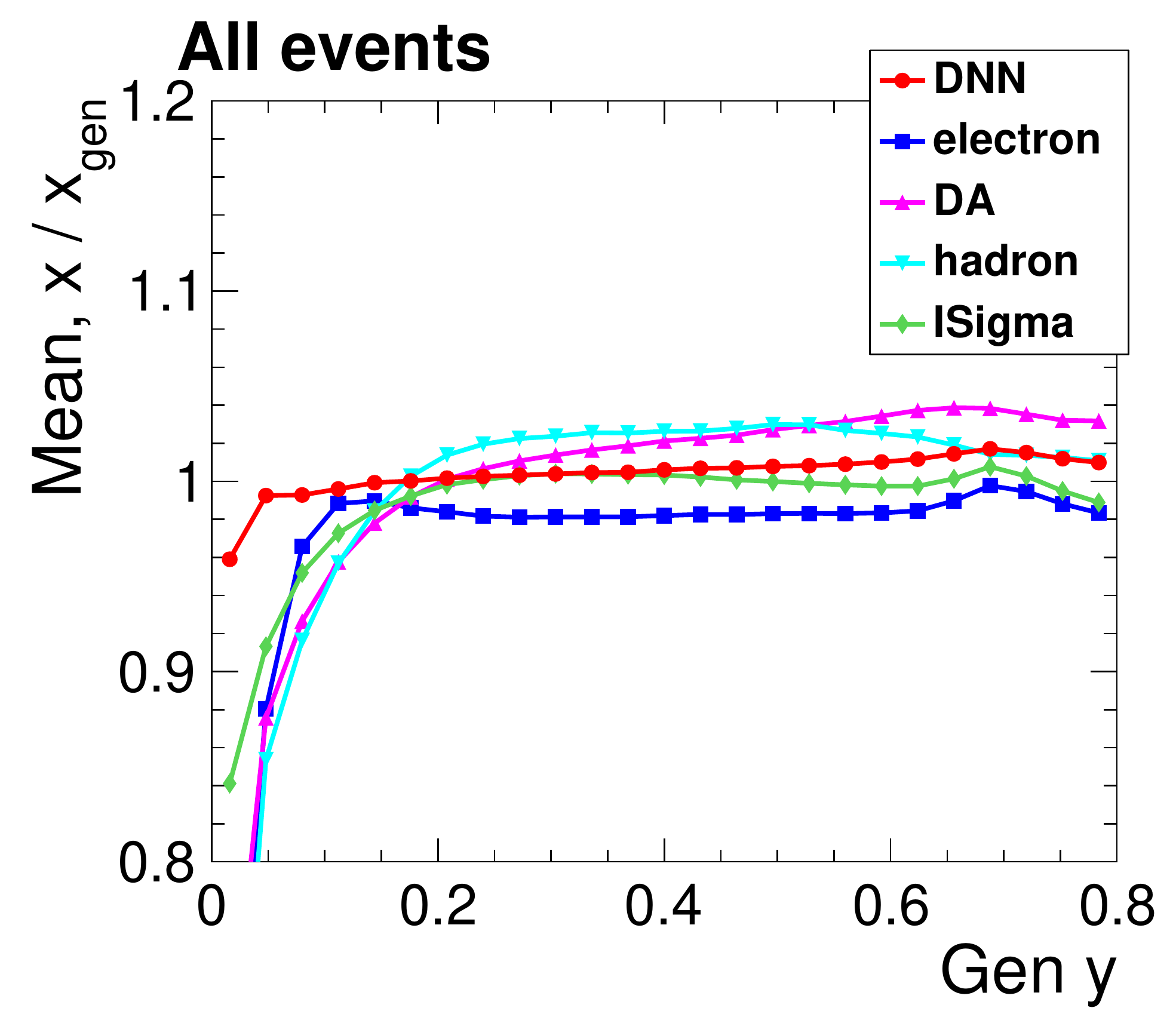} \\  
    \caption{
    Resolution on the reconstruction of $Q^2$ (left), $y$ (middle),
    and $x$ (right) as a function of the generated $y$ for the full simulation of H1.
    The top (bottom) row shows the RMS (mean) of the measured-over-generated distribution
    as a function of the generated $y$.  
        The RMS and mean are calculated using events with the measured-over-generated ratio within the interval 0 to 2.
    }
    \label{fig:h1-xyQ2-resolution-vs-y}
\end{figure}
The mean and RMS of the resolutions are displayed
as a function of $y$ in Figure~\ref{fig:h1-xyQ2-resolution-vs-y}.
The DNN yields the smallest RMS. The mean of the DNN reconstruction is closest to unity over a wide range of $y$. 
Only for \Qsq, or at highest $y$, the electron method achieves comparably good RMS and mean.
In contrast, at lower $y$ the
DNN provides a significant improvement and results in a bias-free
reconstruction of $y$ and \xbj\ with superior resolution\footnote{More detailed representations of the resolutions for the different methods are collected in Appendix~\ref{sec:additionalfigsH1},
where also further reconstruction methods from
Section~\ref{sec:methods} are studied with a real detector simulation.}.

To assess a possible bias in our DNN methodology that may
arise from the details of the MC event generator that is used
to train the DNN, we study the performance of the DNN reconstruction
using two different MC event generators, \textsc{Rapgap} and
\textsc{Djangoh}.
The two event generators differ in the modelling of
higher-order QCD radiation that results in significant differences in the prediction of the HFS.
\textsc{Djangoh} employs the color-dipole model, while \textsc{Rapgap} employs a
matrix-element plus parton shower model, where the parton
shower is in the leading-logarithmic approximation.
In this study, we use H1's full simulation and train the DNN with a
\textsc{Rapgap} event sample.
Subsequently this DNN is applied to a statistically independent
\textsc{Rapgap} event sample and to a simulated \textsc{Djangoh} sample. 

\begin{figure}[tbh!]
    \centering
                {\large \fontfamily{lmss}\selectfont H1 full simulation } \\
    \vspace{2mm} 
    \includegraphics[width=0.32\textwidth]{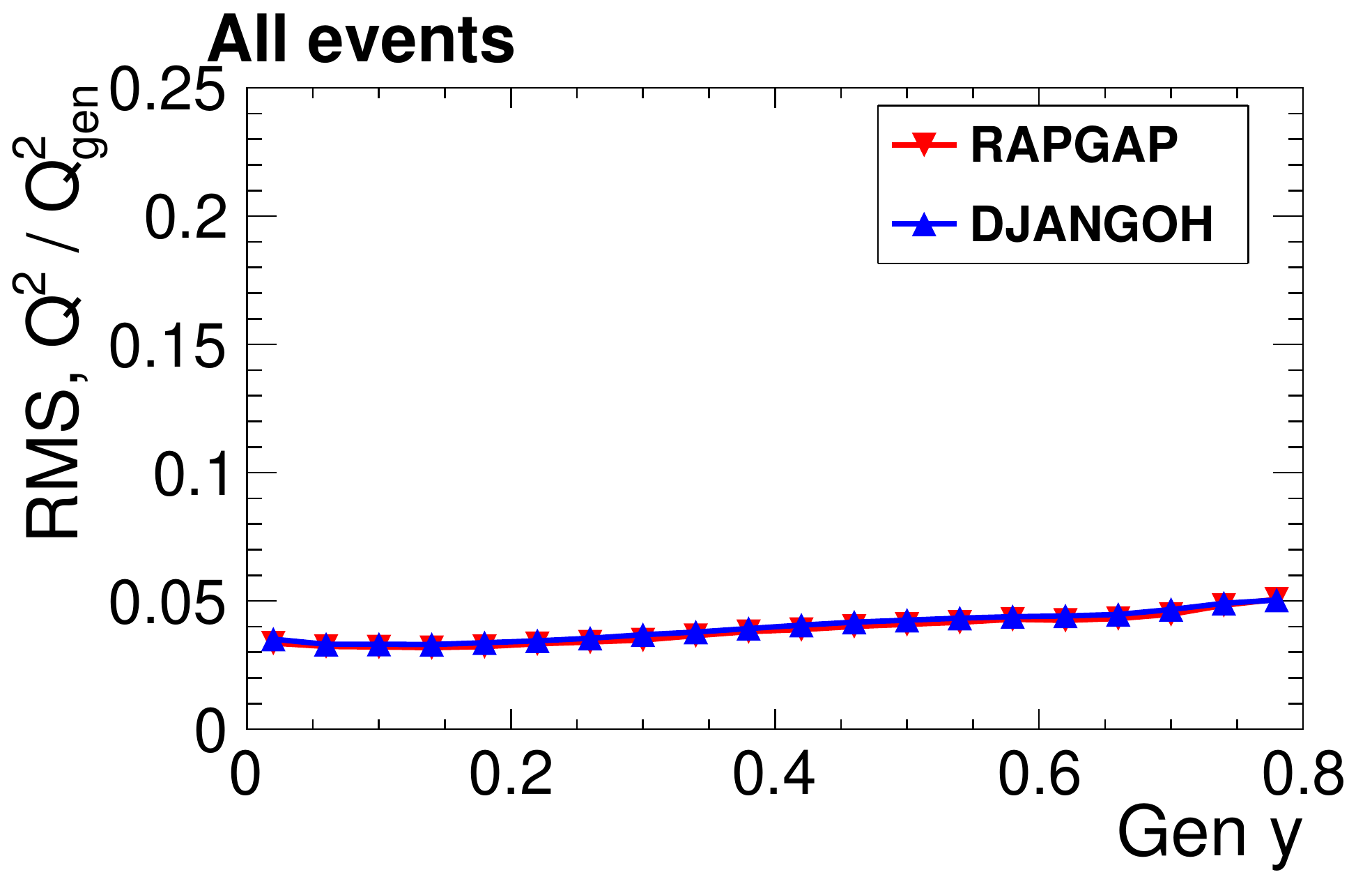}
    \includegraphics[width=0.32\textwidth]{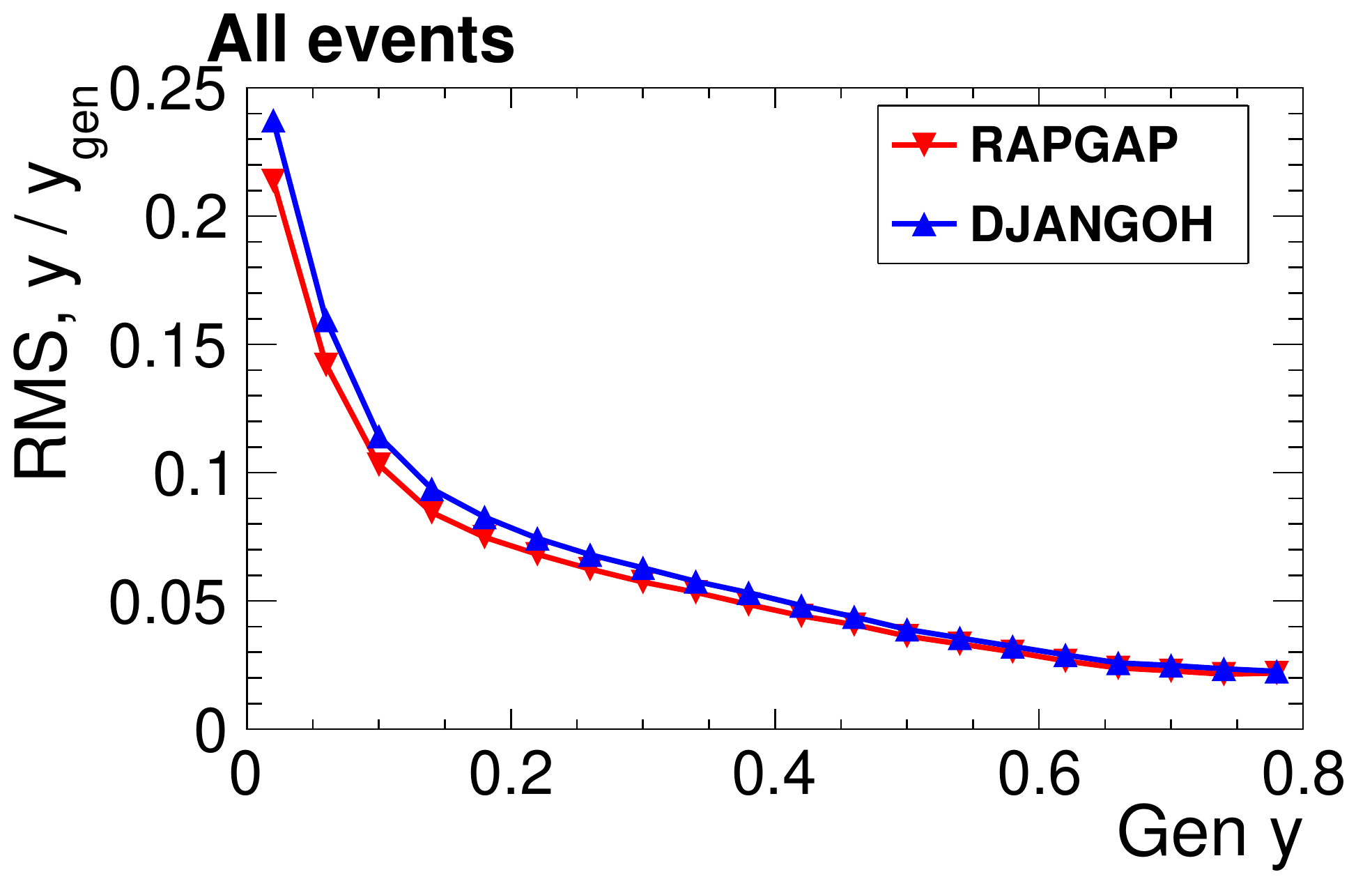}
    \includegraphics[width=0.32\textwidth]{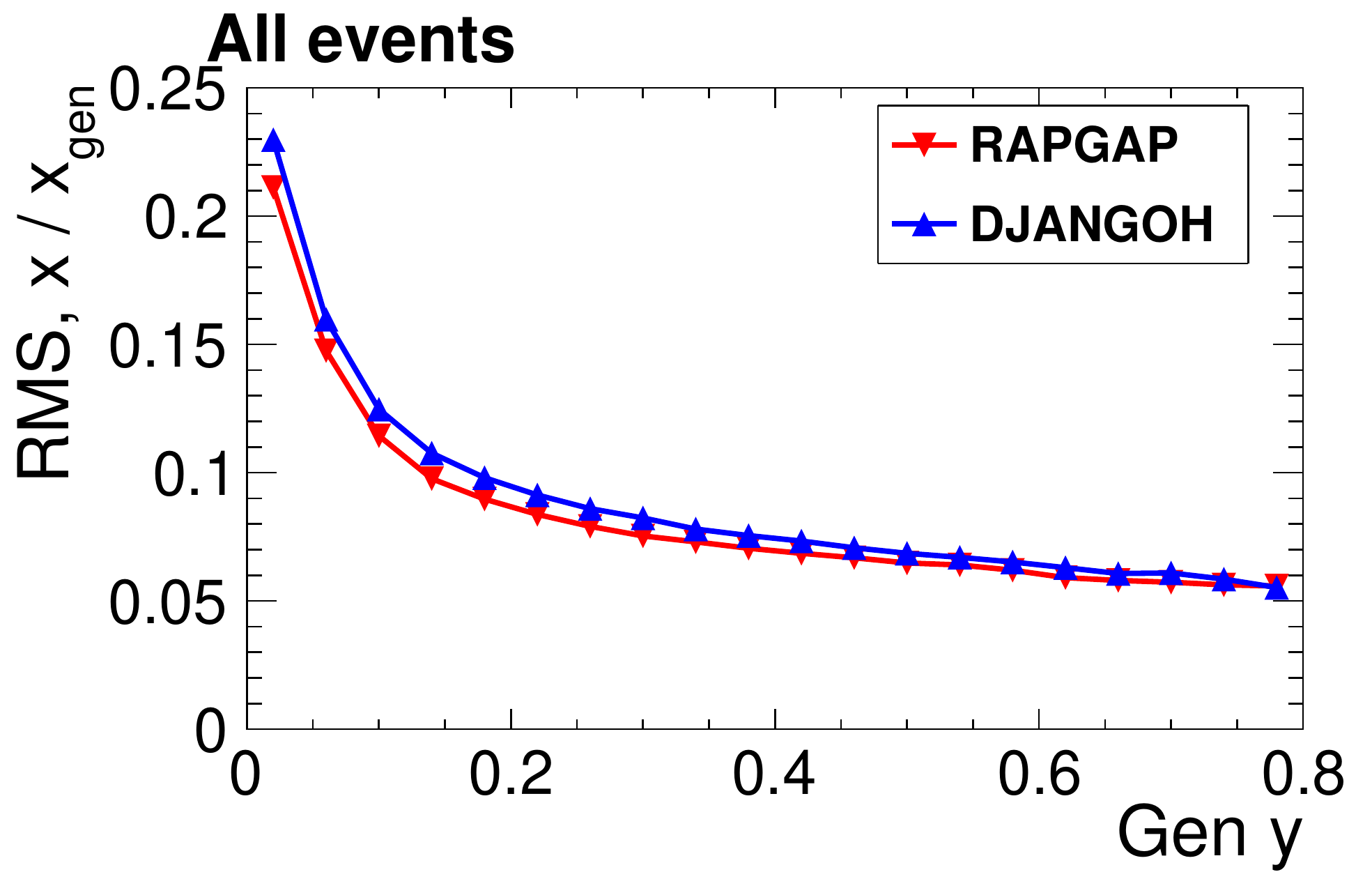} \\
    \includegraphics[width=0.32\textwidth]{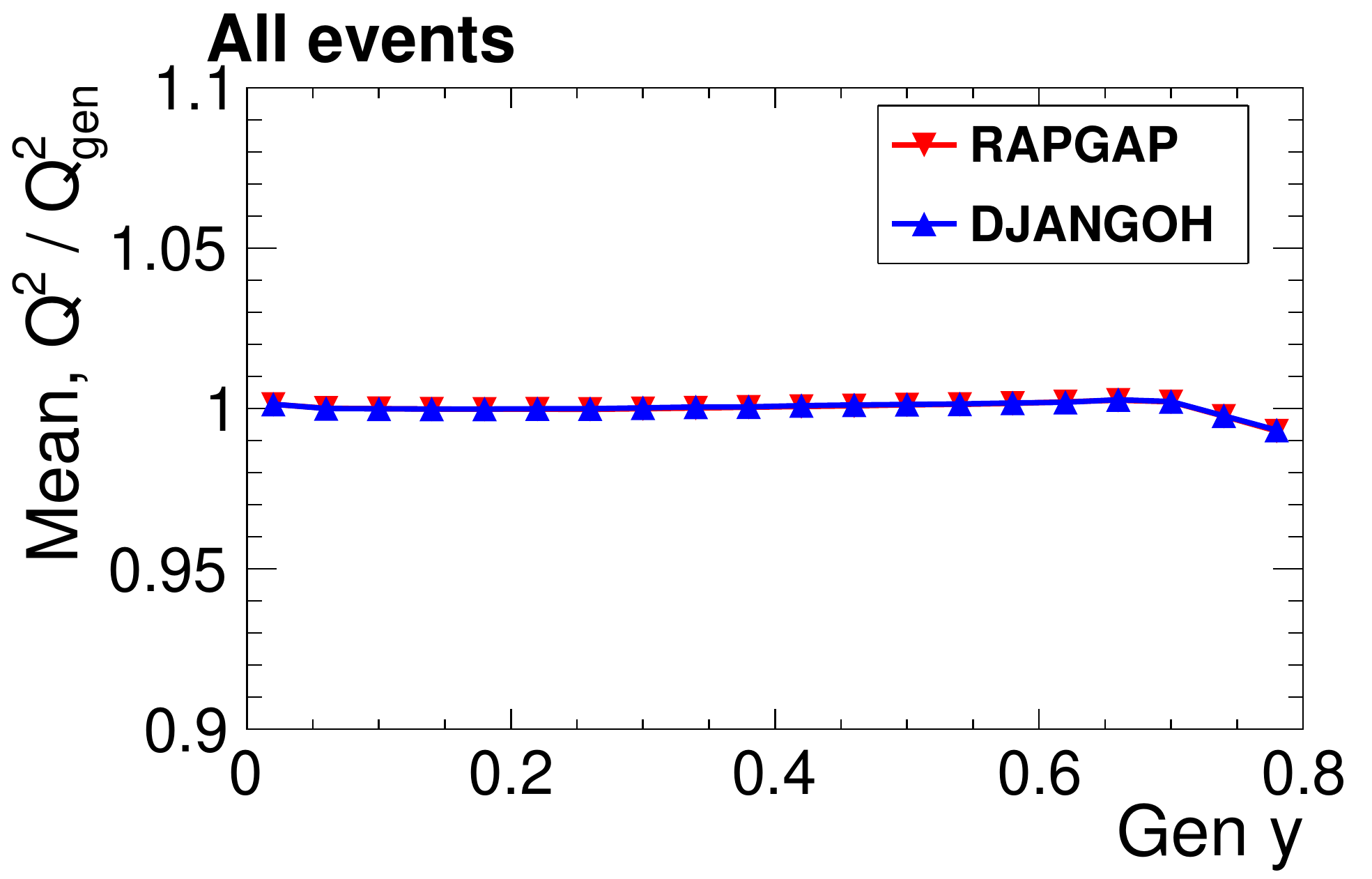} 
    \includegraphics[width=0.32\textwidth]{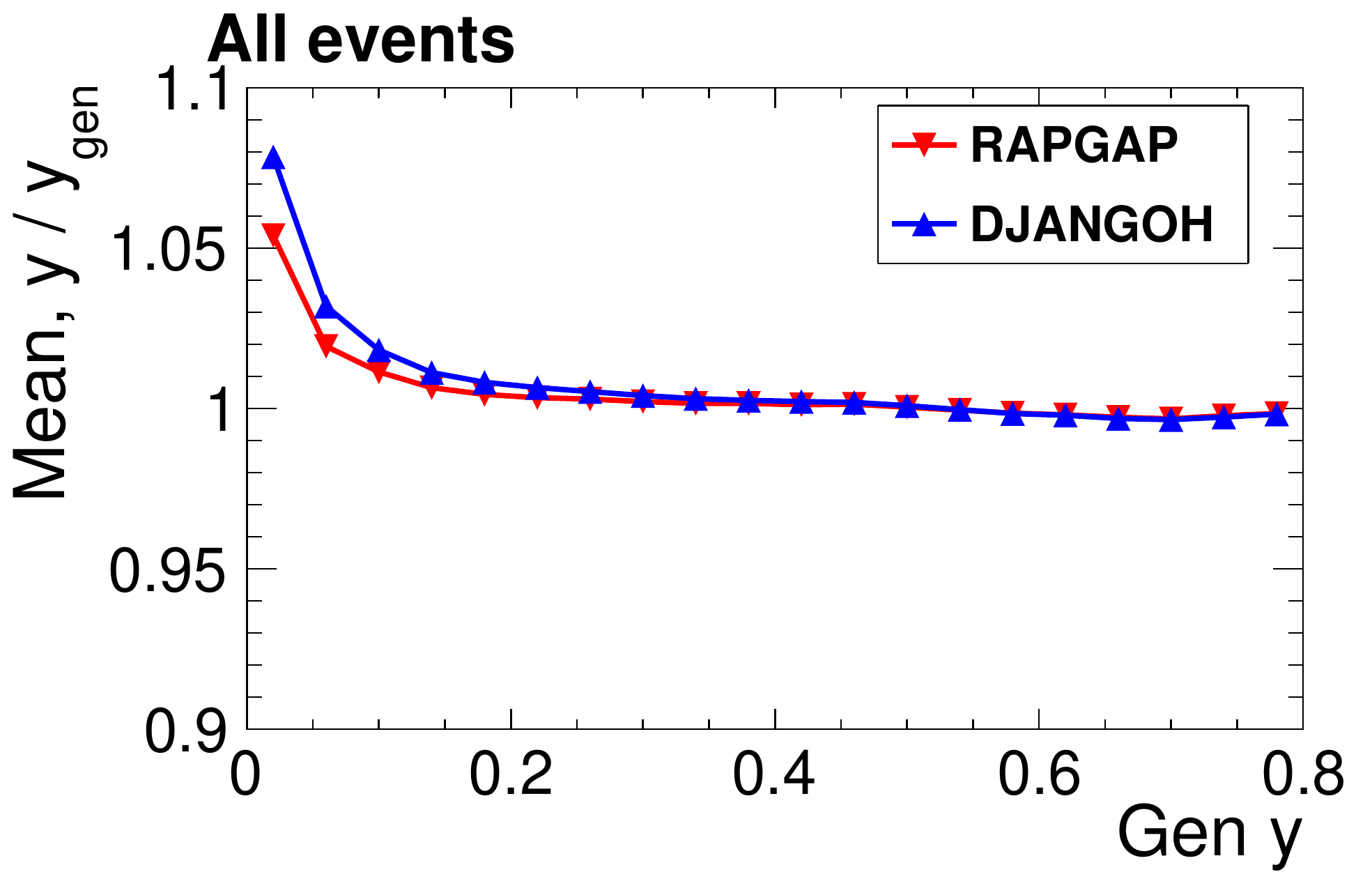}
    \includegraphics[width=0.32\textwidth]{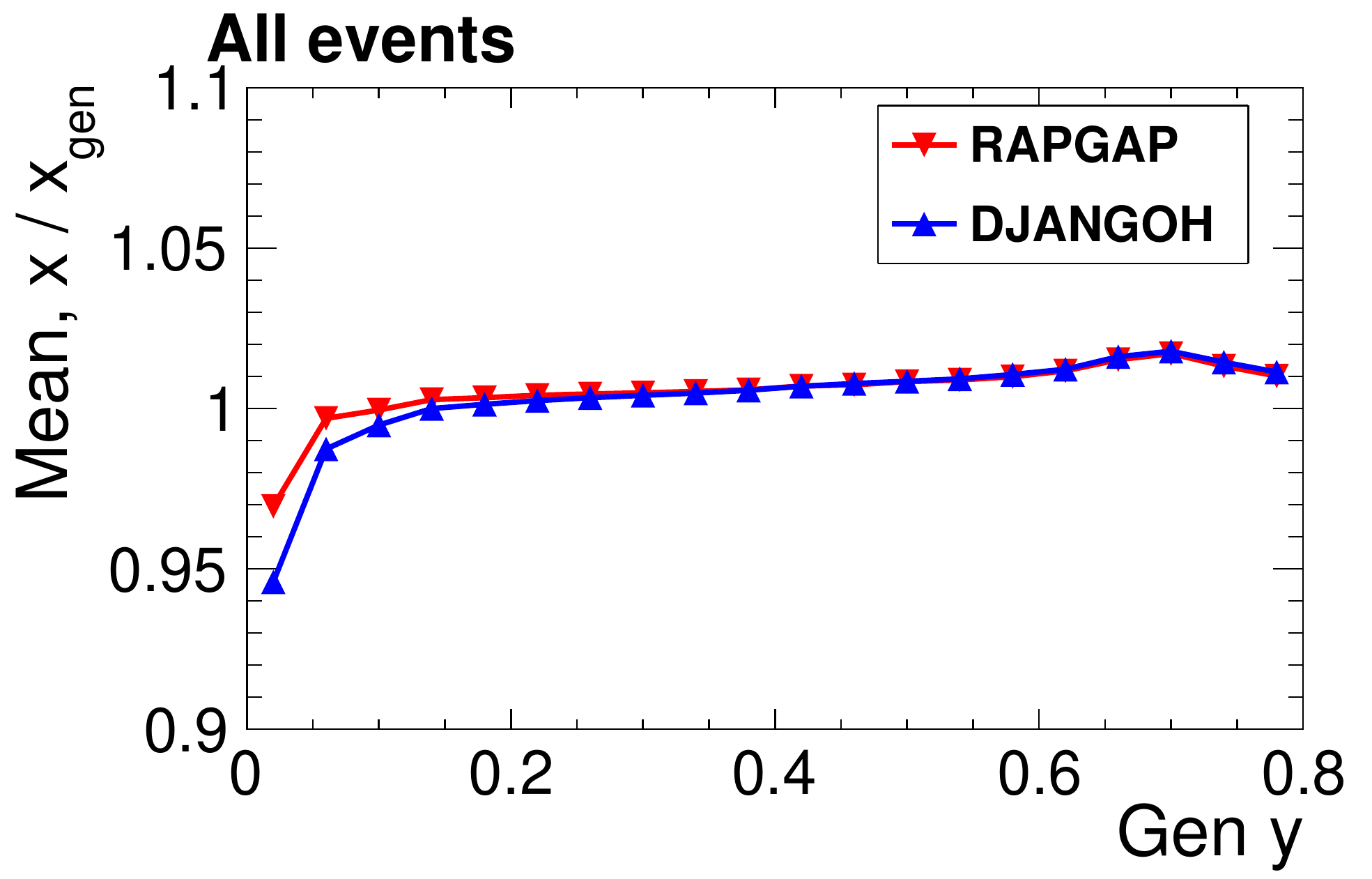} \\
    \caption{
     Resolution comparison of the \textsc{Rapgap} and \textsc{Djangoh}
     generators on the DNN reconstruction of $Q^2$ (left), $y$ (middle),
     and $x$ (right) as a function of the generated $y$ for the full simulation of H1.
    The top (bottom) row shows the RMS (mean) of the
    measured-over-generated distribution as a function of the generated $y$.  
    The red (blue) curves show the results for the \textsc{Rapgap}
    (\textsc{Djangoh}) event sample, while the DNN was trained with
    the \textsc{Rapgap} sample in both cases. 
    }
    \label{fig:h1-xyQ2-resolution-vs-y-comparison}
\end{figure}

Figure~\ref{fig:h1-xyQ2-resolution-vs-y-comparison} shows the \Qxy
resolutions as a function of the generated $y$ for both 
the \textsc{Rapgap} and \textsc{Djangoh} samples when using the same DNN, where the
measured-over-generated ratio is calculated on an event-by-event basis
with the respective generated values.
The results from the \textsc{Djangoh} event sample are
nearly indistinguishable from the \textsc{Rapgap} sample.
In particular, the mean of the distribution is unbiased.
This result suggests that any generator-specific systematic errors in
the DNN predictions is negligible.

The resolution (RMS) for $y$ and \xbj increase at
lower $y$, even for the DNN reconstruction.
Since this pattern is not present in the the ATHENA fast simulation results
and may be attributed to further acceptance, noise, or resolution effects that
deteriorates the measurement of the HFS~\cite{energyflow}. 
A dedicated study using a \textsc{Delphes} fast simulation is presented in the following section.


\section[Impact of further acceptance and resolution effects at low $y$]
        {\boldmath Impact of further acceptance and resolution effects at low $y$}
\label{sec:calo-noise}

At low $y$, the HFS-based methods perform better than the electron method.
The reason is, that the ratio $E(1-\cos\theta)/2E_0$ gets close to one
and cannot be measured accurately because of large values of $E$.
Likewise, however, the HFS momentum balance $\Sigma$ goes to zero as $y$ goes to zero.
Although for kinematic reconstruction  at low $y$ the usage of  $\Sigma$  is
preferred over $\Sigma_e$, the quantity $\Sigma$
is particularly sensitive to resolution 
and acceptance effects. In particular HFS components that are more in
the central region of the calorimeter contribute more, such making
$\Sigma$ at low $y$ especially sensitive resolution effects or
efficiency losses in the central part of the detector or fake
components from calorimetric noise. 

\begin{figure}[tbhp!]
    \centering
    \includegraphics[width=0.9\textwidth]{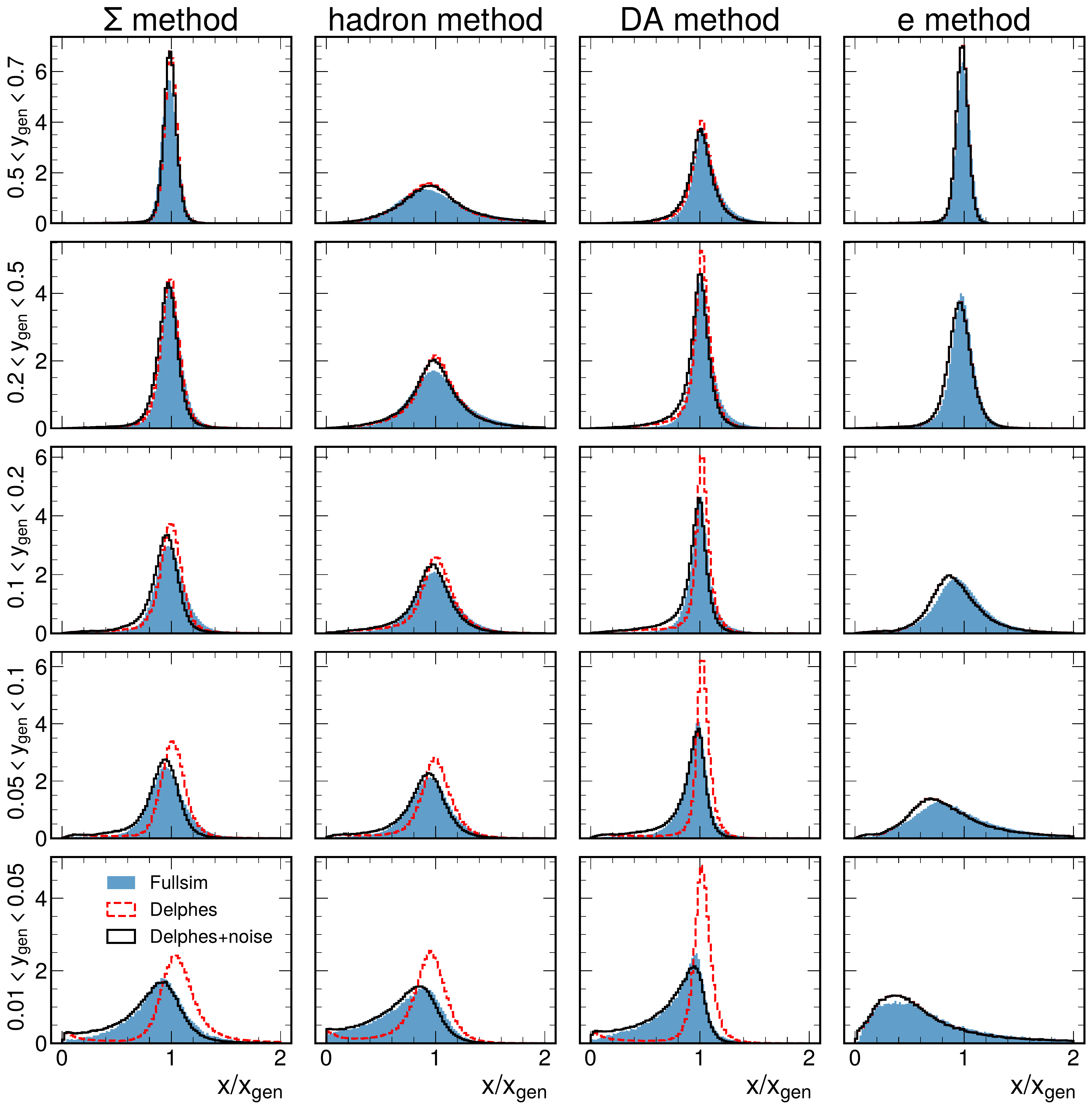}
    \caption{Resolutions for $x$ in various $y$ ranges and reconstruction methods obtained with a full \textsc{Geant} simulation of H1 (shaded area) and a fast simulation based on \textsc{Delphes } with (full line) and without (dashed line) an added ad-hoc noise contribution.}
    \label{fig:H1fastsim-with-noise}
\end{figure}
The \textsc{Delphes} fast simulation does not include calorimeter
noise hits, nor does it account in a full-fledged manner for single-particle
acceptance effects and efficiency losses as they can be present at the boundaries of
calorimeter stacks or because of insensitive material.
To test the hypothesis that such detector effects can be responsible for the resolution
decrease in $x$ and $y$ for hadronic reconstruction methods at low $y$
($y\lesssim0.15$), we have implemented the H1 experiment in \textsc{Delphes}.
Figure~\ref{fig:H1fastsim-with-noise} shows the $x$
resolution for the standard reconstruction methods for our fast simulation
of H1 compared to the full simulation.
The agreement between the fast and full simulation at high $y$ is fairly good.
At low $y$, however, there is a low-side tail for the $\Sigma$, hadron and DA method
for events processed with the full H1 simulation, while that tail is absent in the fast
simulation, and also the mean value is shifted.

We apply an additional additive component with random sign to the HFS
to the fast simulated events, 
which mimics further detector effects, like acceptance or efficiency losses,
reduced resolution or artificial components from electronic noise in the calorimeter.
The model we use to simulate these effects is to generate random
numbers from
\texttt{TRandom::Landau} in \textsc{Root}~\cite{citeulike:363715} with
\texttt{mu=0} and \texttt{sigma=0.05} in units of GeV,
and add it with a random sign to the $p_x$, $p_y$, and
$p_z$ components of the reconstructed HFS four-vector. 

The results are also displayed in Figure~\ref{fig:H1fastsim-with-noise}.
We find that adding such an additional detector effect to the HFS
in the fast simulation brings the fast simulation into good agreement
with the full detector simulation.  Adding these ad-hoc detector effects produces a low-side tail in the
$x$ resolution at low $y$ but does 
not affect the $x$ reconstruction at high $y$.
The electron remains naturally unaltered in that procedure.

This study suggests that further detector effects that are otherwise
not included in the \textsc{Delphes} fast simulation impact the
precise measurement of the HFS and reduce the $x$ and $y$ resolution at
low $y$. We do not currently have an estimate for how large that effect in
ATHENA will be, what is the actual correspondence in the full
simulation (an acceptance, efficiency, resolution or noise effect),
or to which extent it is impacted by calibration or noise-suppression algorithms.
Though, due to a larger coverage of the calorimeter, less dead
material and newer detector technologies of ATHENA as compared to H1,
we expect this additive component to be significantly smaller for
ATHENA than what our ad-hoc model adds to H1 to bring the 
fast and full simulation into agreement. 

\begin{figure}
    \centering
        {\large \fontfamily{lmss}\selectfont ATHENA fast simulation
          with additive resolution effect (Rapgap+Delphes)}
    \vspace{1mm} 
    \includegraphics[width=0.32\textwidth]{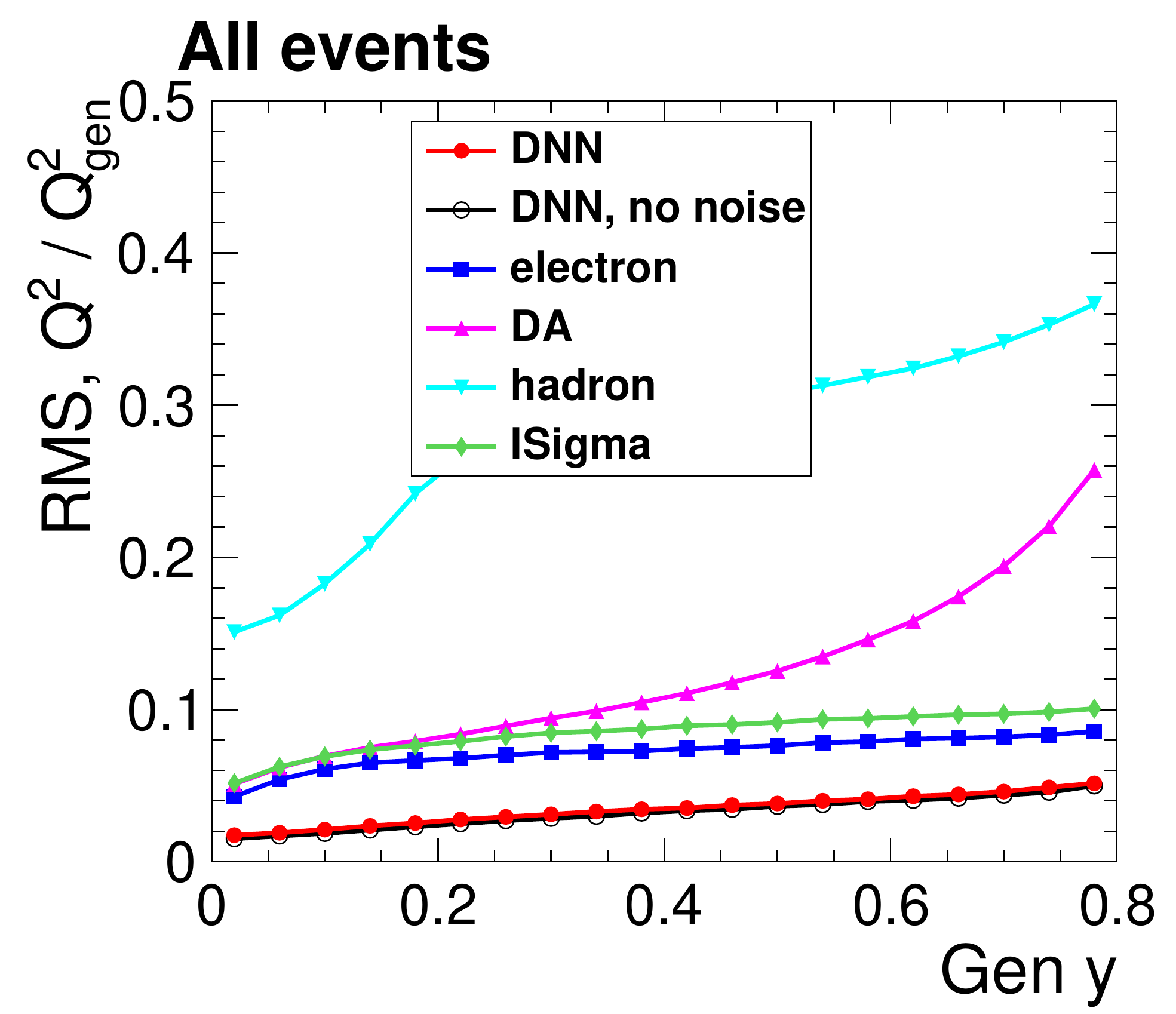}
    \includegraphics[width=0.32\textwidth]{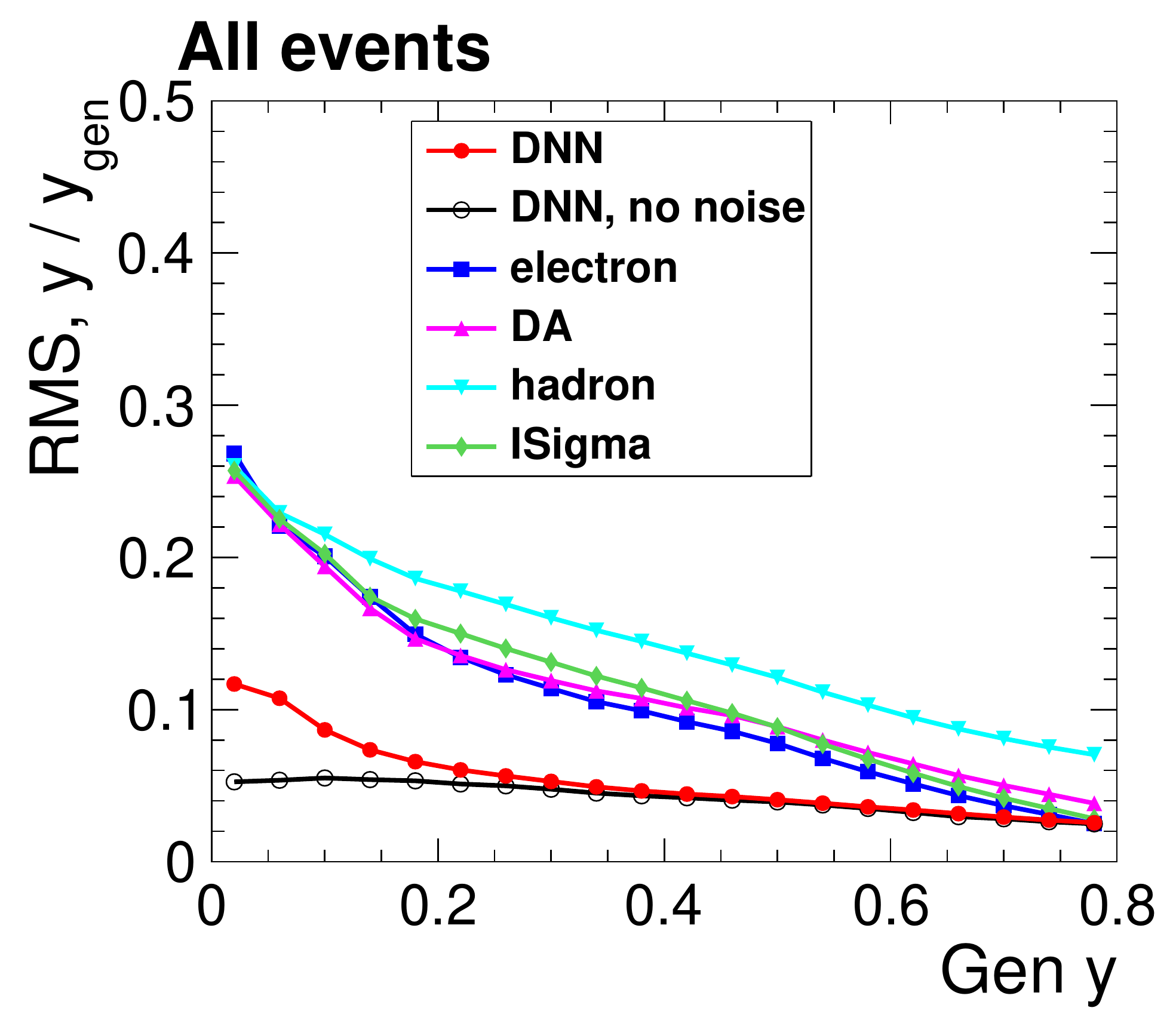}    
    \includegraphics[width=0.32\textwidth]{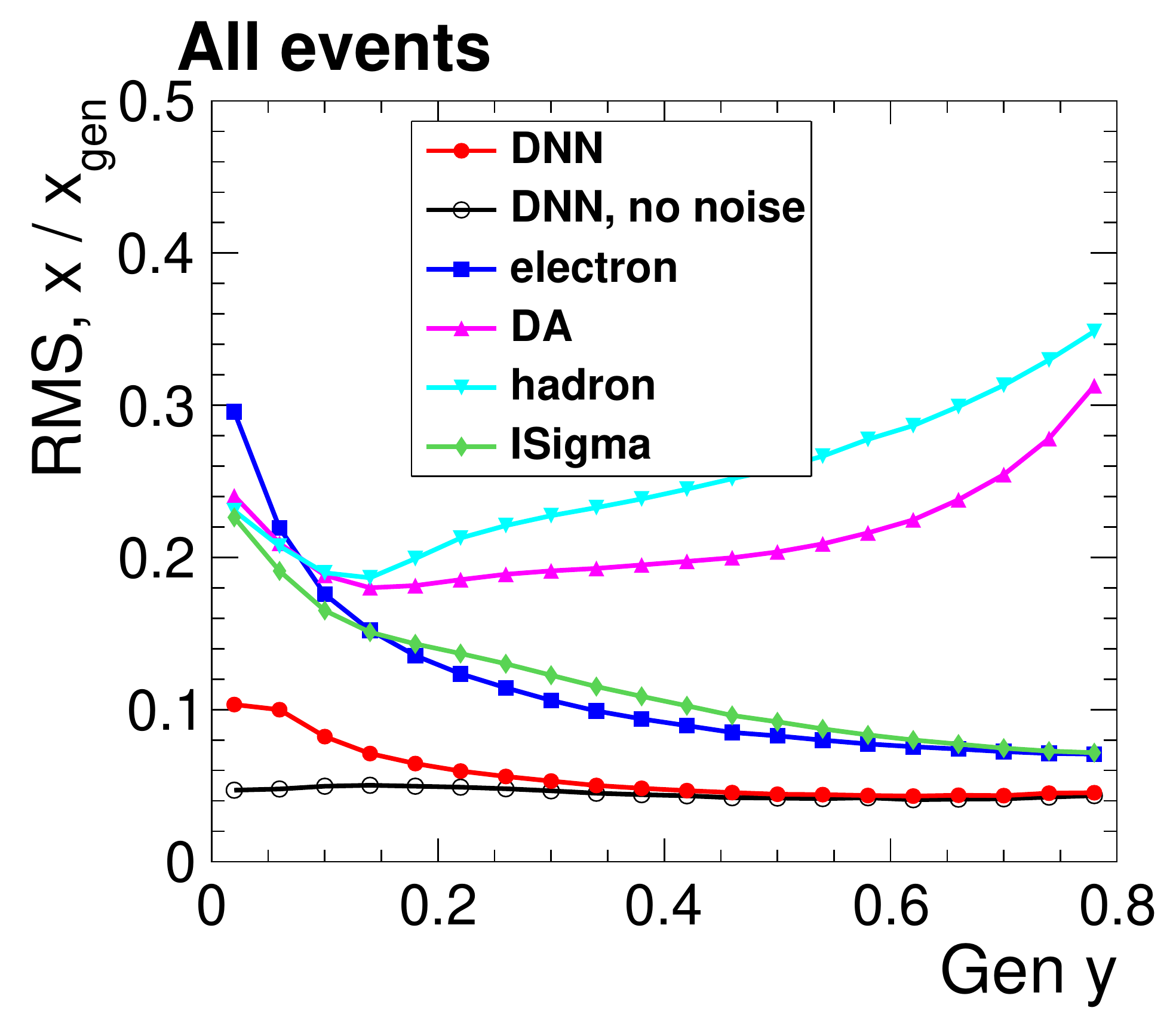} \\    
    \includegraphics[width=0.32\textwidth]{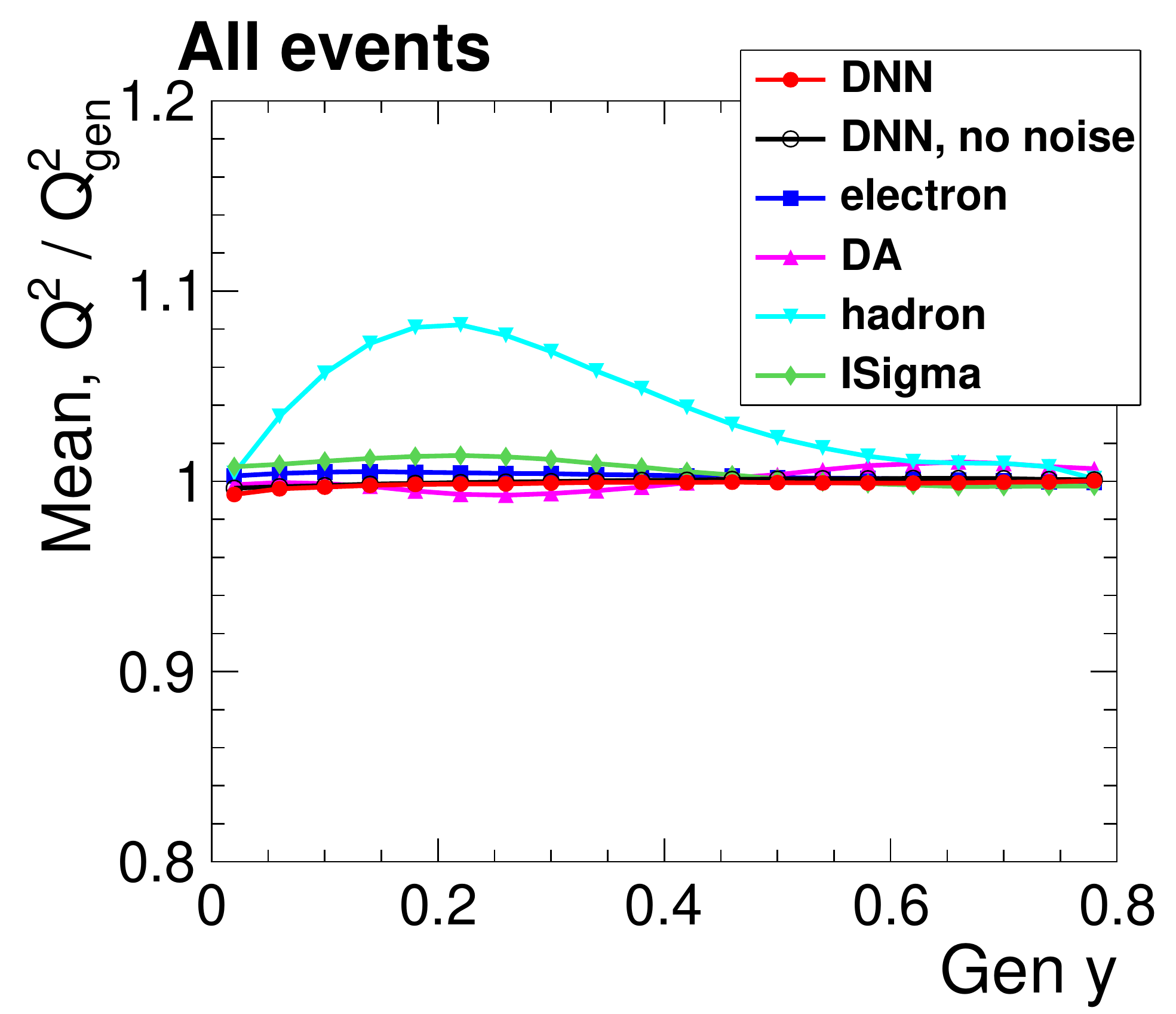}
    \includegraphics[width=0.32\textwidth]{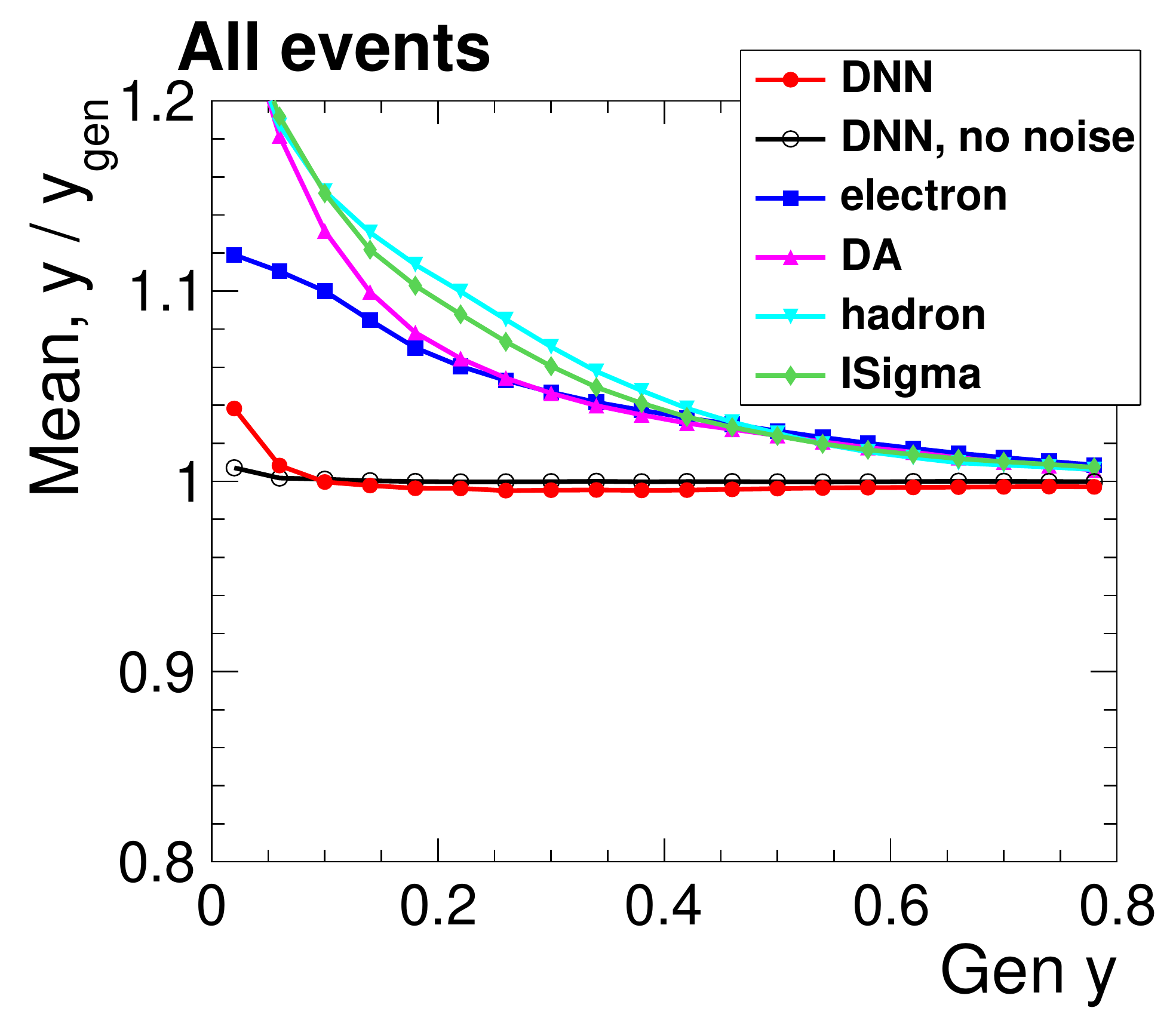}    
    \includegraphics[width=0.32\textwidth]{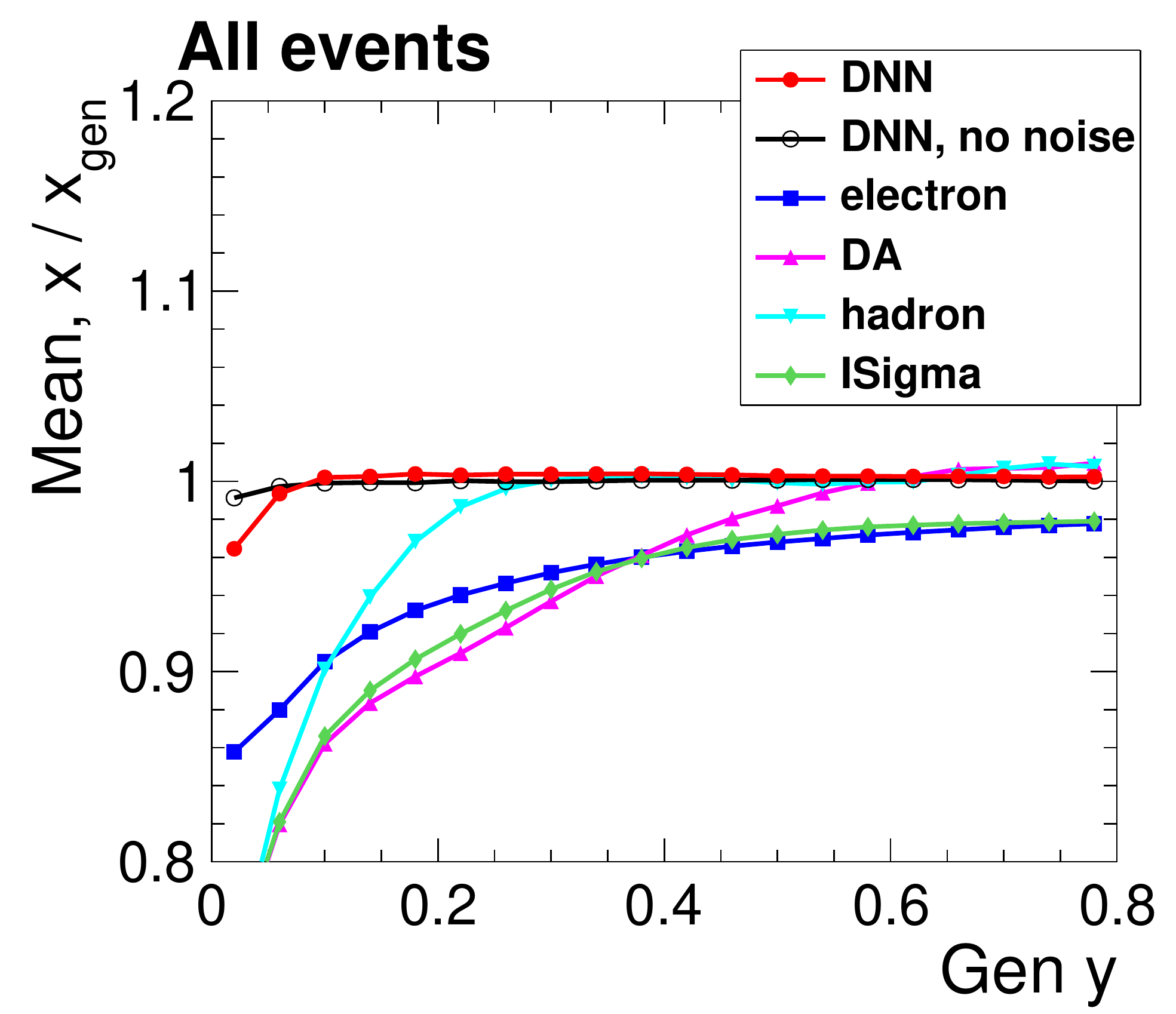} \\  
    \caption{
    Resolution on the reconstruction of $Q^2$ (left), $y$ (middle),
    and $x$ (right) as a function of the generated $y$ for the fast simulation of ATHENA with the ad-hoc additive resolution effects model added.
    The top (bottom) row shows the RMS (mean) of the measured-over-generated distribution
    as a function of the generated $y$.
                For comparison, the DNN curves for that unaltered sample are shown in black for comparison.
    }
    \label{fig:athena-xyQ2-resolution-vs-y-calo-noise-comp}
\end{figure}

To place an upper bound on the impact of these effects in the ATHENA
results, we investigated adding our ad-hoc component to the ATHENA fast simulation.
The full analysis is then repeated, including the DNN training. 
The results are shown in
Figure~\ref{fig:athena-xyQ2-resolution-vs-y-calo-noise-comp} for both
the conventional reconstruction methods and the DNN reconstruction,
where the DNN reconstruction for the unaltered ATHENA sample is
included for comparison. 
The DNN resolution does get worse below $y$ of around 0.2 and there is a small bias at very low $y$, as we expected based on the H1 study.
The $Q^2$ reconstruction is insensitive to that, since it is dominated
by the electron reconstruction.
The results in Figure~\ref{fig:athena-xyQ2-resolution-vs-y-calo-noise-comp} show
that also with a very conservative ad-hoc model, the DNN
outperforms all standard reconstruction methods.

\FloatBarrier

\section{Summary and outlook}
\label{sec:conclusion}
We have presented a novel method to reconstruct the DIS kinematic variables ($Q^2$, $y$ and $x$) using a deep neural network (DNN) that takes as input the electron and hadronic-final-state measurements as well as observables that can indicate the presence of QED radiation. 

We have introduced our methodology using a \textsc{Delphes}-based fast simulation of the ATHENA experiment for the EIC and validated our methods using the well-understood full simulation of the H1 experiment at HERA. 

Our method outperforms traditional methods over a wide kinematic range, improving the resolution, and decreasing the bias.
We validated that our method is independent of the Monte Carlo event generator used to train the DNN. We further performed a study to validate the fast simulation approach by comparing a \textsc{Delphes} model of the H1 detector with the H1 full simulation; these comparisons allowed us to identify key effects for low $y$ events. 

Our DNN-based method shows promise for improving the resolution and extending the kinematic reach of flagship EIC measurements. Given that EIC experiments are being designed, our method offers a way to benchmark detector performance against DIS using an optimal combination of electron and HFS measurements.



\section*{Code availability}
The code in this work can be found in: 
\url{https://github.com/owen234/DIS-reco-paper/}
\section*{Acknowledgments}
We thank our colleagues from the H1 Collaboration for allowing us to use the
simulated MC event samples and for providing valuable feedback to the
analysis an on the manusript, in particular G{\"u}nter Grindhammer, Sergey Levonian, Stefan Schmitt and Zhiqing Zhang.
We also thank members of the ATHENA Collaboration, and in particular Steven Sekula, for help with the ATHENA \textsc{Delphes} model. We thank Hubert Spiesberger for valuable discussions about QED radiative effects and insights into the \textsc{Heracles} routines.
We thank Hannes Jung for providing the \textsc{Rapgap} event generator.
Thanks to DESY-IT and MPI f{\"u}r Physik for providing some computing infrastructure and supporting the data preservation project of the HERA experiments.
M.A was supported through DOE Contract No. DE-AC05-06OR23177 under which JSA operates the Thomas Jefferson National Accelerator Facility, and by the University of California, Office of the President award number 00010100. B.N. was supported by the Department of Energy, Office of Science under contract number DE-AC02-05CH11231.

\newpage
\appendix
\section{Detailed resolution plots for the ATHENA fast simulation}
\label{sec:additionalfigsAthena}

\begin{figure}[hpt!]
    \centering
    \includegraphics[width=0.90\textwidth]{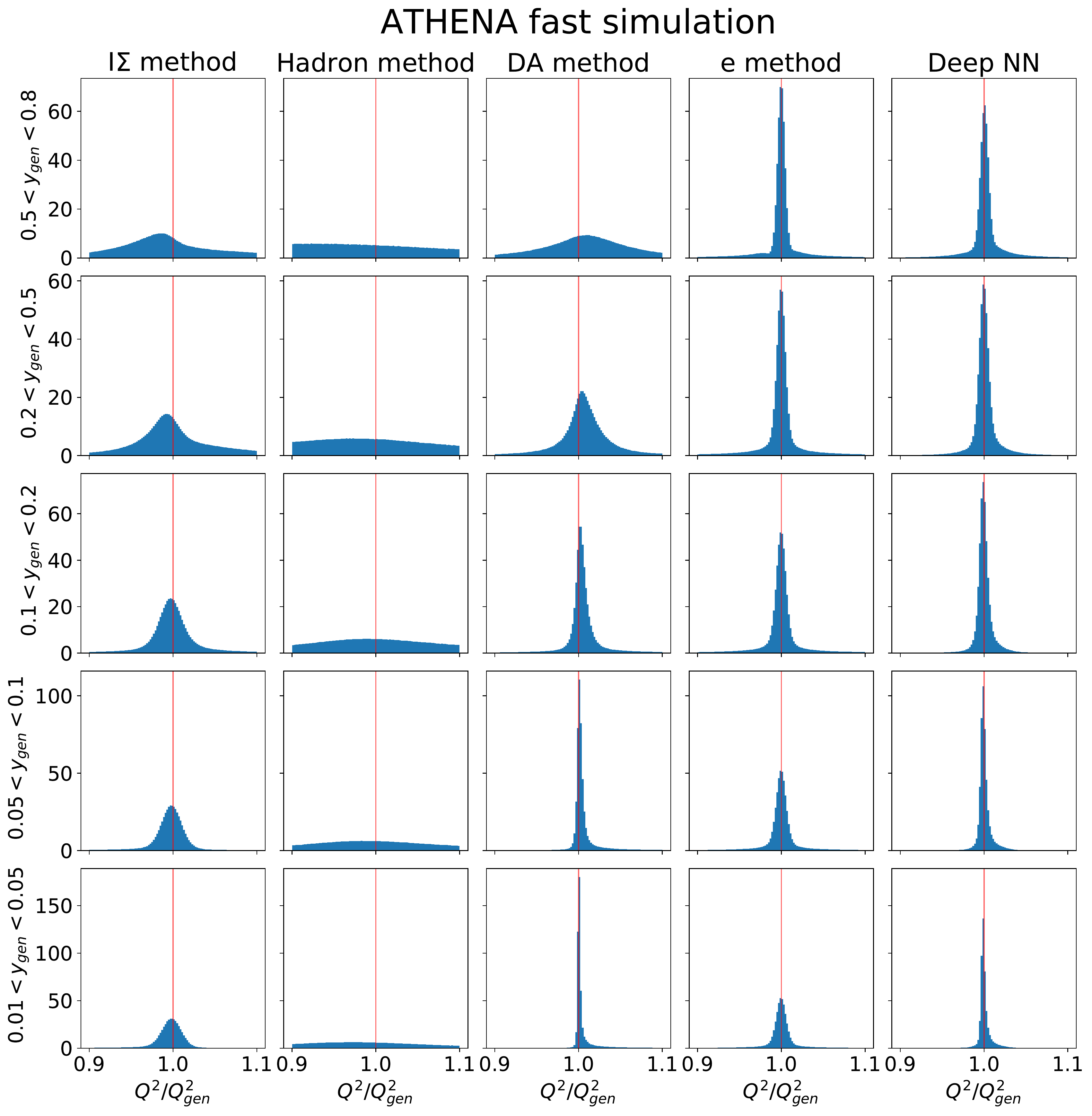}
    \caption{Resolutions for $Q^2$ in various $y$ ranges and reconstruction methods from the  \textsc{Delphes} fast simulation of ATHENA for events with $Q^2>200~{\rm GeV}^2$. The tails in the distributions are from events with ISR and FSR radiation.
    }
    \label{fig:deepNN-Q2-resolution-ATHENA-Q2gt200-allevts}
\end{figure}

\begin{figure}[hpt!]
    \centering
    \includegraphics[width=0.90\textwidth]{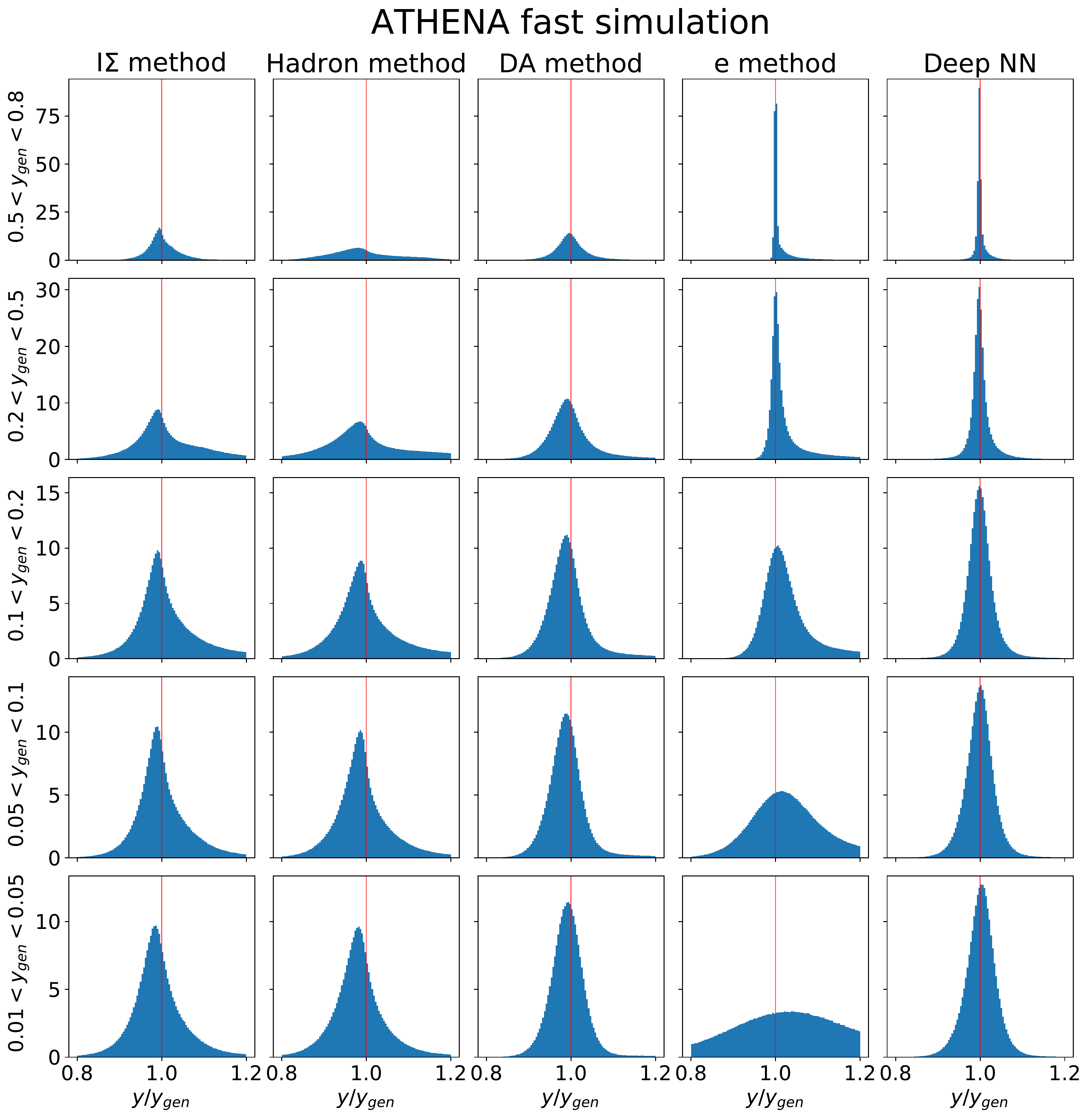}
    \caption{Resolutions for $y$ in various $y$ ranges and reconstruction methods from the  \textsc{Delphes} fast simulation of ATHENA for events with $Q^2>200~{\rm GeV}^2$. The tails in the distributions are from events with ISR and FSR radiation.
    }
    \label{fig:deepNN-y-resolution-ATHENA-Q2gt200-allevts}
\end{figure}

\begin{figure}[hpt!]
    \centering
    \includegraphics[width=0.90\textwidth]{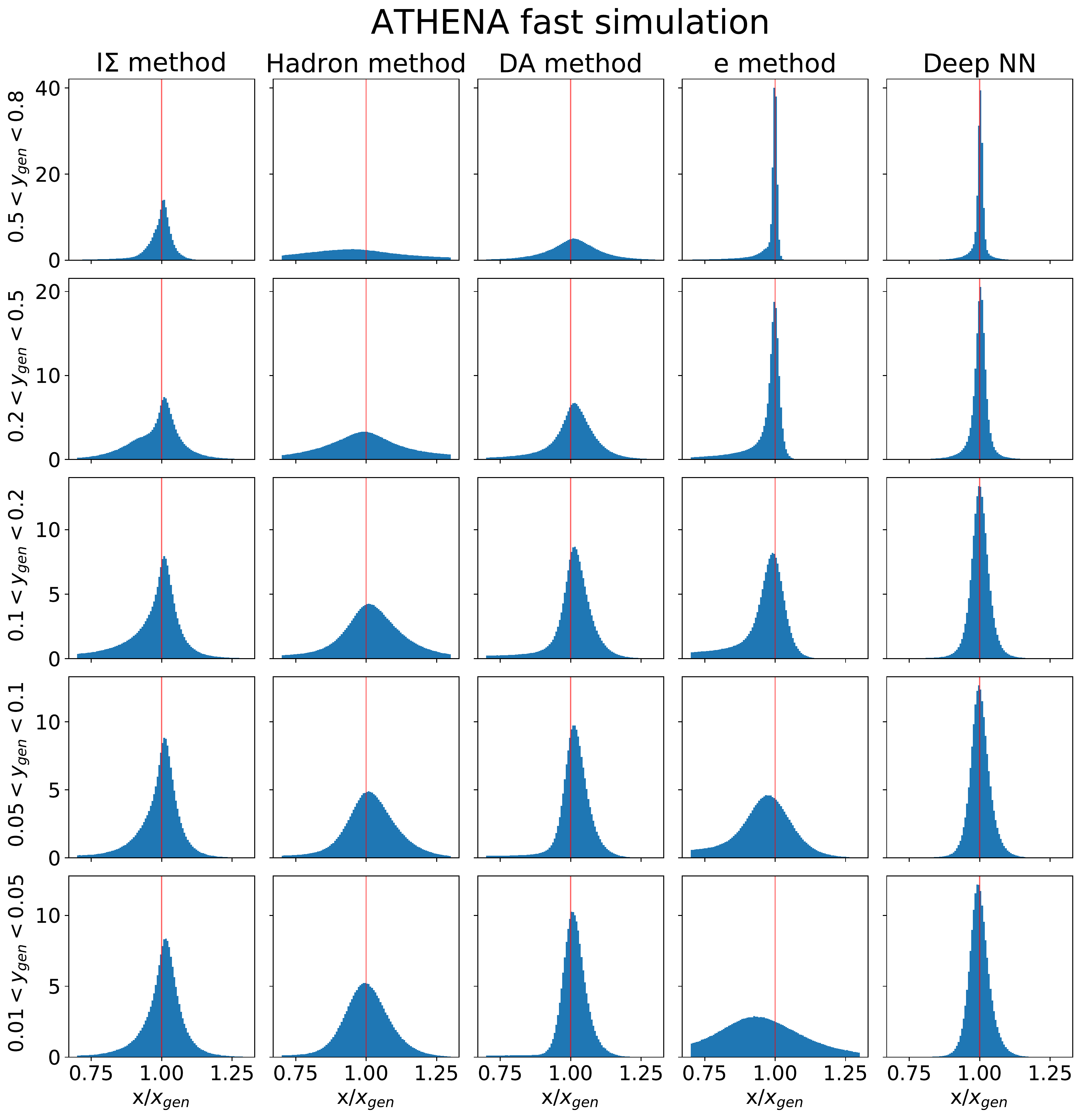}
    \caption{Resolutions for $x$ in various $y$ ranges and reconstruction methods from the  \textsc{Delphes} fast simulation of ATHENA for events with $Q^2>200~{\rm GeV}^2$. The tails in the distributions are from events with ISR and FSR radiation.
    }
    \label{fig:deepNN-x-resolution-ATHENA-Q2gt200-allevts}
\end{figure}

In this section detailed resolution plots for the variables
\Qsq\ (Figure~\ref{fig:deepNN-Q2-resolution-ATHENA-Q2gt200-allevts}),
$y$ (Figure~\ref{fig:deepNN-y-resolution-ATHENA-Q2gt200-allevts})
and $\xbj$
(Figure~\ref{fig:deepNN-x-resolution-ATHENA-Q2gt200-allevts})
for the ATHENA fast simulation at the EIC with $\sqrt{s}=141\,\GeV$  are shown.
Our DNN-reconstruction method is compared  
to four widely used basic reconstruction methods (I$\Sigma$, hadron,
double-angle and electron-method) for $\Qsq>200\,\GeVsq$ in five
kinematic ranges in $y_\text{gen}$.

\clearpage
\section{Detailed resolution plots for the H1 full simulation}
\label{sec:additionalfigsH1}
\begin{figure}[hpt!]
    \centering
    \includegraphics[width=0.90\textwidth]{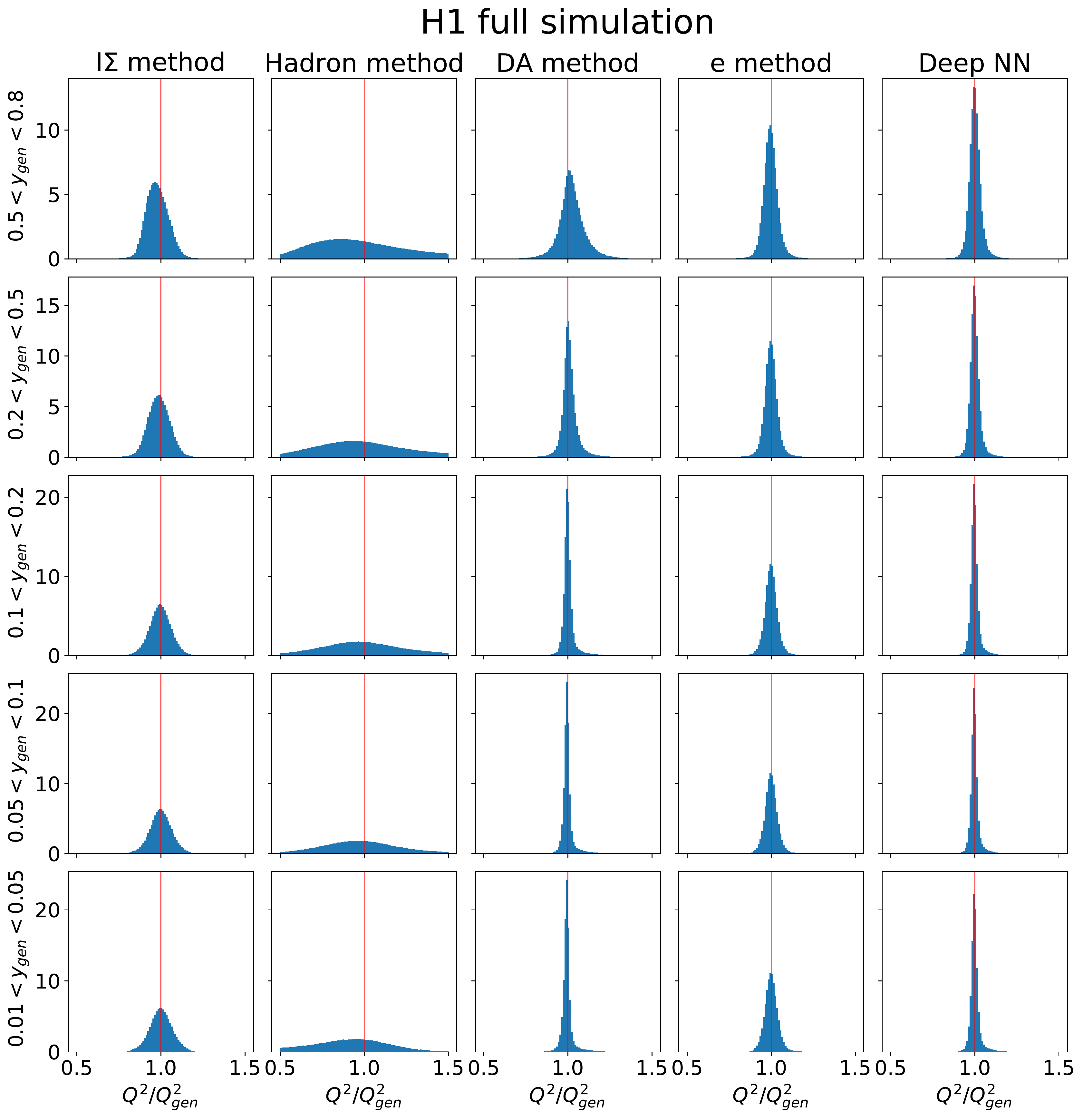}
    \caption{Resolutions for $Q^2$ in various $y$ ranges and reconstruction methods from the full simulation of H1 for events with $Q^2>200~{\rm GeV}^2$.
    }
    \label{fig:deepNN-Q2-resolution-H1-Q2gt200-allevts}
\end{figure}

\begin{figure}[hpt!]
    \centering
    \includegraphics[width=0.90\textwidth]{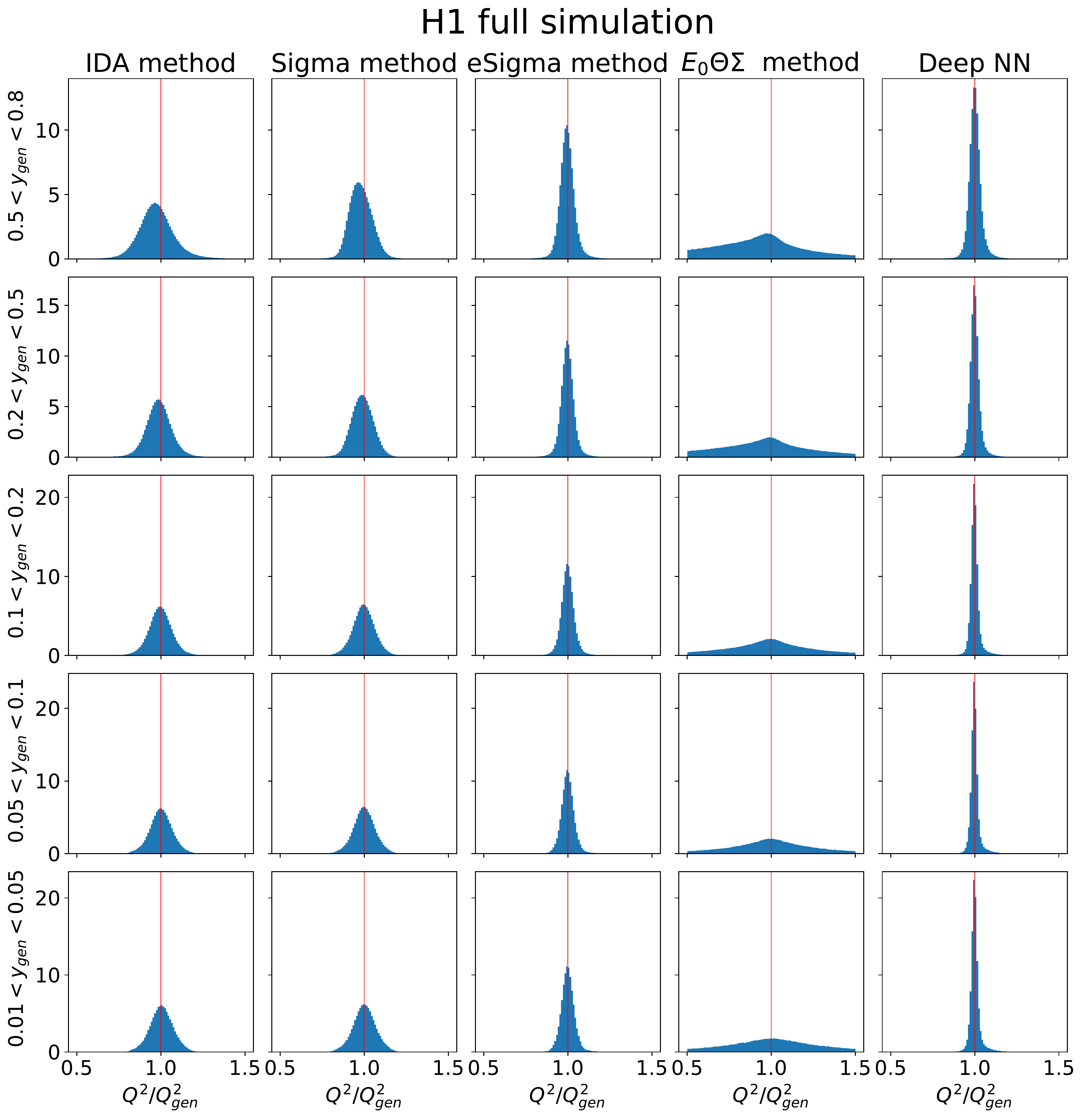}
    \caption{Resolutions for $Q^2$ in various $y$ ranges and reconstruction methods from the full simulation of H1 for events with $Q^2>200~{\rm GeV}^2$.
    }
    \label{fig:deepNN-Q2-resolution-H1-Q2gt200-allevts-set2}
\end{figure}

\begin{figure}[hpt!]
    \centering
    \includegraphics[width=0.90\textwidth]{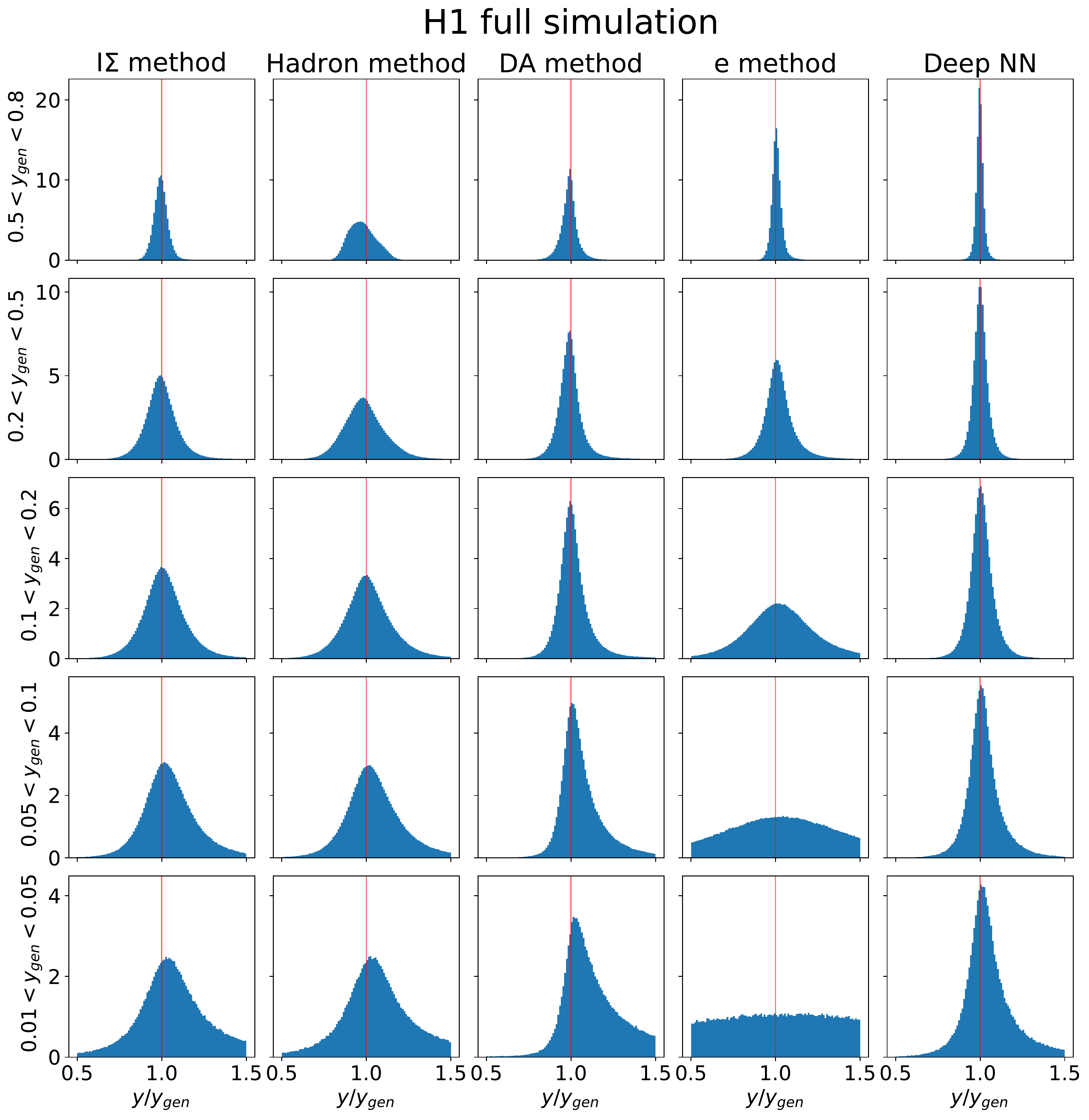}
    \caption{Resolutions for $y$ in various $y$ ranges and reconstruction methods from the full simulation of H1 for events with $Q^2>200~{\rm GeV}^2$.
    }
    \label{fig:deepNN-y-resolution-H1-Q2gt200-allevts}
\end{figure}

\begin{figure}[hpt!]
    \centering
    \includegraphics[width=0.90\textwidth]{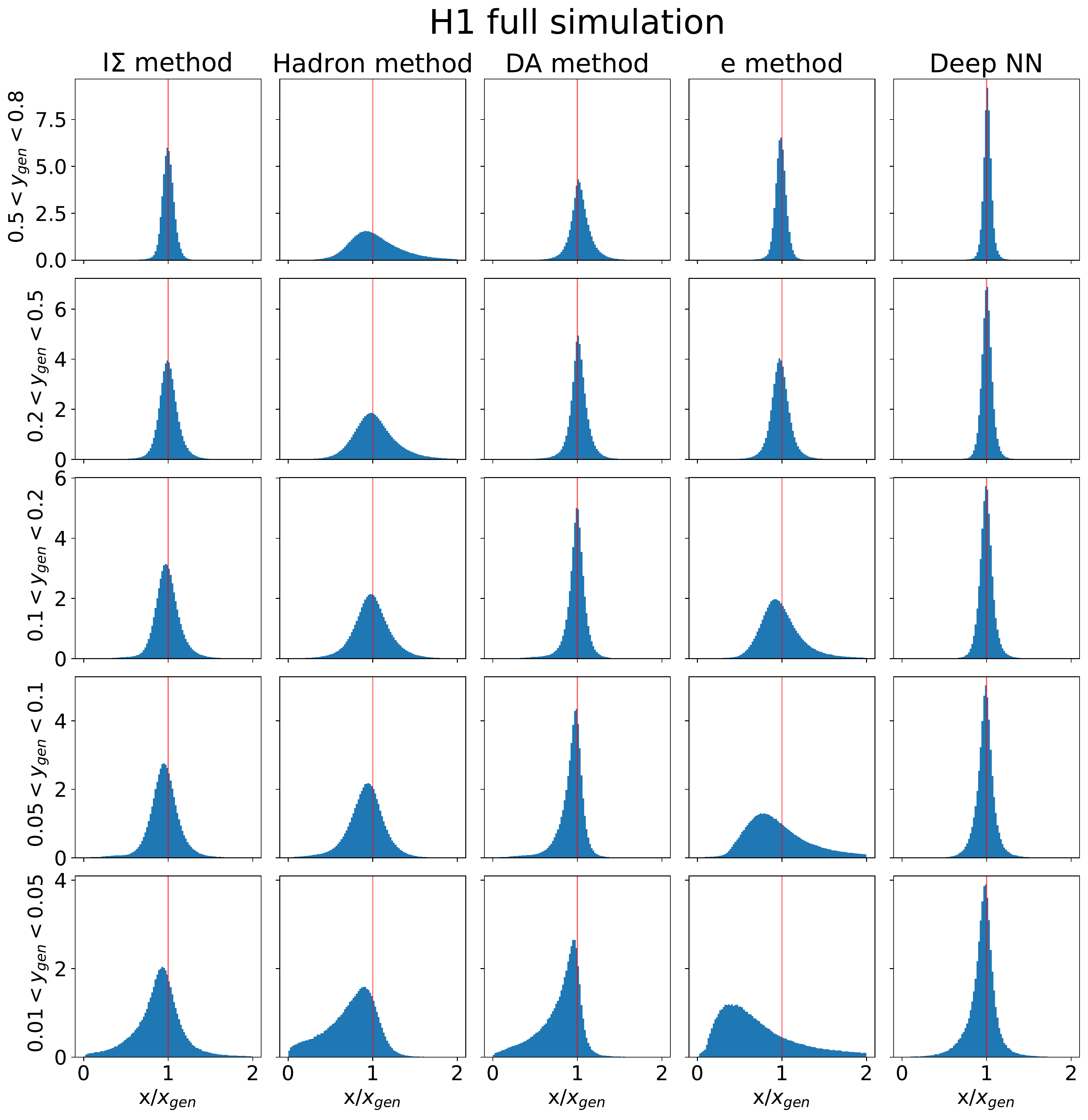}
    \caption{Resolutions for $x$ in various $y$ ranges and reconstruction methods from the full simulation of H1 for events with $Q^2>200~{\rm GeV}^2$.
    }
    \label{fig:deepNN-x-resolution-H1-Q2gt200-allevts}
\end{figure}

\begin{figure}[hpt!]
    \centering
    \includegraphics[width=0.90\textwidth]{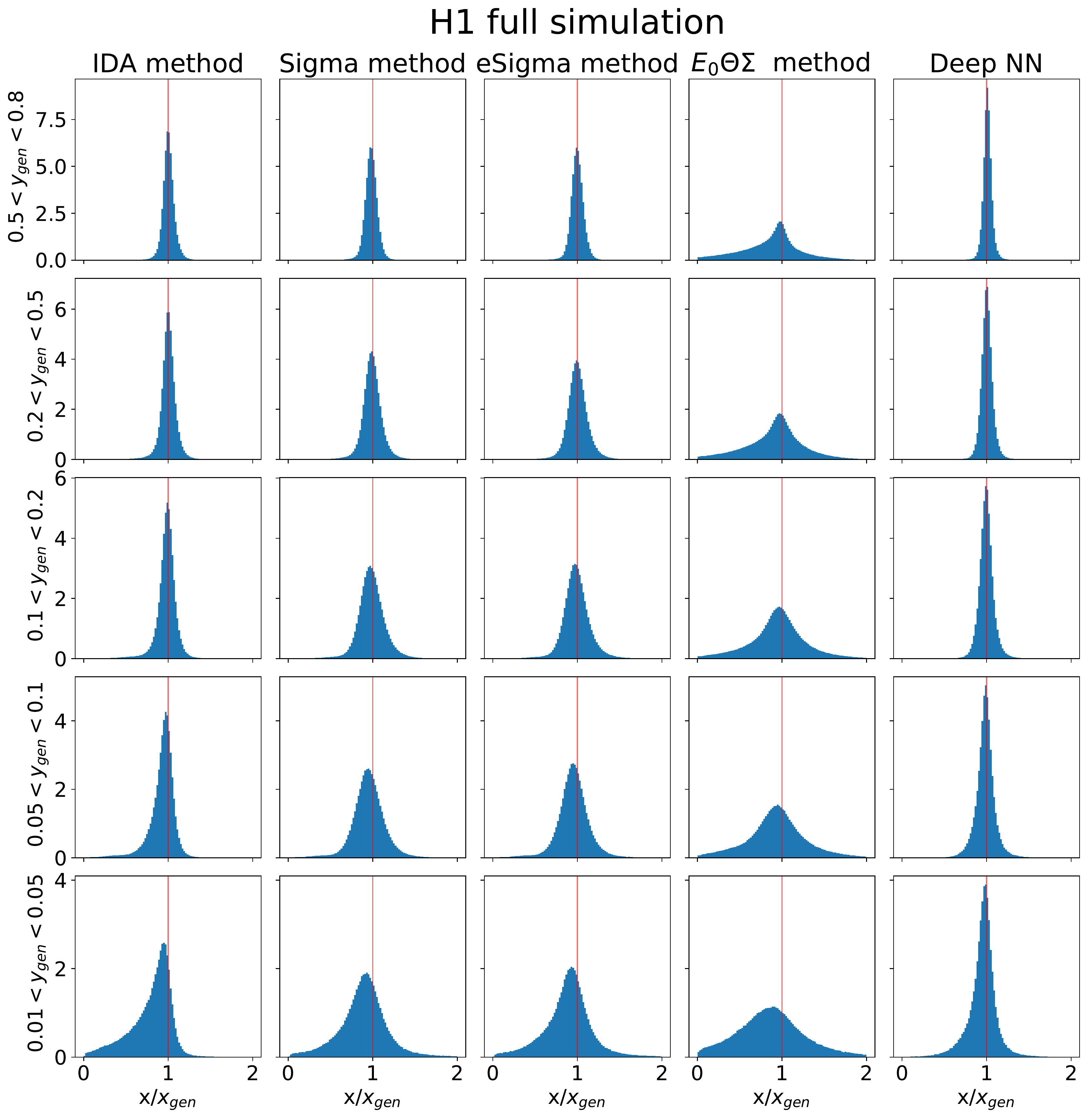}
    \caption{Resolutions for $x$ in various $y$ ranges and reconstruction methods from the full simulation of H1 for events with $Q^2>200~{\rm GeV}^2$.
    }
    \label{fig:deepNN-x-resolution-H1-Q2gt200-allevts-set2}
\end{figure}

In this section detailed resolution plots for the variables
\Qsq\ (Figures~\ref{fig:deepNN-Q2-resolution-H1-Q2gt200-allevts}
and~\ref{fig:deepNN-Q2-resolution-H1-Q2gt200-allevts-set2}), 
$y$ (Figure~\ref{fig:deepNN-y-resolution-H1-Q2gt200-allevts})
and $\xbj$
(Figures~\ref{fig:deepNN-x-resolution-H1-Q2gt200-allevts}
and~\ref{fig:deepNN-x-resolution-H1-Q2gt200-allevts-set2}) 
for H1's full simulation at HERA at $\sqrt{s}=319\,\GeV$ are shown.
Our DNN-reconstruction method is compared  
to the full set of eight basic reconstruction methods (I$\Sigma$, hadron,
double-angle (DA), electron ($e$), IDA, $E_0E\Sigma$, $E_0\theta\Sigma$,
and $\theta\Sigma\gamma$-method) for $\Qsq>200\,\GeVsq$ in five 
kinematic ranges in $y_\text{gen}$.
The double-energy method is omitted since $E_h$ is not measureable du
to relevant acceptance losses through the forward beam-hole.
For $y$ there are only four unambiguous independent basic methods.

\FloatBarrier
 \bibliographystyle{elsarticle-num} 
 \bibliography{cas-refs,HEPML}

\begin{thebibliography}{10}
\expandafter\ifx\csname url\endcsname\relax
  \def\url#1{\texttt{#1}}\fi
\expandafter\ifx\csname urlprefix\endcsname\relax\def\urlprefix{URL }\fi
\expandafter\ifx\csname href\endcsname\relax
  \def\href#1#2{#2} \def\path#1{#1}\fi

\bibitem{ellis_stirling_webber_1996}
R.~K. Ellis, W.~J. Stirling, B.~R. Webber, QCD and Collider Physics, Cambridge
  University Press, 1996.
\newblock \href {https://doi.org/10.1017/CBO9780511628788}
  {\path{doi:10.1017/CBO9780511628788}}.

\bibitem{Devenish:2004pb}
R.~Devenish, A.~Cooper-Sarkar, {Deep inelastic scattering}, Oxford University
  Press, 2004.
\newblock \href {https://doi.org/10.1093/acprof:oso/9780198506713.001.0001}
  {\path{doi:10.1093/acprof:oso/9780198506713.001.0001}}.

\bibitem{ParticleDataGroup:2020ssz}
P.~A. Zyla, et~al., {Review of Particle Physics}, PTEP 2020 (2020) 083C01.
\newblock \href {https://doi.org/10.1093/ptep/ptaa104}
  {\path{doi:10.1093/ptep/ptaa104}}.

\bibitem{JB:1979}
A.~Blondel, F.~Jacquet, {DETECTION AND STUDY OF THE CHARGED CURRENT EVENT },
  in: U.~Amaldi, et~al. (Eds.), {Proceedings of the Study of an $ep$ Facility
  for Europe}, 1979, p. 391.

\bibitem{Blumlein:1990dj}
J.~Bl{\"u}mlein, M.~Klein, {Kinematics and resolution at future $ep$
  colliders}, in: {1990 DPF Summer Study on High-energy Physics: Research
  Directions for the Decade (Snowmass 90)}, 1990, pp. 549--551.

\bibitem{Hoeger:1991wj}
K.~C. Hoeger, {Measurement of $x$, $y$, $Q^2$ in neutral current events}, in:
  {Workshop on Physics at HERA 1991}, 1991.

\bibitem{Bentvelsen:1992fu}
S.~Bentvelsen, J.~Engelen, P.~Kooijman, {Reconstruction of ($x, Q^2$) and
  extraction of structure functions in neutral current scattering at HERA}, in:
  {Workshop on Physics at HERA 1991}, 1992.

\bibitem{Bassler:H1intnote93}
U.~Bassler, G.~Bernardi, {Progress on Kinematical Variables Reconstruction.
  Consequences for D.I.S. Physics Analysis at Low x}, H1 internal note,
  H1-03/93-274 (Mar 1993).

\bibitem{Bassler:1994uq}
U.~Bassler, G.~Bernardi, {On the kinematic reconstruction of deep inelastic
  scattering at HERA: The Sigma method}, Nucl. Instrum. Meth. A 361 (1995)
  197--208.
\newblock \href {http://arxiv.org/abs/hep-ex/9412004}
  {\path{arXiv:hep-ex/9412004}}, \href
  {https://doi.org/10.1016/0168-9002(95)00173-5}
  {\path{doi:10.1016/0168-9002(95)00173-5}}.

\bibitem{ZEUS:1996uid}
M.~Derrick, et~al., {Measurement of the F2 structure function in deep inelastic
  $e^+$ $p$ scattering using 1994 data from the ZEUS detector at HERA}, Z.
  Phys. C 72 (1996) 399.
\newblock \href {http://arxiv.org/abs/hep-ex/9607002}
  {\path{arXiv:hep-ex/9607002}}, \href {https://doi.org/10.1007/s002880050260}
  {\path{doi:10.1007/s002880050260}}.

\bibitem{Bassler:1997tv}
U.~Bassler, G.~Bernardi, {Structure function measurements and kinematic
  reconstruction at HERA}, Nucl. Instrum. Meth. A 426 (1999) 583--598.
\newblock \href {http://arxiv.org/abs/hep-ex/9801017}
  {\path{arXiv:hep-ex/9801017}}, \href
  {https://doi.org/10.1016/S0168-9002(99)00044-3}
  {\path{doi:10.1016/S0168-9002(99)00044-3}}.

\bibitem{Accardi:2012qut}
A.~Accardi, et~al., {Electron Ion Collider: The Next QCD Frontier}:
  {Understanding the glue that binds us all}, Eur. Phys. J. A 52 (2016) 268.
\newblock \href {http://arxiv.org/abs/1212.1701} {\path{arXiv:1212.1701}},
  \href {https://doi.org/10.1140/epja/i2016-16268-9}
  {\path{doi:10.1140/epja/i2016-16268-9}}.

\bibitem{AbdulKhalek:2021gbh}
R.~Abdul~Khalek, et~al., {Science Requirements and Detector Concepts for the
  Electron-Ion Collider: EIC Yellow Report} (Mar 2021).
\newblock \href {http://arxiv.org/abs/2103.05419} {\path{arXiv:2103.05419}}.

\bibitem{Anderle:2021wcy}
D.~P. Anderle, et~al., {Electron-ion collider in China}, Front. Phys. (Beijing)
  16 (2021) 64701.
\newblock \href {http://arxiv.org/abs/2102.09222} {\path{arXiv:2102.09222}},
  \href {https://doi.org/10.1007/s11467-021-1062-0}
  {\path{doi:10.1007/s11467-021-1062-0}}.

\bibitem{LHeCStudyGroup:2012zhm}
J.~L. Abelleira~Fernandez, et~al., {A Large Hadron Electron Collider at CERN:
  Report on the Physics and Design Concepts for Machine and Detector}, J. Phys.
  G 39 (2012) 075001.
\newblock \href {http://arxiv.org/abs/1206.2913} {\path{arXiv:1206.2913}},
  \href {https://doi.org/10.1088/0954-3899/39/7/075001}
  {\path{doi:10.1088/0954-3899/39/7/075001}}.

\bibitem{LHeC:2020van}
P.~Agostini, et~al., {The Large Hadron-Electron Collider at the HL-LHC} (Jul
  2020).
\newblock \href {http://arxiv.org/abs/2007.14491} {\path{arXiv:2007.14491}}.

\bibitem{deOliveira:2018lqd}
L.~De~Oliveira, B.~Nachman, M.~Paganini, {Electromagnetic Showers Beyond Shower
  Shapes}, Nucl. Instrum. Meth. A 951 (2020) 162879.
\newblock \href {http://arxiv.org/abs/1806.05667} {\path{arXiv:1806.05667}},
  \href {https://doi.org/10.1016/j.nima.2019.162879}
  {\path{doi:10.1016/j.nima.2019.162879}}.

\bibitem{Belayneh:2019vyx}
D.~Belayneh, et~al., {Calorimetry with Deep Learning: Particle Simulation and
  Reconstruction for Collider Physics} (Dec 2019).
\newblock \href {http://arxiv.org/abs/1912.06794} {\path{arXiv:1912.06794}},
  \href {https://doi.org/10.1140/epjc/s10052-020-8251-9}
  {\path{doi:10.1140/epjc/s10052-020-8251-9}}.

\bibitem{ATL-PHYS-PUB-2020-018}
{ATLAS Collaboration}, \href{http://cdsweb.cern.ch/record/2724632}{{Deep
  Learning for Pion Identification and Energy Calibration with the ATLAS
  Detector}}, ATL-PHYS-PUB-2020-018 (2020).
\newline\urlprefix\url{http://cdsweb.cern.ch/record/2724632}

\bibitem{MicroBooNE:2020hho}
P.~Abratenko, et~al., {Convolutional neural network for multiple particle
  identification in the MicroBooNE liquid argon time projection chamber}, Phys.
  Rev. D 103 (2021) 092003.
\newblock \href {http://arxiv.org/abs/2010.08653} {\path{arXiv:2010.08653}},
  \href {https://doi.org/10.1103/PhysRevD.103.092003}
  {\path{doi:10.1103/PhysRevD.103.092003}}.

\bibitem{Aurisano:2016jvx}
A.~Aurisano, A.~Radovic, D.~Rocco, A.~Himmel, M.~D. Messier, E.~Niner,
  G.~Pawloski, F.~Psihas, A.~Sousa, P.~Vahle, {A Convolutional Neural Network
  Neutrino Event Classifier}, JINST 11 (2016) P09001.
\newblock \href {http://arxiv.org/abs/1604.01444} {\path{arXiv:1604.01444}},
  \href {https://doi.org/10.1088/1748-0221/11/09/P09001}
  {\path{doi:10.1088/1748-0221/11/09/P09001}}.

\bibitem{CMS:2019uxx}
A.~M. Sirunyan, et~al., {A Deep Neural Network for Simultaneous Estimation of b
  Jet Energy and Resolution}, Comput. Softw. Big Sci. 4 (2020) 10.
\newblock \href {http://arxiv.org/abs/1912.06046} {\path{arXiv:1912.06046}},
  \href {https://doi.org/10.1007/s41781-020-00041-z}
  {\path{doi:10.1007/s41781-020-00041-z}}.

\bibitem{ATL-PHYS-PUB-2018-013}
\href{http://cds.cern.ch/record/2630972}{{Generalized Numerical Inversion: A
  Neural Network Approach to Jet Calibration}}, Tech. Rep.
  ATL-PHYS-PUB-2018-013, CERN, Geneva (Jul 2018).
\newline\urlprefix\url{http://cds.cern.ch/record/2630972}

\bibitem{ATL-PHYS-PUB-2020-001}
\href{http://cds.cern.ch/record/2706189}{{Simultaneous Jet Energy and Mass
  Calibrations with Neural Networks}}, Tech. Rep. ATL-PHYS-PUB-2020-001, CERN,
  Geneva (Jan 2020).
\newline\urlprefix\url{http://cds.cern.ch/record/2706189}

\bibitem{Baldi:2020hjm}
P.~Baldi, L.~Blecher, A.~Butter, J.~Collado, J.~N. Howard, F.~Keilbach,
  T.~Plehn, G.~Kasieczka, D.~Whiteson, {How to GAN Higher Jet Resolution} (Dec
  2020).
\newblock \href {http://arxiv.org/abs/2012.11944} {\path{arXiv:2012.11944}}.

\bibitem{Kasieczka:2020vlh}
F.~O. T.~P. G.~Kasieczka, M.~Luchmann, {Per-Object Systematics using
  Deep-Learned Calibration}, SciPost Phys. 9 (2020) 089.
\newblock \href {http://arxiv.org/abs/2003.11099} {\path{arXiv:2003.11099}},
  \href {https://doi.org/10.21468/SciPostPhys.9.6.089}
  {\path{doi:10.21468/SciPostPhys.9.6.089}}.

\bibitem{1910.03773}
{S. Cheong and A. Cukierman and B. Nachman and M. Safdari and A. Schwartzman},
  {Parametrizing the Detector Response with Neural Networks}, JINST 15 (2020)
  P01030.
\newblock \href {http://arxiv.org/abs/1910.03773} {\path{arXiv:1910.03773}},
  \href {https://doi.org/10.1088/1748-0221/15/01/P01030}
  {\path{doi:10.1088/1748-0221/15/01/P01030}}.

\bibitem{deOliveira:2015xxd}
L.~de~Oliveira, M.~Kagan, L.~Mackey, B.~Nachman, A.~Schwartzman, {Jet-images
  — deep learning edition}, JHEP 07 (2016) 069.
\newblock \href {http://arxiv.org/abs/1511.05190} {\path{arXiv:1511.05190}},
  \href {https://doi.org/10.1007/JHEP07(2016)069}
  {\path{doi:10.1007/JHEP07(2016)069}}.

\bibitem{CMS:2019dqq}
A.~M. Sirunyan, et~al., {A deep neural network to search for new long-lived
  particles decaying to jets}, Mach. Learn. Sci. Tech. 1 (2020) 035012.
\newblock \href {http://arxiv.org/abs/1912.12238} {\path{arXiv:1912.12238}},
  \href {https://doi.org/10.1088/2632-2153/ab9023}
  {\path{doi:10.1088/2632-2153/ab9023}}.

\bibitem{ATLAS:2018wis}
M.~Aaboud, et~al., {Performance of top-quark and $W$-boson tagging with ATLAS
  in Run 2 of the LHC}, Eur. Phys. J. C 79 (2019) 375.
\newblock \href {http://arxiv.org/abs/1808.07858} {\path{arXiv:1808.07858}},
  \href {https://doi.org/10.1140/epjc/s10052-019-6847-8}
  {\path{doi:10.1140/epjc/s10052-019-6847-8}}.

\bibitem{Kasieczka:2019dbj}
A.~Butter, et~al., {The Machine Learning Landscape of Top Taggers}, SciPost
  Phys. 7 (2019) 014.
\newblock \href {http://arxiv.org/abs/1902.09914} {\path{arXiv:1902.09914}},
  \href {https://doi.org/10.21468/SciPostPhys.7.1.014}
  {\path{doi:10.21468/SciPostPhys.7.1.014}}.

\bibitem{Andreassen:2019cjw}
A.~Andreassen, P.~T. Komiske, E.~M. Metodiev, B.~Nachman, J.~Thaler, {OmniFold:
  A Method to Simultaneously Unfold All Observables}, Phys. Rev. Lett. 124
  (2020) 182001.
\newblock \href {http://arxiv.org/abs/1911.09107} {\path{arXiv:1911.09107}},
  \href {https://doi.org/10.1103/PhysRevLett.124.182001}
  {\path{doi:10.1103/PhysRevLett.124.182001}}.

\bibitem{Datta:2018mwd}
K.~Datta, D.~Kar, D.~Roy, {Unfolding with Generative Adversarial Networks}
  (2018).
\newblock \href {http://arxiv.org/abs/1806.00433} {\path{arXiv:1806.00433}}.

\bibitem{Bellagente:2019uyp}
M.~Bellagente, A.~Butter, G.~Kasieczka, T.~Plehn, R.~Winterhalder, {How to GAN
  away Detector Effects}, SciPost Phys. 8~(4) (2020) 070.
\newblock \href {http://arxiv.org/abs/1912.00477} {\path{arXiv:1912.00477}},
  \href {https://doi.org/10.21468/SciPostPhys.8.4.070}
  {\path{doi:10.21468/SciPostPhys.8.4.070}}.

\bibitem{Glazov:2017vni}
A.~Glazov, {Machine learning as an instrument for data unfolding} (Dec 2017).
\newblock \href {http://arxiv.org/abs/1712.01814} {\path{arXiv:1712.01814}}.

\bibitem{Vandegar:2020yvw}
M.~Vandegar, M.~Kagan, A.~Wehenkel, G.~Louppe,
  \href{https://proceedings.mlr.press/v130/vandegar21a.html}{{Neural Empirical
  Bayes: Source Distribution Estimation and its Applications to
  Simulation-Based Inference}}, in: A.~Banerjee, K.~Fukumizu (Eds.),
  {Proceedings of The 24th International Conference on Artificial Intelligence
  and Statistics}, Vol. 130 of Proceedings of Machine Learning Research, PMLR,
  2021, pp. 2107--2115.
\newblock \href {http://arxiv.org/abs/2011.05836} {\path{arXiv:2011.05836}}.
\newline\urlprefix\url{https://proceedings.mlr.press/v130/vandegar21a.html}

\bibitem{Howard:2021pos}
J.~N. Howard, S.~Mandt, D.~Whiteson, Y.~Yang, {Foundations of a Fast,
  Data-Driven, Machine-Learned Simulator} (Jan 2021).
\newblock \href {http://arxiv.org/abs/2101.08944} {\path{arXiv:2101.08944}}.

\bibitem{Baron:2021vvl}
P.~Baro\v{n}, {Comparison of Machine Learning Approach to other Unfolding
  Methods} (Apr 2021).
\newblock \href {http://arxiv.org/abs/2104.03036} {\path{arXiv:2104.03036}}.

\bibitem{Andreassen:2021zzk}
A.~Andreassen, P.~T. Komiske, E.~M. Metodiev, B.~Nachman, A.~Suresh, J.~Thaler,
  {Scaffolding Simulations with Deep Learning for High-dimensional
  Deconvolution} (May 2021).
\newblock \href {http://arxiv.org/abs/2105.04448} {\path{arXiv:2105.04448}}.

\bibitem{H1:2021wkz}
V.~Andreev, et~al., {Measurement of lepton-jet correlation in deep-inelastic
  scattering with the H1 detector using machine learning for unfolding} (Aug
  2021).
\newblock \href {http://arxiv.org/abs/2108.12376} {\path{arXiv:2108.12376}}.

\bibitem{2102.02770}
{M. Feickert and B. Nachman}, {A Living Review of Machine Learning for Particle
  Physics} (2021).
\newblock \href {http://arxiv.org/abs/2102.02770} {\path{arXiv:2102.02770}}.

\bibitem{Larkoski:2017jix}
A.~J. Larkoski, I.~Moult, B.~Nachman, {Jet Substructure at the Large Hadron
  Collider: A Review of Recent Advances in Theory and Machine Learning}, Phys.
  Rept. 841 (2020) 1--63.
\newblock \href {http://arxiv.org/abs/1709.04464} {\path{arXiv:1709.04464}},
  \href {https://doi.org/10.1016/j.physrep.2019.11.001}
  {\path{doi:10.1016/j.physrep.2019.11.001}}.

\bibitem{Guest:2018yhq}
D.~Guest, K.~Cranmer, D.~Whiteson, {Deep Learning and its Application to LHC
  Physics} (2018).
\newblock \href {http://arxiv.org/abs/1806.11484} {\path{arXiv:1806.11484}},
  \href {https://doi.org/10.1146/annurev-nucl-101917-021019}
  {\path{doi:10.1146/annurev-nucl-101917-021019}}.

\bibitem{Albertsson:2018maf}
K.~Albertsson, et~al., {Machine Learning in High Energy Physics Community White
  Paper} (2018).
\newblock \href {http://arxiv.org/abs/1807.02876} {\path{arXiv:1807.02876}},
  \href {https://doi.org/10.1088/1742-6596/1085/2/022008}
  {\path{doi:10.1088/1742-6596/1085/2/022008}}.

\bibitem{Carleo:2019ptp}
G.~Carleo, I.~Cirac, K.~Cranmer, L.~Daudet, M.~Schuld, N.~Tishby,
  L.~Vogt-Maranto, L.~Zdeborová, {Machine learning and the physical sciences},
  Rev. Mod. Phys. 91 (2019) 045002.
\newblock \href {http://arxiv.org/abs/1903.10563} {\path{arXiv:1903.10563}},
  \href {https://doi.org/10.1103/RevModPhys.91.045002}
  {\path{doi:10.1103/RevModPhys.91.045002}}.

\bibitem{Bourilkov:2019yoi}
D.~Bourilkov, {Machine and Deep Learning Applications in Particle Physics},
  Int. J. Mod. Phys. A 34 (2020) 1930019.
\newblock \href {http://arxiv.org/abs/1912.08245} {\path{arXiv:1912.08245}},
  \href {https://doi.org/10.1142/S0217751X19300199}
  {\path{doi:10.1142/S0217751X19300199}}.

\bibitem{Diefenthaler:2021rdj}
M.~Diefenthaler, A.~Farhat, A.~Verbytskyi, Y.~Xu, {Deeply Learning Deep
  Inelastic Scattering Kinematics} (Aug 2021).
\newblock \href {http://arxiv.org/abs/2108.11638} {\path{arXiv:2108.11638}}.

\bibitem{Klein:2008di}
M.~Klein, R.~Yoshida, {Collider Physics at HERA}, Prog. Part. Nucl. Phys. 61
  (2008) 343.
\newblock \href {http://arxiv.org/abs/0805.3334} {\path{arXiv:0805.3334}},
  \href {https://doi.org/10.1016/j.ppnp.2008.05.002}
  {\path{doi:10.1016/j.ppnp.2008.05.002}}.

\bibitem{H1:2015ubc}
H.~Abramowicz, et~al., {Combination of measurements of inclusive deep inelastic
  ${e^{\pm }p}$ scattering cross sections and QCD analysis of HERA data}, Eur.
  Phys. J. C 75 (2015).
\newblock \href {http://arxiv.org/abs/1506.06042} {\path{arXiv:1506.06042}},
  \href {https://doi.org/10.1140/epjc/s10052-015-3710-4}
  {\path{doi:10.1140/epjc/s10052-015-3710-4}}.

\bibitem{H1:2009bcq}
F.~D. Aaron, et~al., {A Precision Measurement of the Inclusive ep Scattering
  Cross Section at HERA}, Eur. Phys. J. C 64 (2009) 561.
\newblock \href {http://arxiv.org/abs/0904.3513} {\path{arXiv:0904.3513}},
  \href {https://doi.org/10.1140/epjc/s10052-009-1169-x}
  {\path{doi:10.1140/epjc/s10052-009-1169-x}}.

\bibitem{H1:2012qti}
F.~D. Aaron, et~al., {Inclusive Deep Inelastic Scattering at High $Q^2$ with
  Longitudinally Polarised Lepton Beams at HERA}, JHEP 09 (2012) 061.
\newblock \href {http://arxiv.org/abs/1206.7007} {\path{arXiv:1206.7007}},
  \href {https://doi.org/10.1007/JHEP09(2012)061}
  {\path{doi:10.1007/JHEP09(2012)061}}.

\bibitem{H1:2013ktq}
V.~Andreev, et~al., {Measurement of inclusive $e p$ cross sections at high
  $Q^2$ at $\sqrt s =$ 225 and 252 GeV and of the longitudinal proton structure
  function $F_L$ at HERA}, Eur. Phys. J. C 74 (2014) 2814.
\newblock \href {http://arxiv.org/abs/1312.4821} {\path{arXiv:1312.4821}},
  \href {https://doi.org/10.1140/epjc/s10052-014-2814-6}
  {\path{doi:10.1140/epjc/s10052-014-2814-6}}.

\bibitem{Spiesberger:237380}
H.~Spiesberger, et~al., \href{https://cds.cern.ch/record/237380}{{Radiative
  corrections at HERA}} (1992) 798--839.
\newline\urlprefix\url{https://cds.cern.ch/record/237380}

\bibitem{Kwiatkowski:1990cx}
A.~Kwiatkowski, H.~Spiesberger, H.~J. Mohring, {Characteristics of radiative
  events in deep inelastic \emph{ep} scattering at HERA}, Z. Phys. C 50 (1991)
  165--178.
\newblock \href {https://doi.org/10.1007/BF01558572}
  {\path{doi:10.1007/BF01558572}}.

\bibitem{Kwiatkowski:1990es}
A.~Kwiatkowski, H.~Spiesberger, H.~J. Mohring, {Heracles: An Event Generator
  for $e p$ Interactions at {HERA} Energies Including Radiative Processes:
  Version 1.0}, Comput. Phys. Commun. 69 (1992) 155--172.
\newblock \href {https://doi.org/10.1016/0010-4655(92)90136-M}
  {\path{doi:10.1016/0010-4655(92)90136-M}}.

\bibitem{Blumlein:1994ii}
J.~Blumlein, {${\cal O} (\alpha^2 L^2)$ radiative corrections to deep inelastic
  e p scattering for different kinematical variables}, Z. Phys. C 65 (1995)
  293--298.
\newblock \href {http://arxiv.org/abs/hep-ph/9403342}
  {\path{arXiv:hep-ph/9403342}}, \href {https://doi.org/10.1007/BF01571886}
  {\path{doi:10.1007/BF01571886}}.

\bibitem{Arbuzov:1995id}
A.~Arbuzov, D.~Y. Bardin, J.~Blumlein, L.~Kalinovskaya, T.~Riemann, {Hector
  1.00: A Program for the calculation of QED, QCD and electroweak corrections
  to e p and lepton+- N deep inelastic neutral and charged current scattering},
  Comput. Phys. Commun. 94 (1996) 128--184.
\newblock \href {http://arxiv.org/abs/hep-ph/9511434}
  {\path{arXiv:hep-ph/9511434}}, \href
  {https://doi.org/10.1016/0010-4655(96)00005-7}
  {\path{doi:10.1016/0010-4655(96)00005-7}}.

\bibitem{tensorflow}
M.~Abadi, P.~Barham, J.~Chen, Z.~Chen, A.~Davis, J.~Dean, M.~Devin,
  S.~Ghemawat, G.~Irving, M.~Isard, et~al., Tensorflow: A system for
  large-scale machine learning., in: OSDI, Vol.~16, 2016, pp. 265--283.

\bibitem{deFavereau:2013fsa}
J.~de~Favereau, C.~Delaere, P.~Demin, A.~Giammanco, V.~Lema\^\i{}tre,
  A.~Mertens, M.~Selvaggi, {DELPHES 3, A modular framework for fast simulation
  of a generic collider experiment}, JHEP 02 (2014) 057.
\newblock \href {http://arxiv.org/abs/1307.6346} {\path{arXiv:1307.6346}},
  \href {https://doi.org/10.1007/JHEP02(2014)057}
  {\path{doi:10.1007/JHEP02(2014)057}}.

\bibitem{miguel_arratia_2021_4592887}
M.~Arratia, S.~Sekula, \href{https://doi.org/10.5281/zenodo.4592887}{A delphes
  card for the eic yellow-report detector} ({Mar} 2021).
\newblock \href {https://doi.org/10.5281/zenodo.4592887}
  {\path{doi:10.5281/zenodo.4592887}}.
\newline\urlprefix\url{https://doi.org/10.5281/zenodo.4592887}

\bibitem{Brun:1987ma}
R.~Brun, F.~Bruyant, M.~Maire, A.~C. McPherson, P.~Zanarini, {GEANT3},
  CERN-DD-EE-84-01 (Sep 1987).

\bibitem{H1:1996prr}
I.~Abt, et~al., {The H1 detector at HERA}, Nucl. Instrum. Meth. A 386 (1997)
  310--347.
\newblock \href {https://doi.org/10.1016/S0168-9002(96)00893-5}
  {\path{doi:10.1016/S0168-9002(96)00893-5}}.

\bibitem{H1:1996jzy}
I.~Abt, et~al., {The Tracking, calorimeter and muon detectors of the H1
  experiment at HERA}, Nucl. Instrum. Meth. A 386 (1997) 348--396.
\newblock \href {https://doi.org/10.1016/S0168-9002(96)00894-7}
  {\path{doi:10.1016/S0168-9002(96)00894-7}}.

\bibitem{Jung:1993gf}
H.~Jung, {Hard diffractive scattering in high-energy e p collisions and the
  Monte Carlo generator RAPGAP}, Comput. Phys. Commun. 86 (1995) 147--161.
\newblock \href {https://doi.org/10.1016/0010-4655(94)00150-Z}
  {\path{doi:10.1016/0010-4655(94)00150-Z}}.

\bibitem{Bengtsson:1987kr}
H.-U. Bengtsson, T.~Sjostrand, {The Lund Monte Carlo for Hadronic Processes:
  Pythia Version 4.8}, Comput. Phys. Commun. 46 (1987) 43.
\newblock \href {https://doi.org/10.1016/0010-4655(87)90036-1}
  {\path{doi:10.1016/0010-4655(87)90036-1}}.

\bibitem{Ingelman:1996mq}
G.~Ingelman, A.~Edin, J.~Rathsman, {LEPTO 6.5: A Monte Carlo generator for deep
  inelastic lepton - nucleon scattering}, Comput. Phys. Commun. 101 (1997)
  108--134.
\newblock \href {http://arxiv.org/abs/hep-ph/9605286}
  {\path{arXiv:hep-ph/9605286}}, \href
  {https://doi.org/10.1016/S0010-4655(96)00157-9}
  {\path{doi:10.1016/S0010-4655(96)00157-9}}.

\bibitem{Dobbs:2001ck}
M.~Dobbs, J.~B. Hansen, {The HepMC C++ Monte Carlo event record for High Energy
  Physics}, Comput. Phys. Commun. 134 (2001) 41--46.
\newblock \href {https://doi.org/10.1016/S0010-4655(00)00189-2}
  {\path{doi:10.1016/S0010-4655(00)00189-2}}.

\bibitem{Charchula:1994kf}
K.~Charchula, G.~A. Schuler, H.~Spiesberger, {Combined QED and QCD radiative
  effects in deep inelastic lepton - proton scattering: The Monte Carlo
  generator DJANGO6}, Comput. Phys. Commun. 81 (1994) 381--402.
\newblock \href {https://doi.org/10.1016/0010-4655(94)90086-8}
  {\path{doi:10.1016/0010-4655(94)90086-8}}.

\bibitem{scikit-learn}
F.~Pedregosa, G.~Varoquaux, A.~Gramfort, V.~Michel, B.~Thirion, O.~Grisel,
  M.~Blondel, P.~Prettenhofer, R.~Weiss, V.~Dubourg, J.~Vanderplas, A.~Passos,
  D.~Cournapeau, M.~Brucher, M.~Perrot, E.~Duchesnay, Scikit-learn: Machine
  learning in {P}ython, Journal of Machine Learning Research 12 (2011)
  2825--2830.

\bibitem{NIPS2017_5d44ee6f}
G.~Klambauer, T.~Unterthiner, A.~Mayr, S.~Hochreiter,
  \href{https://proceedings.neurips.cc/paper/2017/file/5d44ee6f2c3f71b73125876103c8f6c4-Paper.pdf}{Self-normalizing
  neural networks}, in: I.~Guyon, U.~V. Luxburg, S.~Bengio, H.~Wallach,
  R.~Fergus, S.~Vishwanathan, R.~Garnett (Eds.), Advances in Neural Information
  Processing Systems, Vol.~30, Curran Associates, Inc., 2017.
\newline\urlprefix\url{https://proceedings.neurips.cc/paper/2017/file/5d44ee6f2c3f71b73125876103c8f6c4-Paper.pdf}

\bibitem{kingma2017adam}
D.~P. Kingma, J.~Ba, Adam: A method for stochastic optimization (2017).
\newblock \href {http://arxiv.org/abs/1412.6980} {\path{arXiv:1412.6980}}.

\bibitem{10.1214/aoms/1177703732}
P.~J. Huber, \href{https://doi.org/10.1214/aoms/1177703732}{{Robust Estimation
  of a Location Parameter}}, The Annals of Mathematical Statistics 35 (1964) 73
  -- 101.
\newblock \href {https://doi.org/10.1214/aoms/1177703732}
  {\path{doi:10.1214/aoms/1177703732}}.
\newline\urlprefix\url{https://doi.org/10.1214/aoms/1177703732}

\bibitem{Pumplin:2002vw}
J.~Pumplin, D.~R. Stump, J.~Huston, H.~L. Lai, P.~M. Nadolsky, W.~K. Tung, {New
  generation of parton distributions with uncertainties from global QCD
  analysis}, JHEP 07 (2002) 012.
\newblock \href {http://arxiv.org/abs/hep-ph/0201195}
  {\path{arXiv:hep-ph/0201195}}, \href
  {https://doi.org/10.1088/1126-6708/2002/07/012}
  {\path{doi:10.1088/1126-6708/2002/07/012}}.

\bibitem{Andersson:1983ia}
{B. Andersson, G. Gustafson, G. Ingelman, and T. Sj\"ostrand}, {Parton
  fragmentation and string dynamics}, Phys. Rept. 97 (1983) 31--145.
\newblock \href {https://doi.org/10.1016/0370-1573(83)90080-7}
  {\path{doi:10.1016/0370-1573(83)90080-7}}.

\bibitem{Schael:2004ux}
S.~Schael, et~al., {Bose-Einstein correlations in W-pair decays with an
  event-mixing technique}, Phys. Lett. B 606 (2005) 265--275.
\newblock \href {https://doi.org/10.1016/j.physletb.2004.12.018}
  {\path{doi:10.1016/j.physletb.2004.12.018}}.

\bibitem{Fesefeldt:1985yw}
H.~Fesefeldt, {The Simulation of Hadronic Showers: Physics and Applications},
  PITHA-85-02 (Dec 1985).

\bibitem{Grindhammer:1989zg}
G.~Grindhammer, M.~Rudowicz, S.~Peters, {The Fast Simulation of Electromagnetic
  and Hadronic Showers}, Nucl. Instrum. Meth. A 290 (1990) 469.
\newblock \href {https://doi.org/10.1016/0168-9002(90)90566-O}
  {\path{doi:10.1016/0168-9002(90)90566-O}}.

\bibitem{Gayler:1991cr}
J.~Gayler, {Simulation of H1 calorimeter test data with GHEISHA and FLUKA}, in:
  {Workshop on Detector and Event Simulation in High-energy Physics (MC '91)},
  1991, p. 312.

\bibitem{Kuhlen:1992ey}
M.~Kuhlen, {The Fast H1 detector Monte Carlo}, in: {26th International
  Conference on High-energy Physics}, 1992, pp. 1787--1790.
\newblock \href {https://doi.org/10.1063/1.43288} {\path{doi:10.1063/1.43288}}.

\bibitem{Grindhammer:1993kw}
G.~Grindhammer, S.~Peters, {The Parameterized simulation of electromagnetic
  showers in homogeneous and sampling calorimeters}, in: {International
  Conference on Monte Carlo Simulation in High-Energy and Nuclear Physics (MC
  '93)}, 1993.
\newblock \href {http://arxiv.org/abs/hep-ex/0001020}
  {\path{arXiv:hep-ex/0001020}}.

\bibitem{Glazov:2010zza}
A.~Glazov, N.~Raicevic, A.~Zhokin, {Fast simulation of showers in the H1
  calorimeter}, Comput. Phys. Commun. 181 (2010) 1008--1012.
\newblock \href {https://doi.org/10.1016/j.cpc.2010.02.004}
  {\path{doi:10.1016/j.cpc.2010.02.004}}.

\bibitem{energyflowthesis}
M.~Peez, {Search for deviations from the standard model in high transverse
  energy processes at the electron proton collider HERA. (Thesis, Univ. Lyon)},
  Ph.D. thesis (Jun 2003).

\bibitem{energyflowthesis2}
S.~Hellwig, {Untersuchung der $D^*$ - $\pi$ slow Double Tagging Methode in
  Charmanalysen}, {Diploma thesis, Univ. Hamburg} (Jun 2004).

\bibitem{energyflowthesis3}
B.~Portheault, {First measurement of charged and neutral current cross sections
  with the polarized positron beam at HERA II and QCD-electroweak analyses.
  (Thesis, Univ. Paris XI)}, Ph.D. thesis (Mar 2005).

\bibitem{Andreev:2014wwa}
V.~Andreev, et~al., {Measurement of multijet production in $ep$ collisions at
  high $Q^2$ and determination of the strong coupling $\alpha _s$}, Eur. Phys.
  J. C 75 (2015) 65.
\newblock \href {http://arxiv.org/abs/1406.4709} {\path{arXiv:1406.4709}},
  \href {https://doi.org/10.1140/epjc/s10052-014-3223-6}
  {\path{doi:10.1140/epjc/s10052-014-3223-6}}.

\bibitem{Kogler:2011zz}
R.~Kogler, {Measurement of jet production in deep-inelastic e p scattering at
  HERA}, Phd thesis (Feb 2011).
\newblock \href {https://doi.org/10.3204/DESY-THESIS-2011-003}
  {\path{doi:10.3204/DESY-THESIS-2011-003}}.

\bibitem{Britzger:2021xcx}
D.~Britzger, S.~Levonian, S.~Schmitt, D.~South, {Preservation through
  modernisation: The software of the H1 experiment at HERA}, EPJ Web Conf. 251
  (2021) 03004.
\newblock \href {http://arxiv.org/abs/2106.11058} {\path{arXiv:2106.11058}},
  \href {https://doi.org/10.1051/epjconf/202125103004}
  {\path{doi:10.1051/epjconf/202125103004}}.

\bibitem{energyflow}
M.~Peez, et~al., An energy flow algorithm for hadronic reconstruction in oo:
  Hadroo2, H1-Internal Note, vol. H1-01/05-616, DESY, 2005 (2005).

\bibitem{citeulike:363715}
R.~Brun, F.~Rademakers, Root - an object oriented data analysis framework, in:
  AIHENP'96 Workshop, Lausane, Vol. 389, 1996, pp. 81--86.

\end{thebibliography}

\appendix




\end{document}